\documentclass{emulateapj}

\pdfoutput=1

\newcommand {\apgt} {\ {\raise-.5ex\hbox{$\buildrel>\over\sim$}}\ }
\newcommand {\aplt} {\ {\raise-.5ex\hbox{$\buildrel<\over\sim$}}\ }

\shorttitle{Measurement of Galactic Logarithmic Spiral Arm Pitch Angle}
\shortauthors{Davis et al. (2012)}

\usepackage{color}
\usepackage{url}
\usepackage[latin1]{inputenc}
\usepackage{tikz}
\usetikzlibrary{shapes,arrows}

\submitted{Published in \apjs, 199, 33}

\begin{document}

\title{Measurement of Galactic Logarithmic Spiral Arm Pitch Angle Using Two-Dimensional Fast Fourier Transform Decomposition}

\author{Benjamin L. Davis}
\affil{Arkansas Center for Space and Planetary Sciences, 202 Field House, University of Arkansas, Fayetteville, AR 72701, USA}
\author{Joel C. Berrier, Douglas W. Shields, Julia Kennefick, and Daniel Kennefick}
\affil{Department of Physics, University of Arkansas, 835 West Dickson Street, Fayetteville, AR 72701, USA and Arkansas Center for Space and Planetary Sciences, 202 Field House, University of Arkansas, Fayetteville, AR 72701, USA}
\author{Marc S. Seigar}
\affil{Department of Physics and Astronomy, University of Arkansas at Little Rock, 2801 South University Avenue, Little Rock, AR 72204, USA and Arkansas Center for Space and Planetary Sciences, 202 Field House, University of Arkansas, Fayetteville, AR 72701, USA}
\author{Claud H. S. Lacy}
\affil{Department of Physics, University of Arkansas, 835 West Dickson Street, Fayetteville, AR 72701, USA and Arkansas Center for Space and Planetary Sciences, 202 Field House, University of Arkansas, Fayetteville, AR 72701, USA}
\author{Iv\^anio Puerari}
\affil{Instituto Nacional de Astrof\'isica, Optica y Electr\'onica, Calle Luis Enrique Erro 1, 72840 Santa Mar\'ia Tonantzintla, Puebla, Mexico}

\begin{abstract}
  A logarithmic spiral is a prominent feature appearing in a majority
  of  observed galaxies.   This feature has  long been
  associated  with the traditional  Hubble classification  scheme, but
  historical quotes of pitch  angle  of   spiral  galaxies  have  been  almost
  exclusively   qualitative.    We   have developed a methodology, utilizing two-dimensional  fast Fourier  transformations of images  of spiral
  galaxies, in order  to isolate and measure the  pitch angles of their
  spiral arms.  Our technique provides  a quantitative way  to measure
  this  morphological  feature. This will allow comparison of spiral galaxy pitch angle to other galactic parameters and test spiral arm genesis theories. In  this  work,  we  detail our  image
  processing  and  analysis  of  spiral  galaxy images  and  discuss  the
  robustness of our analysis techniques.
\end{abstract}


\section{Introduction}

Approximately 60\% of galaxies in the local Universe are spiral \citep{Buta:1989}.  A considerable number of these spiral galaxies show Grand Design
structure, where the spiral pattern is uniform and spans the entire disc
of the galaxy.  In these galaxies, the spiral pattern is often
logarithmic in nature \citep{Seigar:James:1998}, and so their appearance is scale independent.
The best geometric measure for logarithmic spirals is the pitch angle,
and this can be measured for any galaxy in which spiral structure can be
discerned, independently of the distance to the galaxy.

It is worth noting that spiral structure has been observed for over 150 years\footnote{Spiral structure in galaxies was observed as early as 1845 \citep[Lord Rosse's sketch of M51;][]{Herschel:1859}.} with no clear standard for quantitative measurement having emerged, even though it
correlates well with other important features of galaxies, such as
central supermassive black hole (SMBH) mass \citep{Seigar:2008}.
Furthermore, spiral arm pitch angle could serve as a means to
discriminate between rival theories for the formation of spiral
structure in galactic discs. The aim of this paper is to present one such method, based on a
Two-Dimensional (2-D) Fast Fourier Transform (FFT) algorithm\footnote{This code is publicly available for use at \url{http://dafix.uark.edu/~ages/downloads.html} and \url{http://astro.host.ualr.edu/2DFFT/}.}, which decomposes images into
spirals of different pitch angles and numbers of arms.

A long-standing and quite successful theory of spiral structure in
galaxies is the quasi-stationary density wave model \citep{Lin:Shu:1964}.
As gas enters this density wave, it is compressed to a density at which
stars can form \citep{Roberts:1969,Shu:1972}.  These star forming
regions, along with stars and gas, conglomerate together into spiral
arms in the disc regions of spiral galaxies, with star forming regions
found on the leading edges of arms, and dust (in the form of dust lanes)
seen on their trailing edges.

A recently formulated rival theory proposes that spiral arms are
composed of identifiable groups of stars in highly eccentric and chaotic
orbits, which originate near the ends of galactic bars.  These orbits,
though chaotic, keep the stars grouped in relatively narrow tubes known
as manifolds, which are responsible for the observed spiral structure
\citep{Athanassoula:2009a, Athanassoula:2009b, Athanassoula:2010}.  According to this theory,
galaxies with stronger bar potentials should have more open spiral
structure. In a recent study of 27 galaxies, \citet{Martinez-Garcia:2011}
found that $\approx60\%$ of galaxies corroborate this theory and that galaxies in which the spiral arms maintain a 
logarithmic shape for azimuthal ranges greater than $70^{\circ}$ seem to corroborate the 
predicted trend.

Spiral galaxies are classified into three main types of spiral structure: grand
design, flocculent \citep{Elmegreen:1981}, and multi-armed. Grand design spirals are well-defined two-armed galaxies and theoretical efforts have naturally focused on explaining these very striking patterns. Flocculent spirals are less regular with sporadic spiral arm segments. It has been proposed that the origins of this kind of spiral are quite different from grand design spirals, the products of stochastic self-propagating star formation being acted upon by the differential rotation of the disc to create segments with the appearance of spiral arms \citep{Seiden:Gerola:1982}. Multi-arm spirals have distinct spiral arms, not necessarily symmetrically spaced. It is likely that they formed as a result of galaxy harassment \citep[frequent high speed galaxy encounters within clusters;][]{Moore:1996}. Spiral arm generation from external forces has been proposed to explain the genesis of multi-arm spiral structure in our own Milky Way \citep[e.g.,][]{Purcell:2011}.

In general,  logarithmic spirals are good approximations  of the shape
of   galactic  spiral  arms   \citep{Seigar:James:1998}.   Logarithmic
spirals are defined in polar coordinates as
\begin{equation}
r = r_0e^{\theta\tan(\phi)}
\label{eqn5}
\end{equation}
where $r$ is  the radius, $\theta$ is the central  angle, $r_0$ is the
initial radius  when $\theta  = 0^{\circ}$,  and $-90^{\circ}  \leq  \phi \leq
90^{\circ}$  is  the pitch  angle.  The limits for  the
absolute value  of the pitch  angle are $0^{\circ}$  and $90^{\circ}$,
which produce a circle and a line, respectively. Pitch angle is defined
as the angle  between the line tangent to a circle  and the line tangent
to a logarithmic spiral at a specified radius. 
Small pitch angle absolute values are associated with  tightly wound spirals and high absolute values with loosely  wound spirals.  The sign  of the
pitch angle indicates the chirality of the spiral, with positive pitch
angles indicating  clockwise outward  winding and negative  pitch angles
indicating counterclockwise outward winding (as seen from a given observer's position, i.e., above or below the galactic plane).

In this paper we present  a method for determining reliable galactic spiral arm
pitch angles. Given sufficient quality images, our software can reliably measure pitch angles by iterative 2-D FFT analyses. The paper is  outlined as follows:  \S \ref{sect2}
describes the observations of the  images we use and our procedure to prepare those images for measurement through our software. 
\S \ref{subsect2.2} details the need for and the nature of our iterative adaptation to the FFT software, along with interpretation of its results. \S  \ref{sect3}  discusses how  we  determine
errors  on  our  measured   pitch  angles. \S \ref{sect4} describes our image
analysis and related tools for further image refinement and evaluation. 
Finally, in \S \ref{sect5} we present a discussion of our results and a few possible applications of the code.

\section{Observations and Data Analysis Techniques}\label{sect2}

The  galaxy  images we  use in  this  paper (unless mentioned otherwise) come  from  the
Carnegie-Irvine                      Galaxy                     Survey
\citep[CGS\footnote{\url{http://cgs.obs.carnegiescience.edu/}};][]{Ho:2011}. This is  a statistically complete, magnitude-limited
sample  of  605  bright  ($B_T  <  12.9$  mag),  Southern  ($\delta  <
0^{\circ}$)   galaxies observed using the SITe2k CCD camera (with a pixel scale of 0.259\arcsec$\,$pixel$^{-1}$) on the 2.5 m du Pont telescope at the Las Campanas Observatory in Chile. The overall quality of  the images is high, both in
terms of  resolution (median  seeing $\sim 1 \arcsec$),  field-of-view ($8.9 \arcmin
\times 8.9 \arcmin$),  and depth  (median limiting surface  brightness $\sim$
27.5, 26.9, 26.4, and 25.3 mag arcsec$^{-2}$ in the B, V, R, and
I bands, respectively). All CGS images  have been oriented to have up as
North and left as East (before we subsequently rotate images for deprojection purposes).  In this paper, we use a subset of galaxies from the CGS sample in order to test our methods. For images not included in the CGS sample, we use the NASA/IPAC Extragalactic  Database (NED)\footnote{\url{http://ned.ipac.caltech.edu/}} to acquire images.


\subsection{Two-Dimensional Fast Fourier Transformations of Galaxy Images}\label{2DFFT}

\begin{figure*}
\begin{center}
\includegraphics[width=8.97cm]{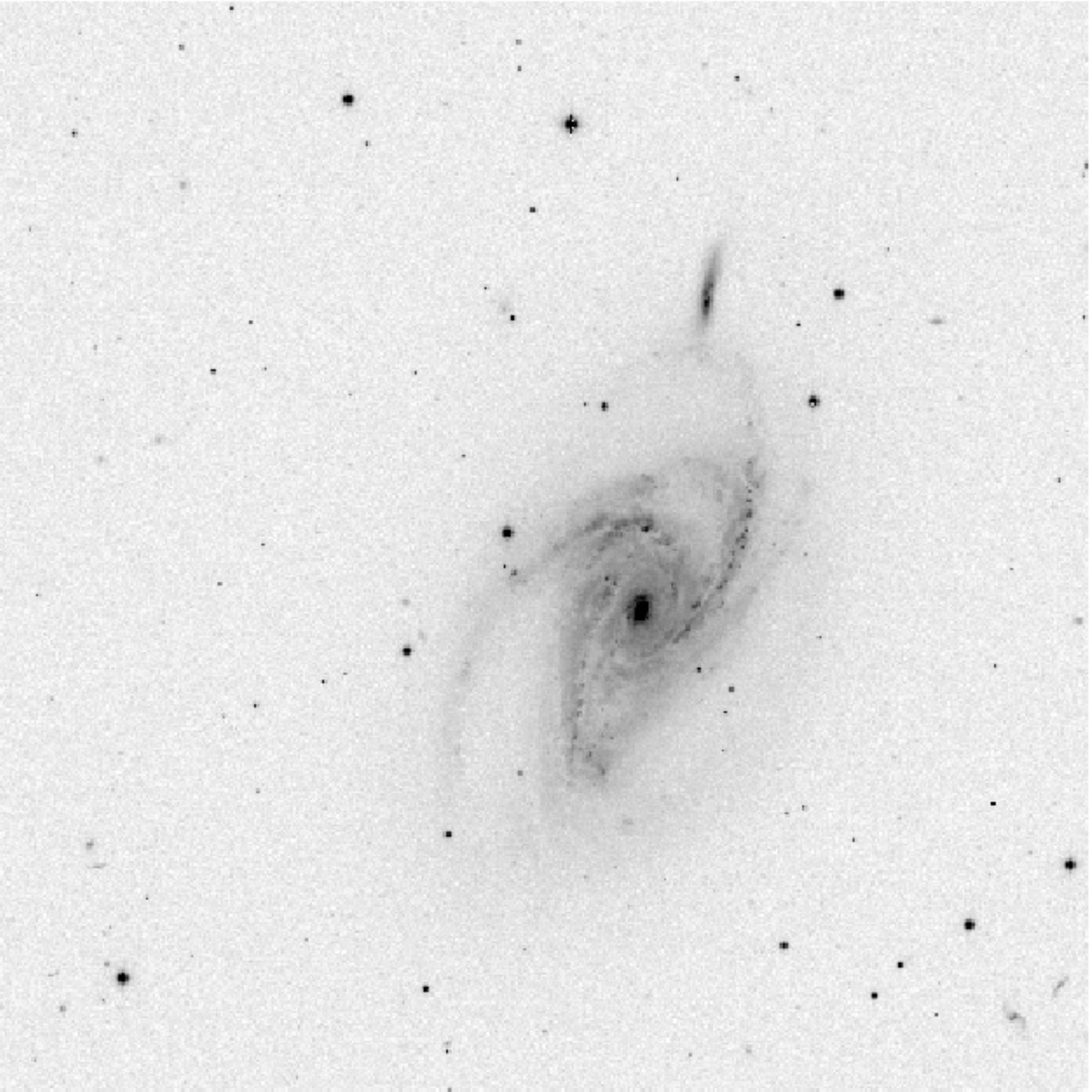}
\includegraphics[trim = 0mm 4mm 0mm 0mm, clip, width=8.97cm]{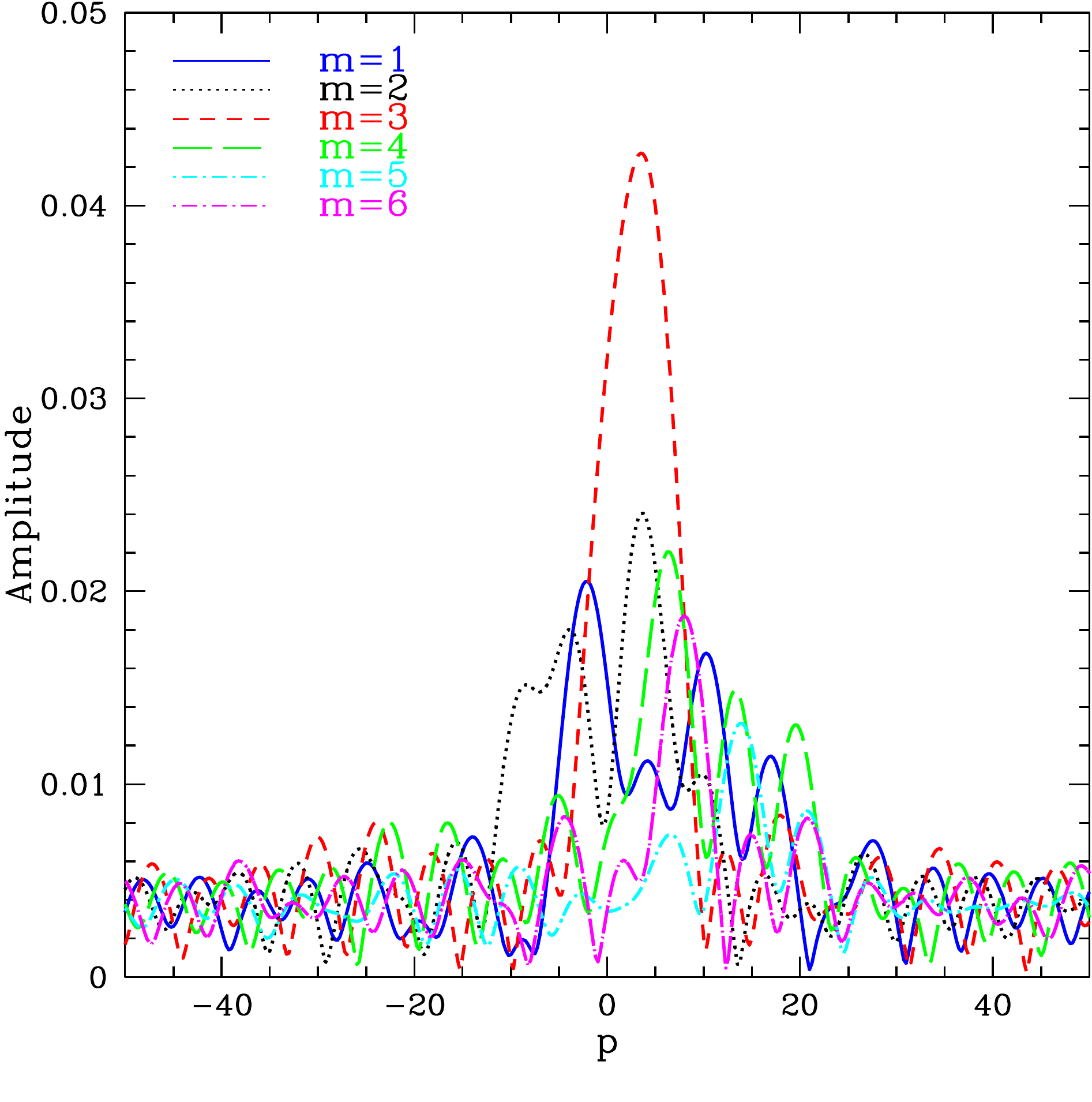}
\caption{{\it Fig. \ref{fig1}a (left)} - B-band (inverted color) image of NGC 5054. NGC 5054 is  measured to have a
  position  angle  ($PA$)  of  $160^{\circ}$ and  an  inclination  angle
  $(\alpha)$  of  $53.84^{\circ}$. {\it  Fig.  \ref{fig1}b (right)}  -
  $A(m,p)$ values for a deprojected B-band image of NGC 5054 with a
  measurement annulus defined by an  inner radius of 160 pixels (41.4\arcsec) and an
  outer radius of 508 pixels (132\arcsec). This indicates a peak in the three-armed
  spiral  harmonic mode  at $p_{max}  = 3.50$.  The  equivalent single
  value     pitch     angle     via     Equation     \ref{eqn2}     is
  $-40.60^{\circ}$ ({\it Note} - subsequent sections and figures will revise this measurement with improved methods).\label{fig1}}
\end{center}
\end{figure*}

A   program    called   {\it   2DFFT} \citep{Schroeder:1994}  accomplishes the 2-D FFT decomposition of images. {\it 2DFFT} itself is an adaptation
of  the  {\it  FOURN}  routine  from  {\it  Numerical  Recipes  in  C}
\citep{Press:1989}   that  deals   with  CCD   (charge-coupled   device)
images.  The program  is intended  to  analyze  face-on or  deprojected
galaxy  orientations.  The  decomposition  is modeled  on  logarithmic
spirals. As  pointed out by  \citet{Considere:Athanassoula:1988}, this
method does  not  assume that  observed  spiral structures  are
logarithmic. It  only decomposes the observed distributions  into a superposition of  logarithmic spirals of different pitch angles and number of arms, which  can be thought  of as
building  blocks.  This is analogous to the usual Fourier method of decomposing signals into a superposition of sinusoidal functions of different frequency. As  per  \citet{Puerari:2000},  the amplitude  of  each
Fourier component is given by
\begin{equation}
A(p, m) = \frac{1}{D}\int_{-\pi}^{+\pi}\int_{r_{min}}^{r_{max}}I(u, \theta)e^{-i(m\theta + pu)}dud\theta\label{eqn1}
\end{equation}
where $u \equiv \ln r$,  $r$ (radius) and $\theta$ (central angle) are
in polar coordinates,  $r_{min}$ is the inner radius, $r_{max}$ is
the outer radius of  the user-defined calculation  annulus, and $D$  is a
normalization factor written as
\begin{equation}
D = \int_{-\pi}^{+\pi}\int_{r_{min}}^{r_{max}}I(u, \theta)dud\theta.
\label{eqn8}
\end{equation}
$I(u, \theta)$  is the  distribution of light  of a  given deprojected
galaxy, in a $(u, \theta)$ plane, $m$ represents the number of arms or
harmonic  modes, and  $p$ is  the variable  associated with  the pitch
angle $(\phi)$, defined by
\begin{equation}\label{eqn2}
\tan(\phi) = \frac{-m}{p_{max}}
\end{equation}
with $p_{max}$ being the value of $p$ with the highest amplitude for a
given harmonic  mode (see  Figure \ref{fig1}). 
As  currently defined,
the code calculates Equation \ref{eqn1} for $0  \leq m \leq
6$.  Additionally, the  code reports  a phase  angle ($\Phi$)  for the
orientation of the spiral arm pattern, calculated as
\begin{equation}
\Phi = arctan\frac{Im[A]}{Re[A]}
\label{eqn10}
\end{equation}
where  $Im[A]$ and  $Re[A]$ are  the imaginary  and the  real  part of
$A(p, m)$, respectively.

\subsection{Image Preprocessing}\label{Image Preprocessing}

\subsubsection{Deprojection}\label{subsect2.1}

An important step in measuring the  pitch angle of  a galaxy,
regardless  of the method,  is to  deproject the  galaxy to  a face-on
orientation. This process assumes that  a galaxy with the plane of its
disc parallel to  the plane of the sky will  be circular. A circular
galaxy   with  random   inclination   appears  on   the   sky  as   an
ellipse. Thus, a circular galaxy can  be described by its  position angle
($PA$;  orientation  of  the  semi-major  axis  in  degrees  East of
North) and  its axis ratio.   In turn, the  axis ratio can  be further
incorporated to characterize the  angle of inclination $(\alpha)$ from
the plane of the sky defined by
\begin{equation}\label{eqn3}
\alpha=\arccos(b/a)
\end{equation}
where $a$ is the semi-major axis and $b$ is the semi-minor axis. Thus,
an  inclination  angle  of  $0^{\circ}$ and  $90^{\circ}$  describes  a
face-on and  an edge-on galaxy, respectively. The position angle  and axis
ratio  can be determined  easily from  images using  various programs,
e.g.,  {\it SExtractor} \citep[Source  Extractor;][]{Bertin:Arnouts:1996} or
the {\it  ELLIPSE} routine in IRAF\footnote{IRAF  (Image Reduction and
  Analysis Facility) is distributed  by the National Optical Astronomy
  Observatory  (NOAO),  which  is   operated  by  the  Association  of
  Universities  for  Research in  Astronomy  (AURA) under  cooperative
  agreement   with    the   National   Science    Foundation   (NSF).}
\citep{Tody:1986,Jedrzejewski1987}. {\it ELLIPSE} works by iteratively
fitting  isophotes  interactively  to  a galaxy  image  and  reporting
various parameters; most importantly, position angle and ellipticity
($1 - (b/a)$).

\begin{figure*}
\begin{center}
\includegraphics[width=5.95cm]{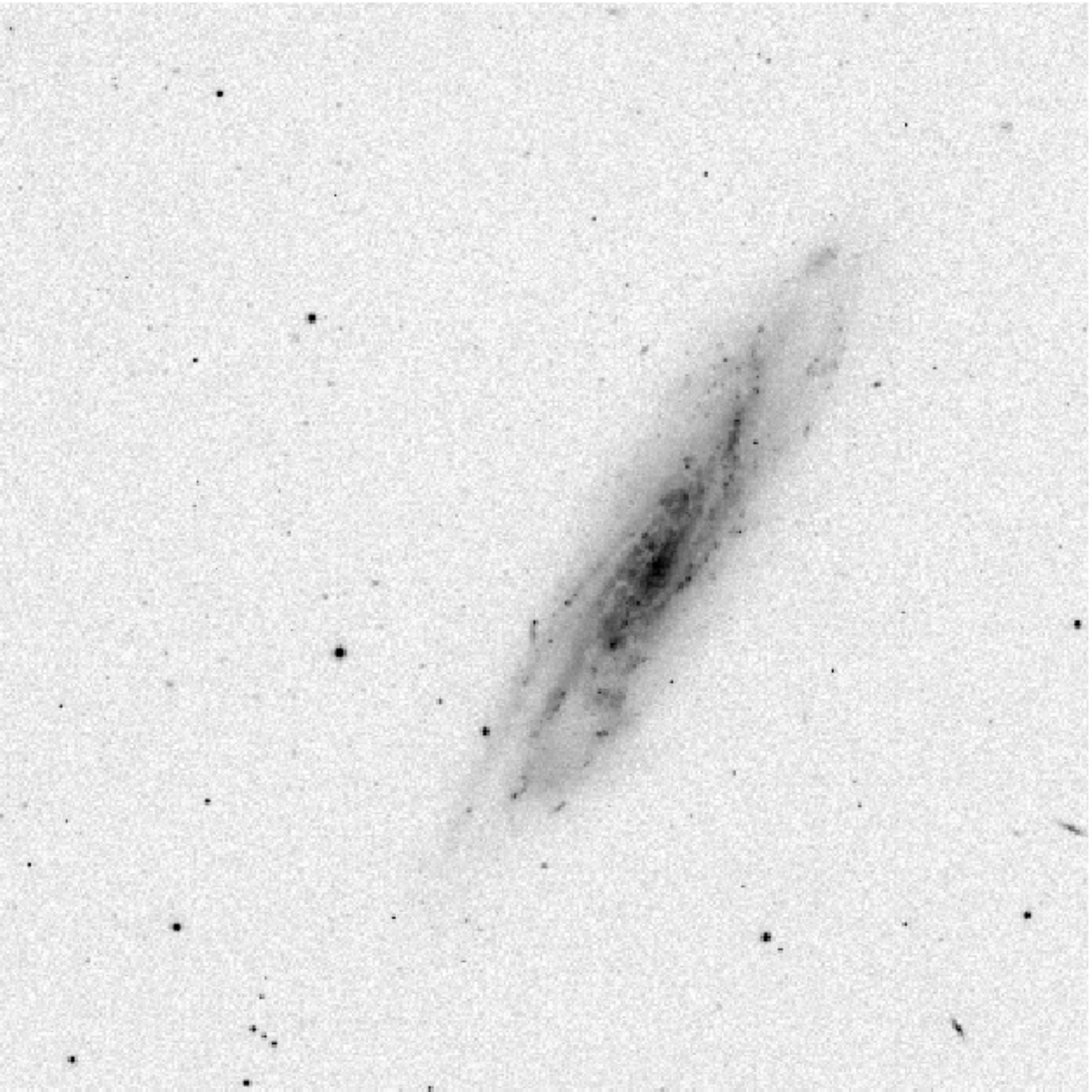} 
\includegraphics[width=5.95cm]{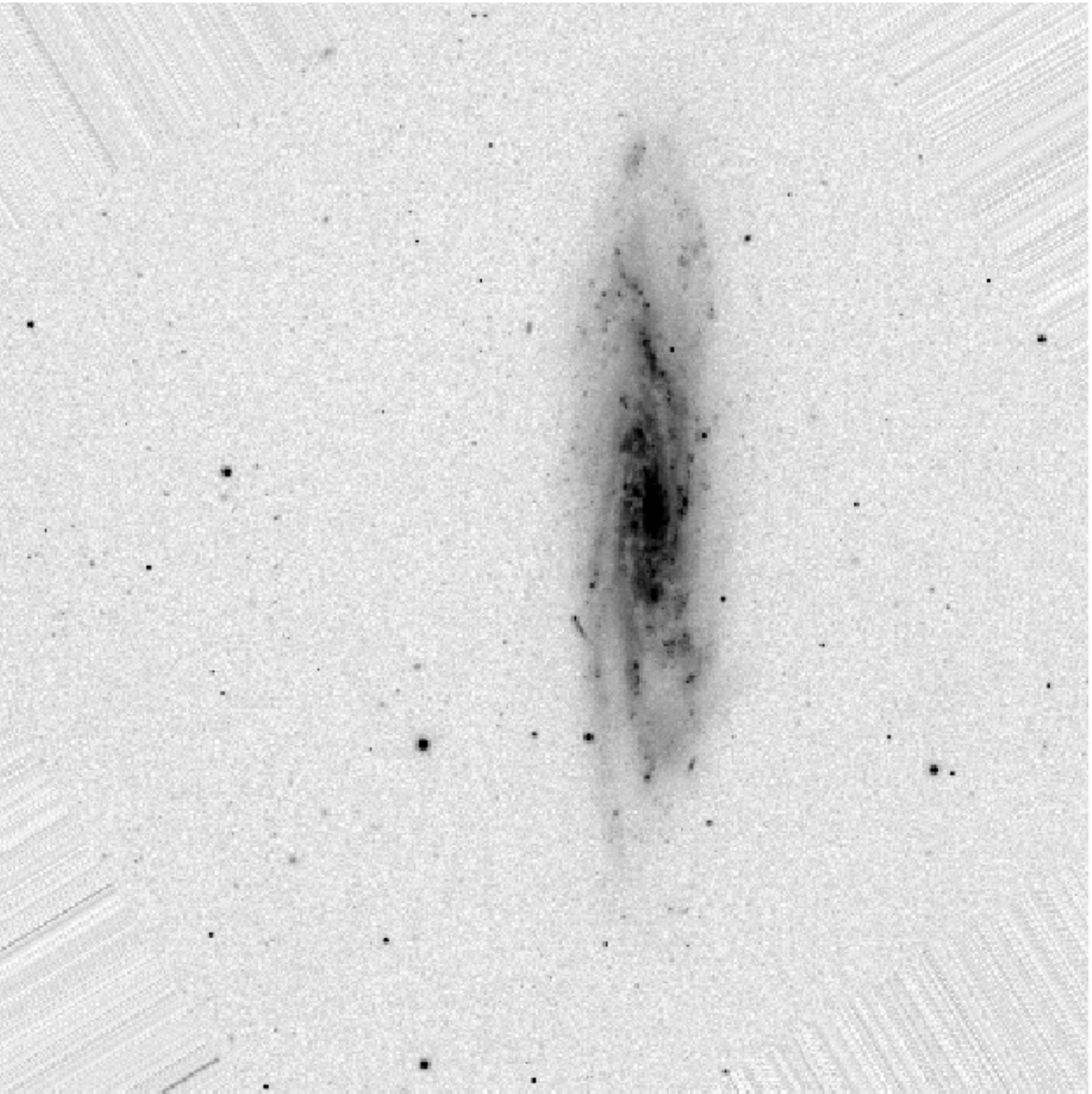}
\includegraphics[width=5.95cm]{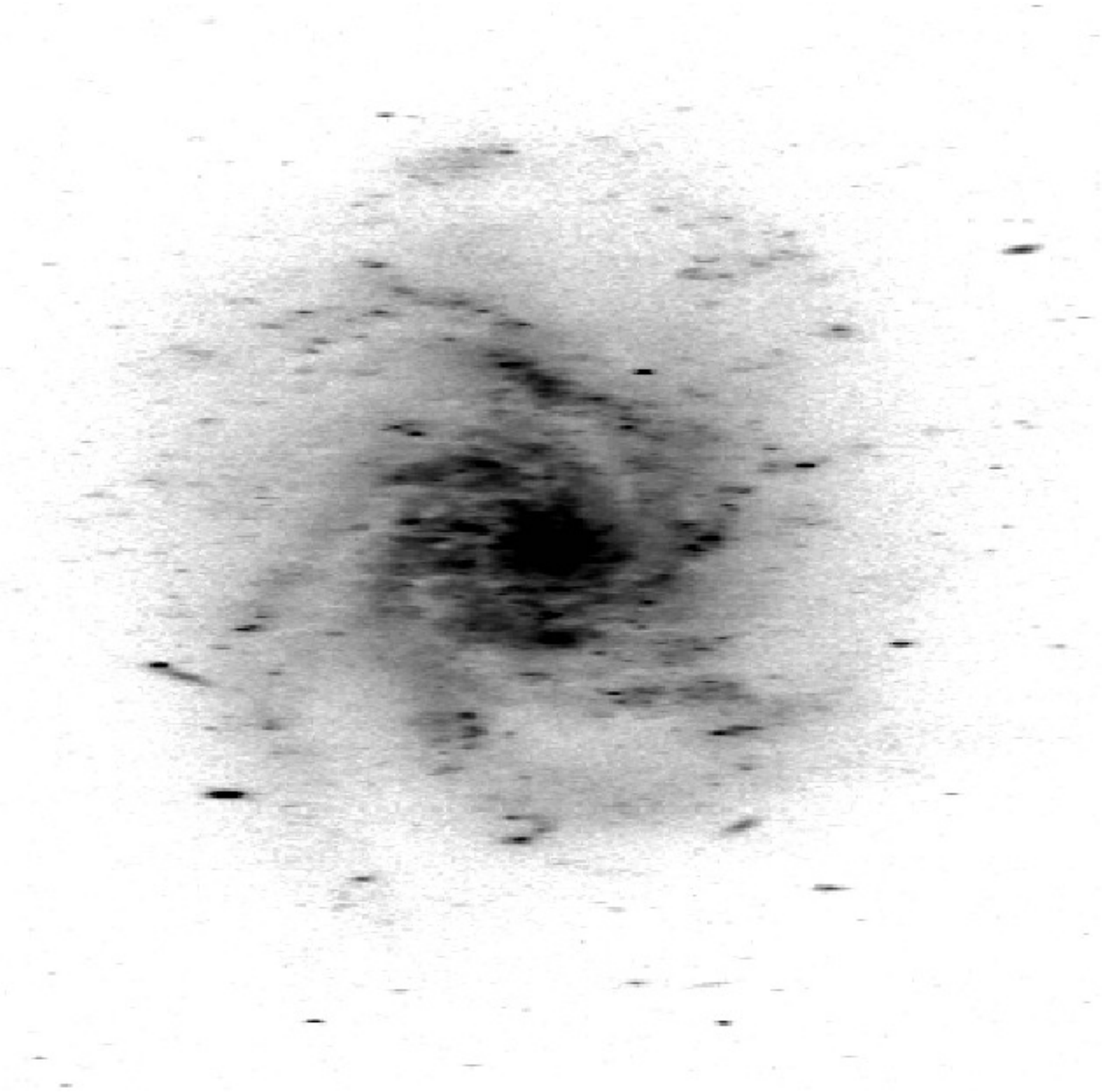}
\caption{{\it  Fig.  \ref{fig2}a (left)}  -  B-band (inverted color) image of  NGC
  1337.  NGC 1337  displays a  position angle  of  $-30.97^{\circ}$, a
  semi-minor  to  semi-major  $(b/a)$  axis  ratio  of  0.218,  and  a
  corresponding  angle of  inclination from  the plane  of the  sky of
  $77.41^{\circ}$   as  determined   by   Equation  \ref{eqn3}.   {\it
    Fig. \ref{fig2}b (middle)}  - B-band (inverted color) image of  NGC 1337 rotated
  by  $30.97^{\circ}$   ($-PA$)  to  align  the   semi-major  axis  with
  the $y$-axis. {\it Fig. \ref{fig2}c (right)}  - B-band (inverted color) image of NGC 1337:
  the result  of stretching  the $x$-axis of  Fig. \ref{fig2}b  by the
  $a/b$  axis ratio  (4.59), effectively  deprojecting the  image by
  circularizing the once elliptical shape of the galaxy. Subsequently,
  the image is  cropped and made square with the  center of the galaxy
  at the center of the image.\label{fig2}}
\end{center}  
\end{figure*}

With the position angle and axis ratio well-defined, the galaxy can be
readily deprojected. This is accomplished by rotating the image of the
galaxy (see Figure  \ref{fig2}a) 
by $-PA$ (see Figure \ref{fig2}b)  and then  stretching the  $x$-axis   by the $a/b$ axis  ratio (see Figure  \ref{fig2}c).  We use  the  IRAF
routines  {\it  ROTATE} and  {\it  MAGNIFY}  to  accomplish the  image
rotation  and  stretching,  respectively.  This  procedure  effectively
aligns  the semi-major axis  of the  galaxy with  the $y$-axis  on the
image and then stretches the semi-minor axis to an equal length as the
semi-major axis,  thus creating one  unique radius for the  galaxy and
turning what was an ellipse into a circle. The process of deprojection
is conducted in order to  minimize errors in the resulting measurement
of   pitch  angle.   However,   as  discussed   later  in   \S
\ref{subsect3.2}, precise  deprojection is not necessary for the measurement of the pitch angle. Deprojection increases the range of galactic radii over which valid pitch angles may be measured, and thus decreases error  in those measurements.   As  a result,  our  assumption  of
galaxies being  intrinsically circular is not especially critical to
the measurement of spiral arm pitch angle. In the case of highly inclined (i.e., nearly edge-on) galaxies, much of the spiral arms are hidden from sight and recovery of the intrinsic geometry via deprojection becomes increasingly difficult with higher inclination angles. However, we have had success with deprojection on galaxies up to $\alpha = 77.47^{\circ}$ for the case of IC 4831. Of course, the ability to extract meaningful information from a deprojected image will strongly depend on the resolution of the image. For images of low to moderate resolution, it is unlikely to be able to meaningfully analyze galaxies with $\alpha > 60^{\circ}$.


\subsubsection{Image Cropping}\label{Image Cropping}

After deprojection, the next step is to determine the center of the galaxy. We assume that the center of
the galaxy  is the brightest  region within the galactic  nucleus using the IRAF  routine {\it IMCNTR} to determine the  brightest pixel location  within a  specified search  region. The apparent center of a galaxy is strongly affected by the interstellar extinction. As a result, different wavebands may yield slightly different center coordinates based on this routine. We have conducted a test of 10 randomly selected galaxies from the CGS sample and have determined that, on average, the positions of the central coordinates vary by a distance of 1.986 pixels (0.514\arcsec) between B and I band images. This insignificant discrepancy is made even less important by further findings in \S\ref{subsect3.2}, which show that measurement of pitch angle does not critically depend on location of the precise center of the galaxy. The {\it IMCNTR}-determined coordinates are then adopted as the center of the galaxy. The image  is then cropped  about the galaxy,  with the center  of the
galaxy as the center of the cropped image (see Figure \ref{fig2}c), and
the resulting  image made to be  a perfect square, as  required by the
{\it 2DFFT} code, with odd-numbered pixel-sized edges in  order to allow a unique median
pixel location  as the exact center  of the image\footnote{As  required by the
{\it 2DFFT} code, the  input image file must be in the  form of a text
file. We  use the IRAF routine  {\it WTEXT} to convert {\it .fits} images to {\it .txt} files.}.

\subsubsection{Star Subtraction}\label{subsect3.1}

FFT image analysis is widely used for its mimicry of the human eye's
ability to pick out symmetries and repetitions even in noisy or cluttered
images. The {\it 2DFFT} code can measure pitch angles, in spite of the presences
of many non-spiral features in a galaxy's image. An interesting feature
is that  though the code models the spiral image as a superposition of
spirals of different numbers of arms and different pitch angles, it measures
the correct pitch angle even for harmonic modes where the number of spiral arms is
incorrect. Thus, for a sufficiently low-noise image, it is not even necessary
to correctly infer the number of spiral arms in the galaxy in order to
accurately measure its pitch angle. 

\begin{figure*}
\begin{center}
\includegraphics[width=8.97cm]{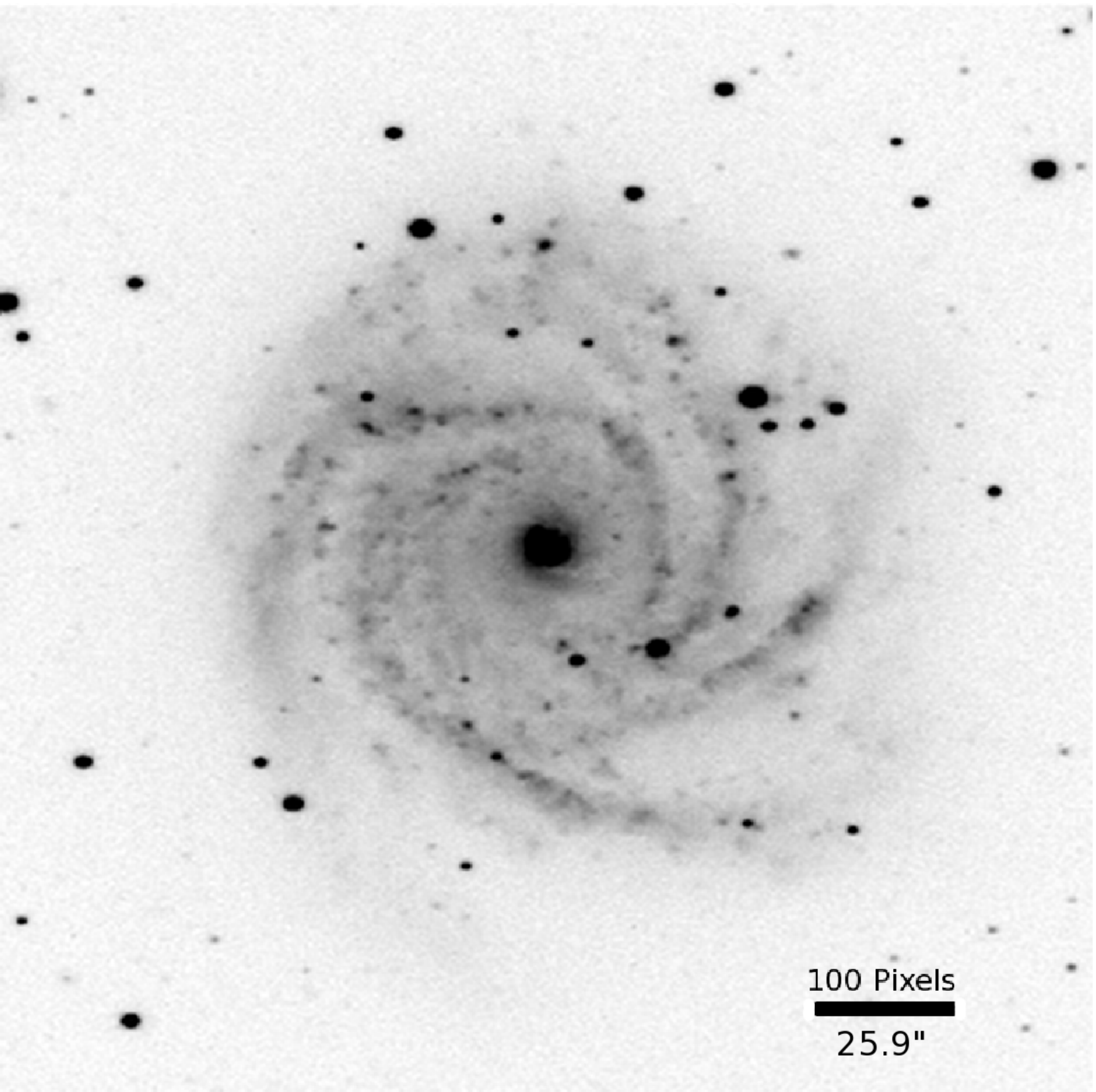}
\includegraphics[width=8.97cm]{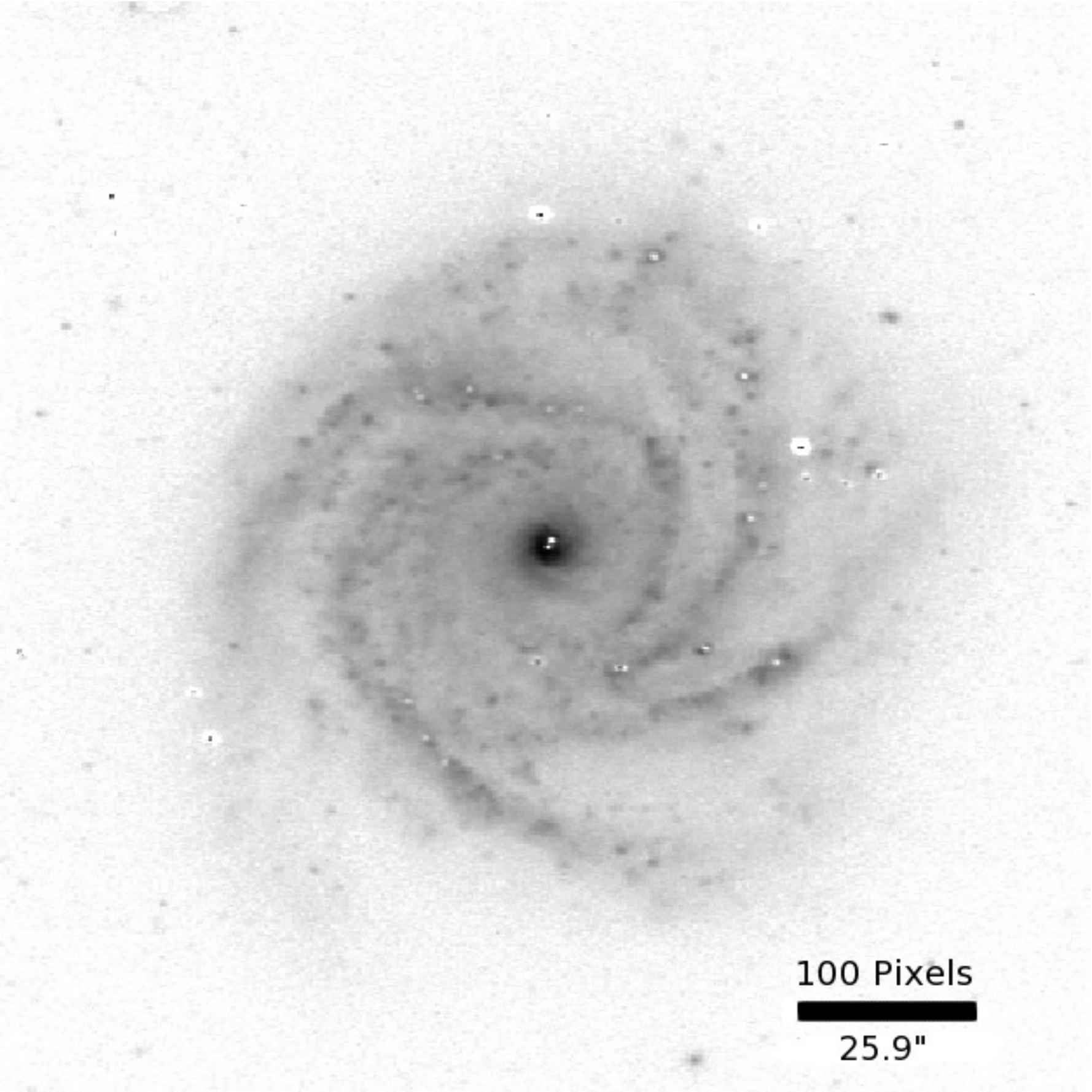}
\caption{{\it Fig. \ref{fig18}a  (left)} - Deprojected B-band image of IC 4538 before star subtraction. IC
  4538  is  measured   to  have  $PA  =  50^{\circ}$   and  $\alpha  =
  39.65^{\circ}$. {\it  Fig. \ref{fig18}b (right)}  - Deprojected B-band  (inverted color) image  of  IC  4538  after a  Gaussian  star  subtraction  was
  performed.\label{fig18}}
\end{center}
\end{figure*}

When noise is introduced, this agreement in pitch angle measurement between the different harmonic modes is the most obvious casualty. One important source of noise is
the presence of bright foreground stars, especially when they are superimposed
on the disc of the galaxy itself (see Figure \ref{fig18}). 
Nevertheless, the 2-D FFT of the harmonic mode with
the correct number of arms (most commonly, $m = 2$) will still usually give a stable value, which seems to correspond
to the correct pitch angle even when other harmonic modes show no reliable measure.
In order to increase confidence in our pitch angle measurement, we reduce the noise by subtracting the foreground stars. Since the IRAF {\it ELLIPSE} function does not always work when analyzing star-subtracted images (it fails to start if light from the center of a galaxy has been removed and it cannot locate the center), it is best to measure the ellipticity before performing star subtraction\footnote{We fit a Gaussian Point Spread Function to the bright stars and subtract them using the IRAF package {\it DAOPHOT}.}. Often, this results in several of the
harmonic modes coming into at least rough agreement with the one harmonic mode previously
selected as the best single example. 
This suggests that foreground
star contamination is a leading source of noise in the images and that
star subtraction is a useful step. At the same time, it may not always
be required for an accurate measurement. This is corroborated by \citet{Martinez-Garcia:2011}, who finds from a study of 27 galaxies, that the presence of foreground stars does not affect the value of pitch angle in general.


\vspace{7mm}

\subsection{Image Measurement}\label{Image Processing}

Following pre-processing, the first step  is to  specify an  inner and
outer radius of the galactic disc. The inner radius is the location where the
spiral arms begin,  i.e., where the galactic bar  or bulge terminates;
the outer  radius is the location  where the spiral  arms cease, usually
the outermost visible radius of the galaxy.  Thus, an annulus  is defined within which the  {\it 2DFFT}  code conducts  its Fourier Decomposition. We then take this previously established procedure a step further by automating the code to measure many annuli so the final quoted pitch angle is not determined solely by one user-determined annulus. Our modifications and motivations behind the modifications to the code are detailed in the following section.

\section{Pitch Angle as a Function of Inner Radii}\label{subsect2.2}

The  greatest source  of  human  error lies in  choosing an  inner
radius. Whereas it is seemingly easy for the user to visually identify
the edge of  the galaxy (i.e., the outer  radius), it is significantly
more difficult for the user to accurately specify the cessation of the
bar/bulge feature  of a galaxy (i.e., the  inner radius). Furthermore,
slight  error in  specification of  the  outer radius  has little  ill
effect, whereas slight error in  specification of the inner radius may 
have significant effect. To illustrate this result, consider different 
values in outer  radii: underestimation results in the  full length of
spiral  arms not  being measured;  overestimation results  in  the sky
being measured at  the edge of the spiral  arms. Since our FFT computations  are  luminosity  biased,  sky
inclusion does not significantly affect  the computations. On the other hand, consider
different values  in inner radii:  overestimation results in  the full
length of spiral arms not being measured; underestimation results in a
bright bar/bulge  feature being measured  in addition to  spiral arms,
this last case being the worst possible scenario.

As a result of this observed sensitivity to inner radius selection, we
run {\it 2DFFT} iteratively  at different
inner radii. This allows the user to specify an outer radius
and calculate pitch angles  for all  possible inner
radii  within the defined  outer radius.  A scripting utility has been created to
calculate pitch  angles at  all possible inner  radii, given  an outer
radius,  for  a galaxy (see Figure \ref{flow}). 
\tikzstyle{decision} = [diamond, draw, fill=blue!20, 
    text width=10em, text badly centered, node distance=3.3cm, inner sep=0pt]
\tikzstyle{block} = [rectangle, draw, fill=blue!20, 
    text width=10em, text centered, rounded corners, minimum height=4em]
\tikzstyle{line} = [draw, -latex']
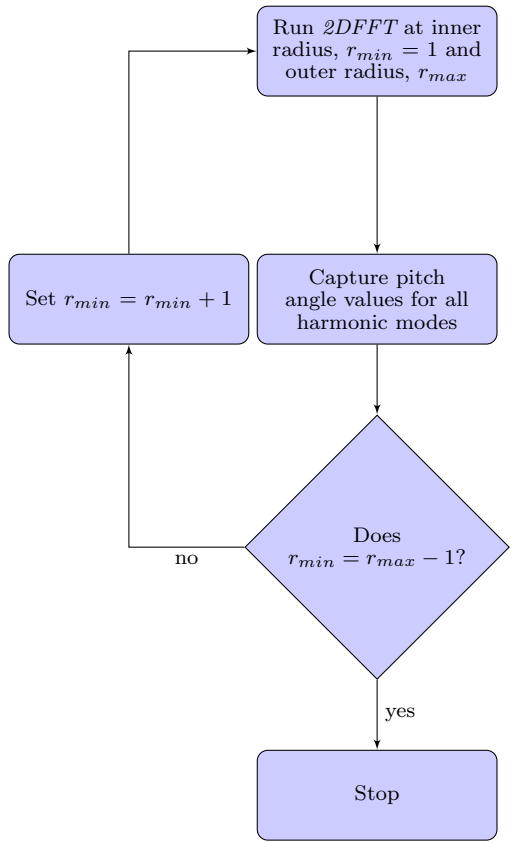
\begin{figure}
\begin{center}
\begin{tikzpicture}[node distance = 3.3cm, auto]
    \node [block] (2DFFT) {Run {\it 2DFFT} at inner radius, $r_{min} = 1$ and outer radius, $r_{max}$};
    \node [block, below of=2DFFT] (capture) {Capture pitch angle values for all harmonic modes};
    \node [block, left of=capture] (iterate) {Set $r_{min} = r_{min} + 1$};
    \node [decision, below of=capture] (decide) {Does $r_{min} = r_{max} - 1$?};
    \node [block, below of=decide, node distance=3.3cm] (stop) {Stop};
    \path [line] (2DFFT) -- (capture);
    \path [line] (capture) -- (decide);
    \path [line] (decide) -| node [near start] {no} (iterate);
    \path [line] (iterate) |- (2DFFT);
    \path [line] (decide) -- node {yes}(stop);
\end{tikzpicture}
\caption{Flow chart representing our iterative method for measuring pitch angle as a function of inner radii.\label{flow}}
\end{center}
\end{figure}
Additionally, we have modified the memory allocation of the original code to allow for input image sizes up to $2048 \times 2048$ pixels.

\begin{figure*}
\begin{center}
\includegraphics[width=4.44cm]{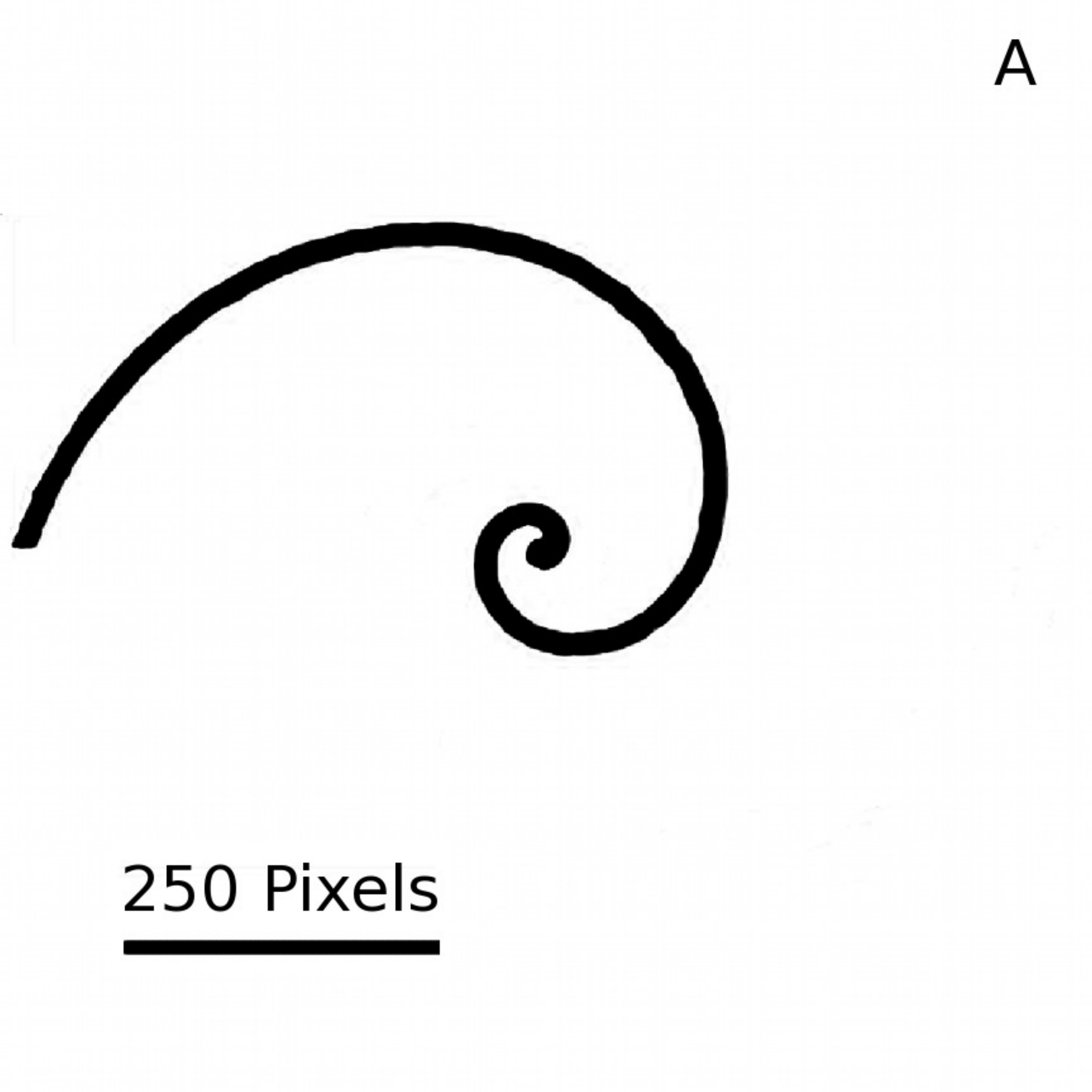} 
\includegraphics[width=4.44cm]{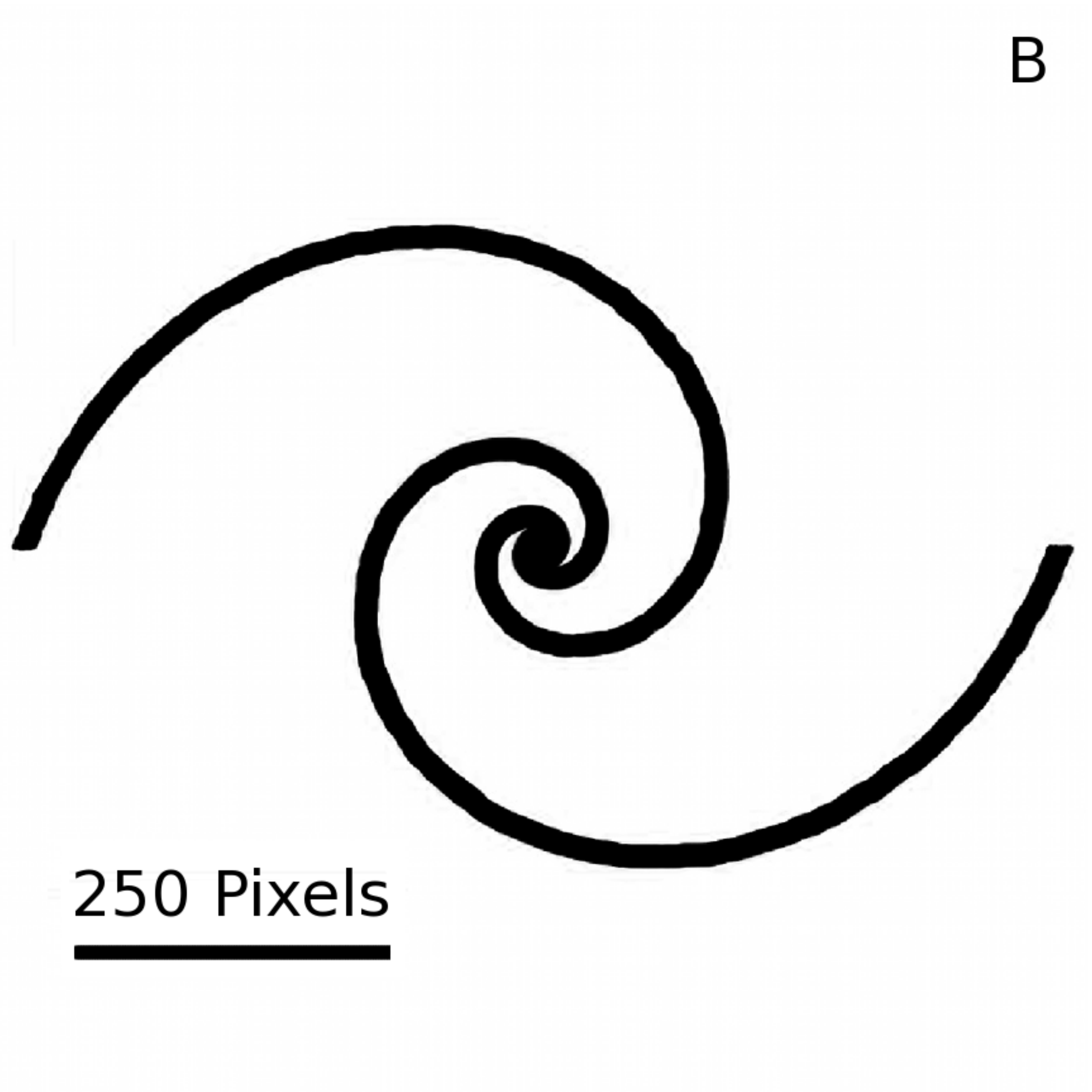}
\includegraphics[width=4.44cm]{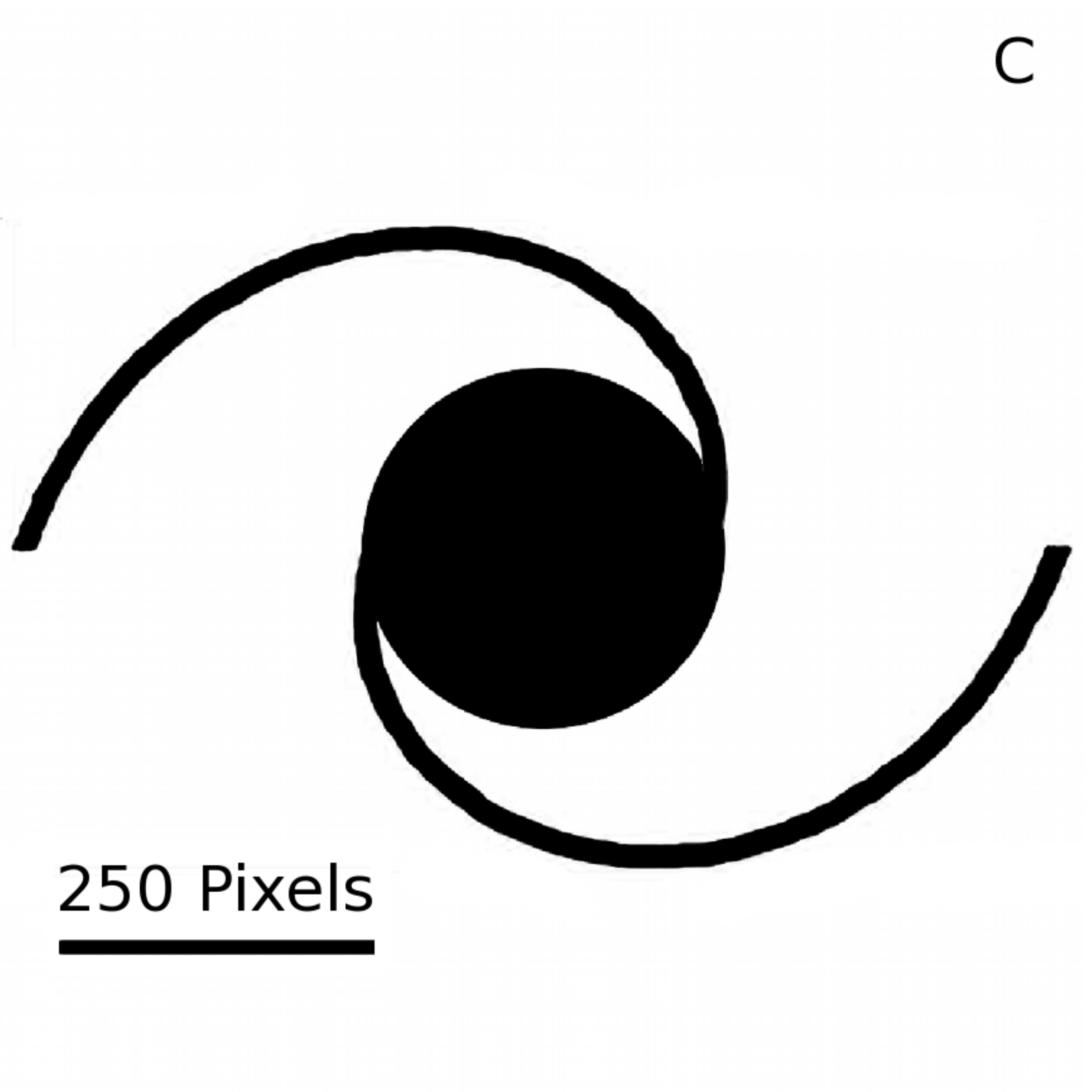}
\includegraphics[width=4.44cm]{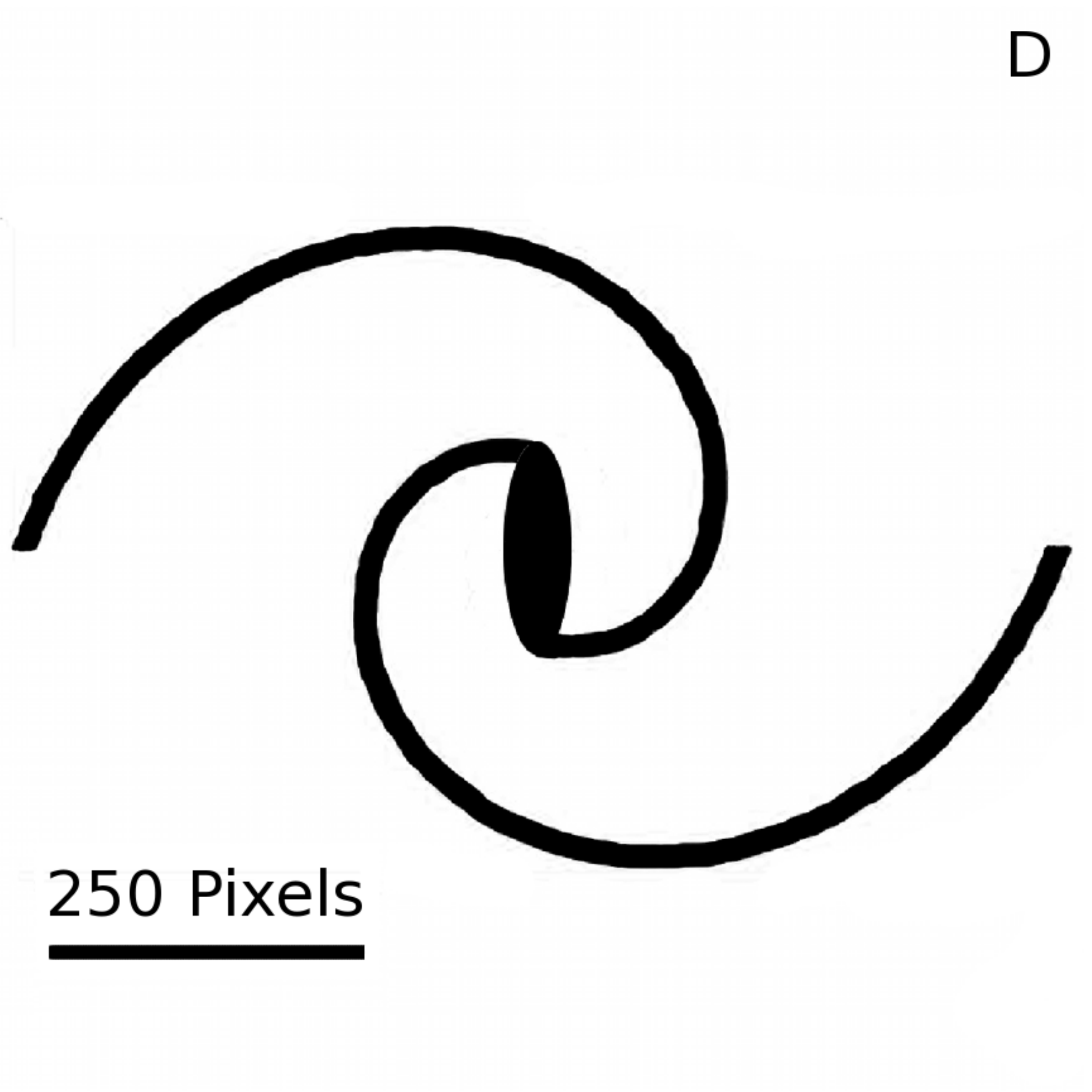}
\includegraphics[width=4.44cm]{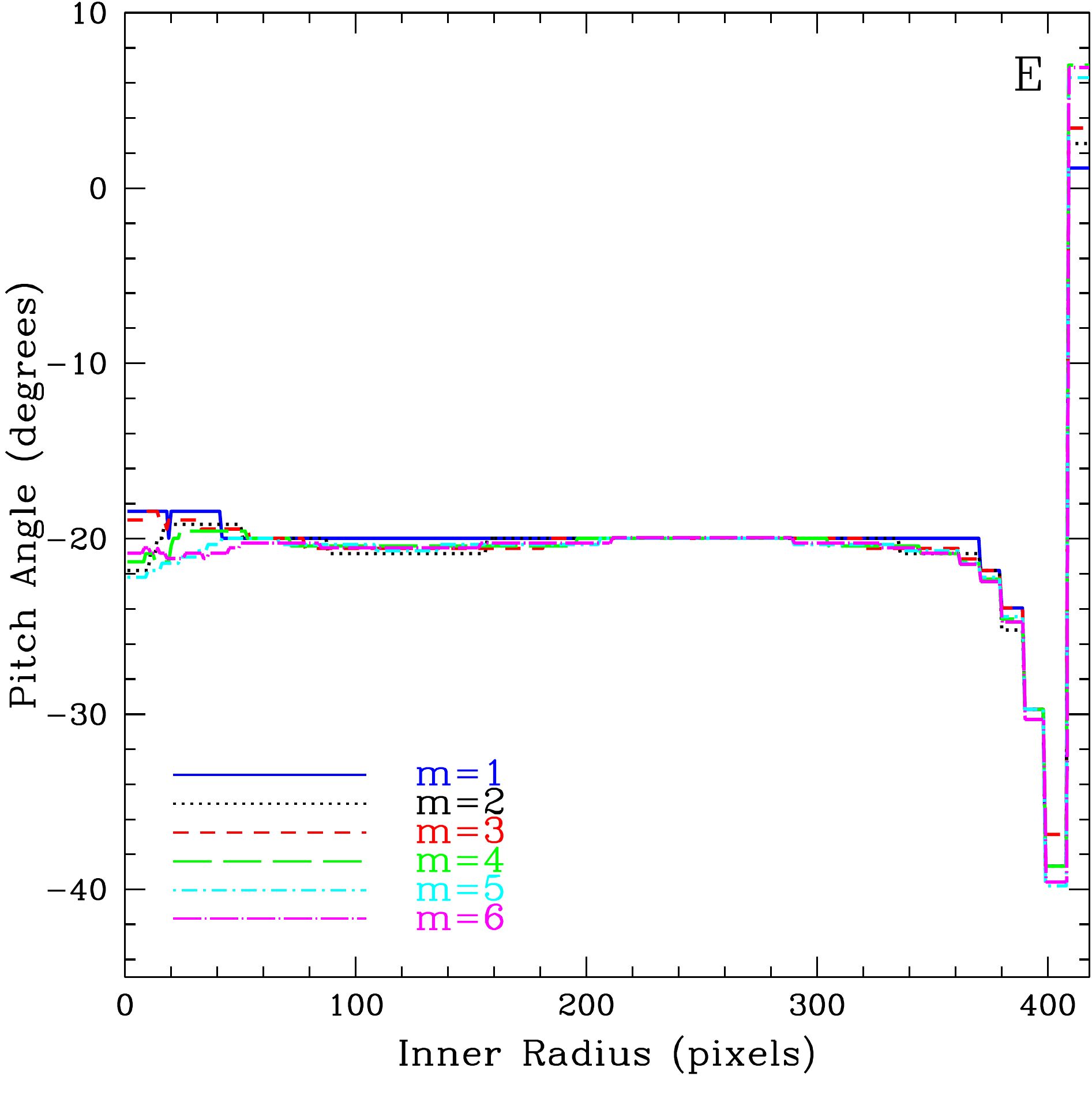}
\includegraphics[width=4.44cm]{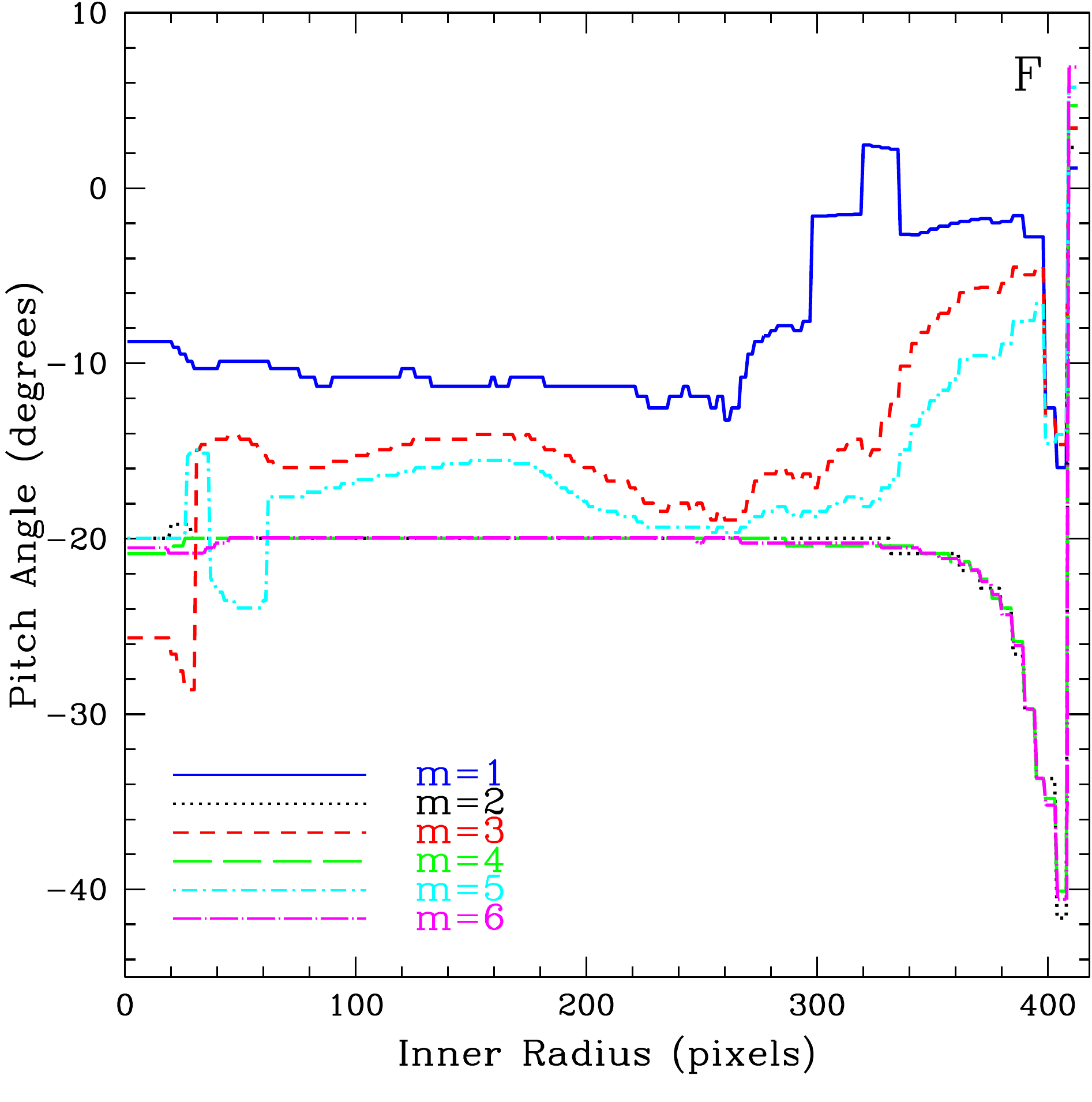}
\includegraphics[width=4.44cm]{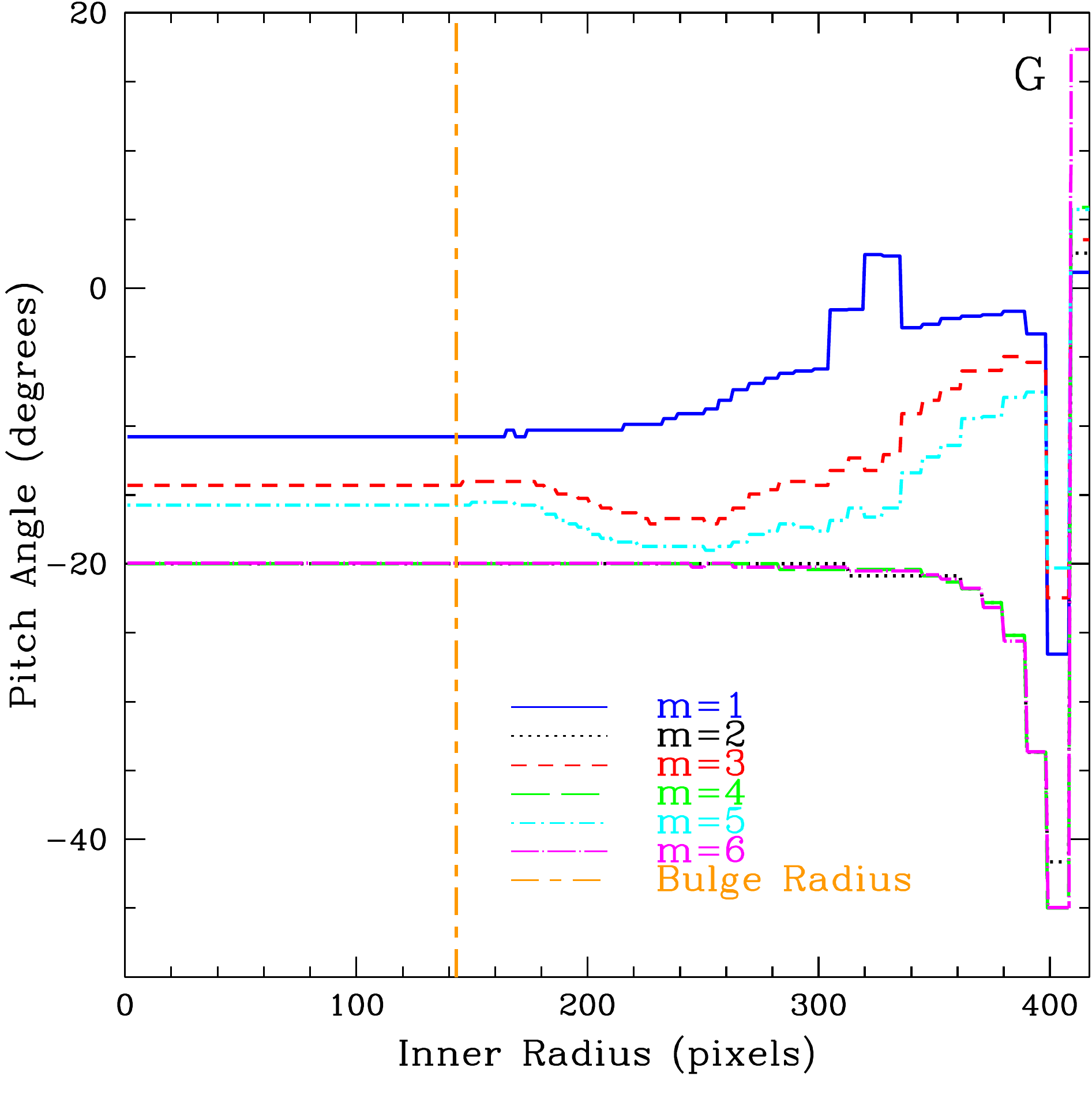}
\includegraphics[width=4.44cm]{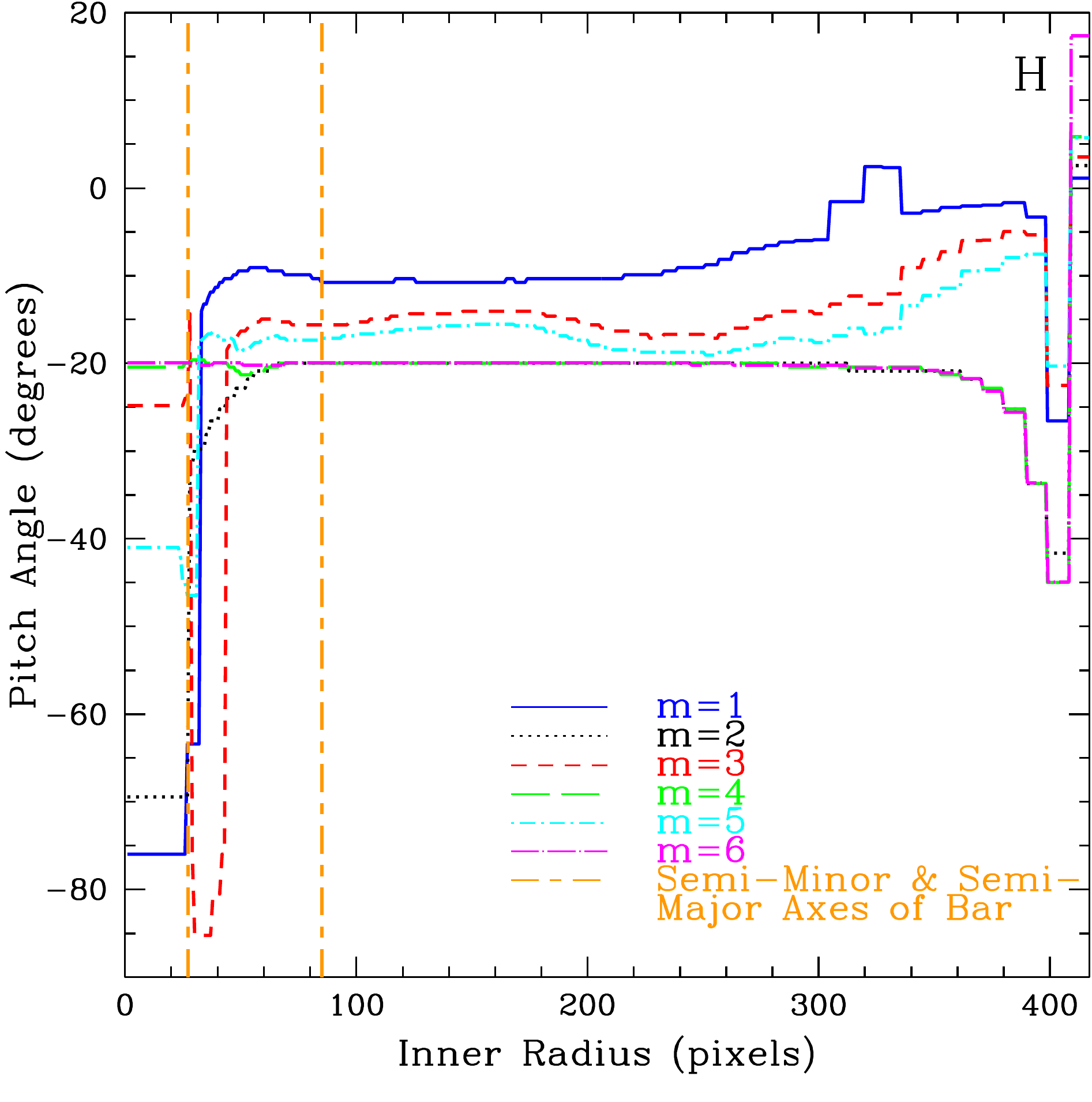}
\caption{Synthetic spirals and their corresponding pitch angles as a function of inner radius below. The eight individual panels will be identified and discussed in the following top to bottom, left to right fashion. {\it Fig. \ref{Syn_figs}a} - One-armed synthetic logarithmic spiral (inverted color) with $\phi = -20^{\circ}$. {\it Fig. \ref{Syn_figs}b} - Synthetic two-armed logarithmic spiral  (inverted color) with a
  constant   $m   =   2$    pitch   angle   of   $-20^{\circ}$. {\it Fig. \ref{Syn_figs}c} - $\phi = -20^{\circ}$ synthetic two-armed logarithmic spiral (inverted color) with a circular bulge component added. {\it Fig. \ref{Syn_figs}d} - $\phi = -20^{\circ}$ synthetic two-armed logarithmic spiral (inverted color) with a barred nuclear component added. {\it Fig. \ref{Syn_figs}e} - Pitch angle results for Fig. \ref{Syn_figs}a. All harmonic modes display the correct pitch angle until $\approx 90\%$ of the outer radius is reached. {\it Fig. \ref{Syn_figs}f} - Pitch angle results for Fig. \ref{Syn_figs}b. The results  for the even harmonic  modes are
  essentially the same and accurately measure the correct pitch angle until $\approx 90\%$ of the outer radius is reached. {\it Fig. \ref{Syn_figs}g} - Pitch angle results for Fig. \ref{Syn_figs}c. The vertical line at 143 pixels represents the radius of the circular bulge. The odd numbered harmonic modes have a
  systematically  lower absolute value  of pitch  angle, but  the even
  harmonic modes   are  unchanged  by   the  addition   of  a   circular  bulge
  component. {\it Fig. \ref{Syn_figs}h} - Pitch angle results for Fig. \ref{Syn_figs}d. The vertical lines at 27 and 85 pixels represent the semi-minor and semi-major axes of the bar, respectively. The odd numbered  harmonic modes have
  systematically lower absolute values of pitch angle, just as in the
  case of the circular bulge  component, but the innermost inner radii
  demonstrate    intuitively   high    absolute   values    of   pitch
  angle.\label{Syn_figs}}
\end{center}
\end{figure*}

The inner radius is a numerical artifact which should not affect the measurement of pitch angle. Therefore, we seek a harmonic mode in which we find a range of inner radius over which the measured pitch angle appears to be the most stable and consistent with the observed appearance of the galaxy. We aim, typically, for a measurement of pitch angle with an associated error of $2^{\circ}$ to $4^{\circ}$. The resulting pitch angles can be plotted vs. inner radius 
in order to visually identify stable pitch angle regions as a function
of  inner radius beyond  the influence  of a  potential bar  or oblate
bulge feature.  Stable regions are  selected by several  criteria; the
stable region  must be  of the same  sign (chirality) as  the observed
spiral arm windings in the image, it should be of the same harmonic mode  as the visually observed  number of spiral arms, there must  not be any erratic  fluctuation in pitch angle, and the region of stable pitch angles must be contiguous. In certain cases, the resultant pitch angle agrees in multiple harmonic modes, 
therefore, selection of the harmonic mode is not critically important. This allows us to focus on picking a stable pitch angle, even when the correct $m$ value is ambiguous. To understand the code's behavior, we have conducted tests with very low noise images, artificially
created logarithmic spirals. Not surprisingly, the code finds it trivial to measure the pitch angle of such an image. For these synthetic spirals (see Figure \ref{Syn_figs}), 
the apparent stable regions are easily perceptible. However, in real galaxies, careful image inspection and other techniques (see \S\ref{sect4} and its subsections) are sometimes required to pick out more visually elusive stable regions amidst the range of harmonic modes available. 

To be clear, our method does not avoid having to inspect each image individually. This is a necessary and not totally undesirable requirement. Computer vision methods are currently under development to measure galactic spiral arm pitch angle \citep[e.g.,][]{Davis:Hayes:2012}. The human eye has proven itself as the most reliable tool for the geometric classification of galaxies \citep[e.g., Galaxy Zoo;][]{Galaxy:Zoo:2008}. The human operator is required only to inspect the image qualitatively for signs of gross error, not to re-perform any measurements quantitatively.


An unavoidable side effect of deprojection is an oblate distortion of a galaxy's nuclear region. Initially, a nuclear galactic  bulge is considered  to be spheroidal.  Therefore, a
perfectly  face-on galaxy  (no deprojection  required) with  a nuclear
bulge does  not hinder the selection  of an inner  radius. This can be seen in the case of a synthetic logarithmic spiral with a nuclear bulge component added (see Figures
\ref{Syn_figs}c  and   \ref{Syn_figs}g). Since  a   galaxy  is  deprojected
according to the outer region of  the galaxy and not the inner region,
the nucleus can be distorted to an oblate spheroid. This creation of a
non-spherically symmetric feature can negatively affect the calculated
pitch angle in the innermost regions of a galaxy.


The largest  and likeliest source of  error due to  inner radii
determination is when barred  galaxies are measured. Galactic bars are
linear features and therefore have high pitch angles $(\phi \simeq
90^{\circ})$.  Inclusion  of  a  high  pitch angle  feature  into  the
measurement annulus of  the {\it 2DFFT} code results  in a significant
overall biasing  of the  resulting pitch angle  towards the  high side. This is
always the  case, because  the highest practical  limit for  spiral arm
pitch angle  is significantly lower than  the pitch angle produced by a galactic bar. The effect of galactic bars are illustrated in the example of a synthetic two-armed spiral with a bar component added (see Figures \ref{Syn_figs}d and \ref{Syn_figs}h).

\section{Error Determination}\label{sect3}

The most obvious error  is the variance about the mean pitch
angle over  the selected  stable region in  inner radii. The  error is
found by calculating  the mean and  standard deviation  of the
sample of pitch angles over  the selected stable region. This standard
deviation  of pitch  angle over  the  selected stable  region is  then
weighted by the length of the stable region compared to the total length from the innermost spiral structure to the edge of the galaxy. Based on our observation from running synthetic logarithmic spirals through our code (see subsequent subsections), reliable pitch angles are not measurable for inner radii selected beyond $\approx 90\%$ of the selected outer radius. At this point, too much information has been ignored for the code to accurately measure a pitch angle.

\begin{figure}
\begin{center}
\includegraphics[width=8.6cm]{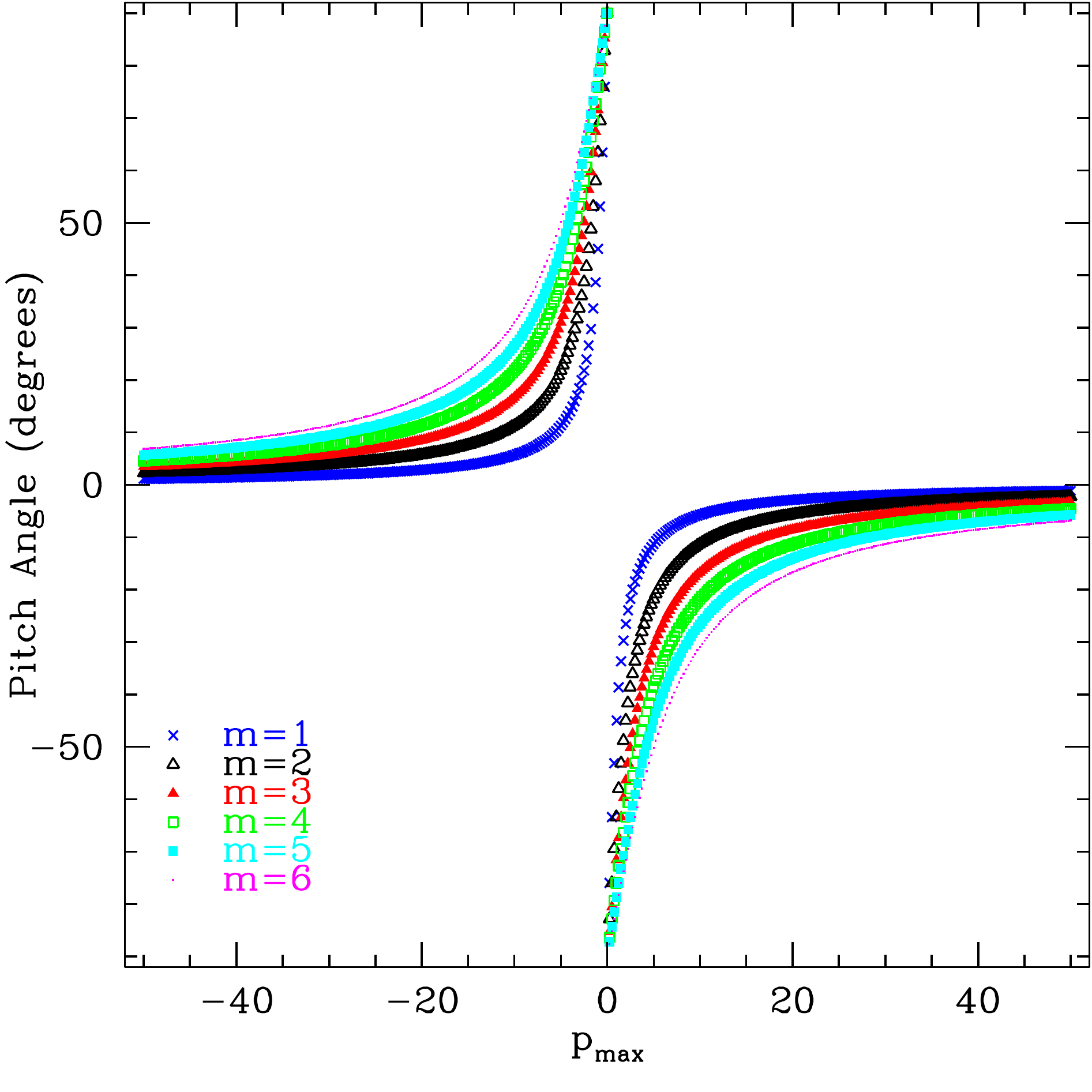}
\caption{Every  possible value of pitch angle
  calculable by the {\it 2DFFT} code. Pitch angles are determined from
  $p_{max}$ according to Equation \ref{eqn2}.\label{fig6}}
\end{center}
\end{figure}

\begin{figure}
\begin{center}
\includegraphics[width=8.6cm]{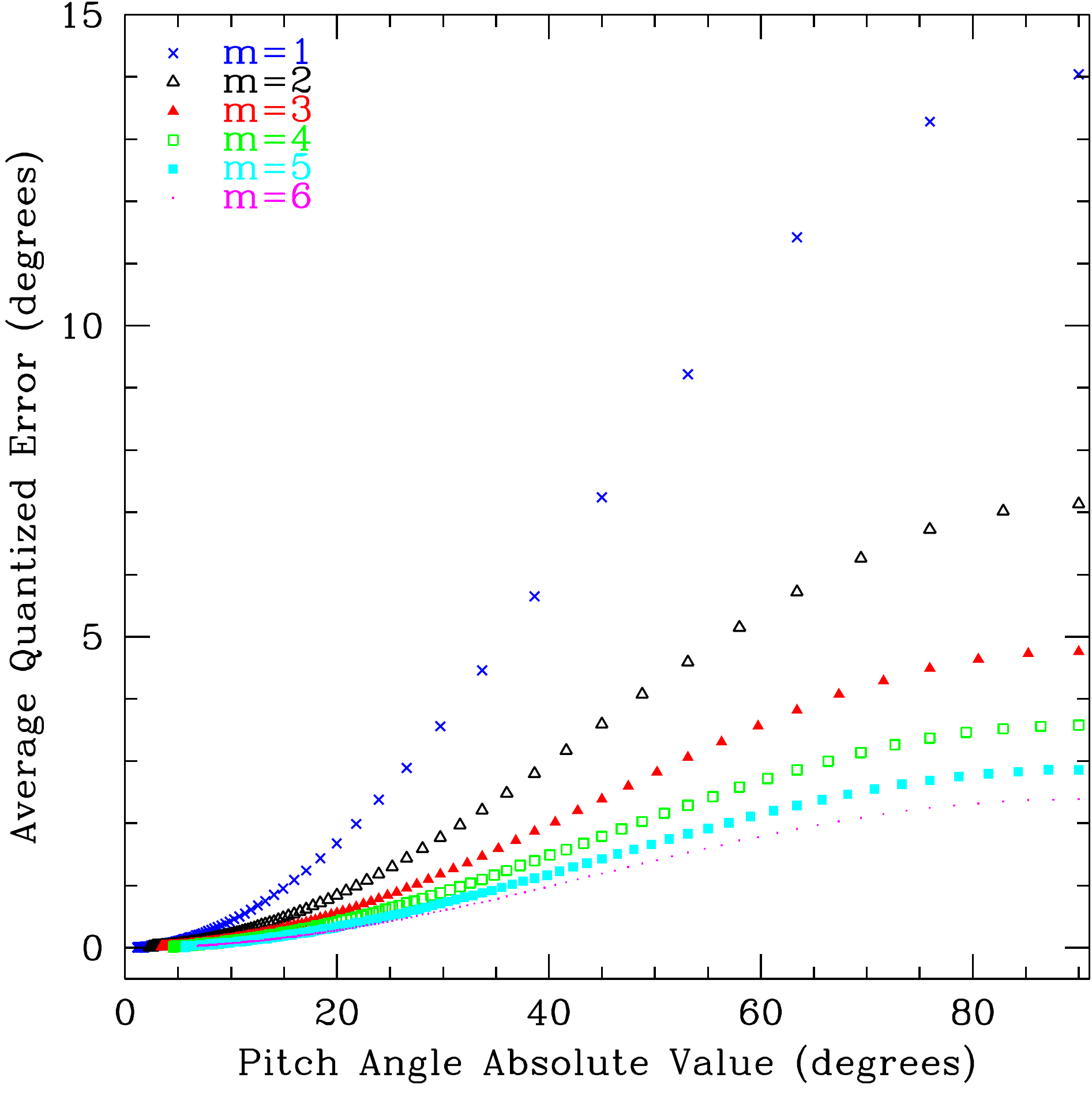}
\caption{Every possible  absolute  value of
  pitch angle calculable  by the {\it 2DFFT} code  and their resulting
  average  quantized error  due  to the  discrete  step resolution.  A
  third-ordered  best fit  polynomial  is fit  to  each harmonic  mode
  $(m)$ to interpolate error values at all points: \textcolor{blue}{$y=-4\times10^{-5}x^3+0.0058x^2-0.0137x+0.0234$; $R^2=0.99994$ for $m = 1$}, $y=-2\times10^{-5}x^3+0.0029x^2-0.0084x+0.0222$; $R^2=0.99997$ for $m = 2$, \textcolor{red}{$y=-1\times10^{-5}x^3+0.002x^2-0.0064x+0.0214$; $R^2=0.99998$ for $m = 3$}, \textcolor{green}{$y=-1\times10^{-5}x^3+0.0015x^2-0.0054x+0.0207$; $R^2=0.99998$ for $m = 4$}, \textcolor{cyan}{$y=-9\times10^{-6}x^3+0.0012x^2-0.0046x+0.02$; $R^2=0.99999$ for $m = 5$}, and \textcolor{magenta}{$y=-7\times10^{-6}x^3+0.001x^2-0.0041x+0.0191$; $R^2=0.99999$ for $m = 6$}.
\label{fig7}} 
\end{center}
\end{figure}

\begin{figure*}
\begin{center}
\includegraphics[width=8.97cm]{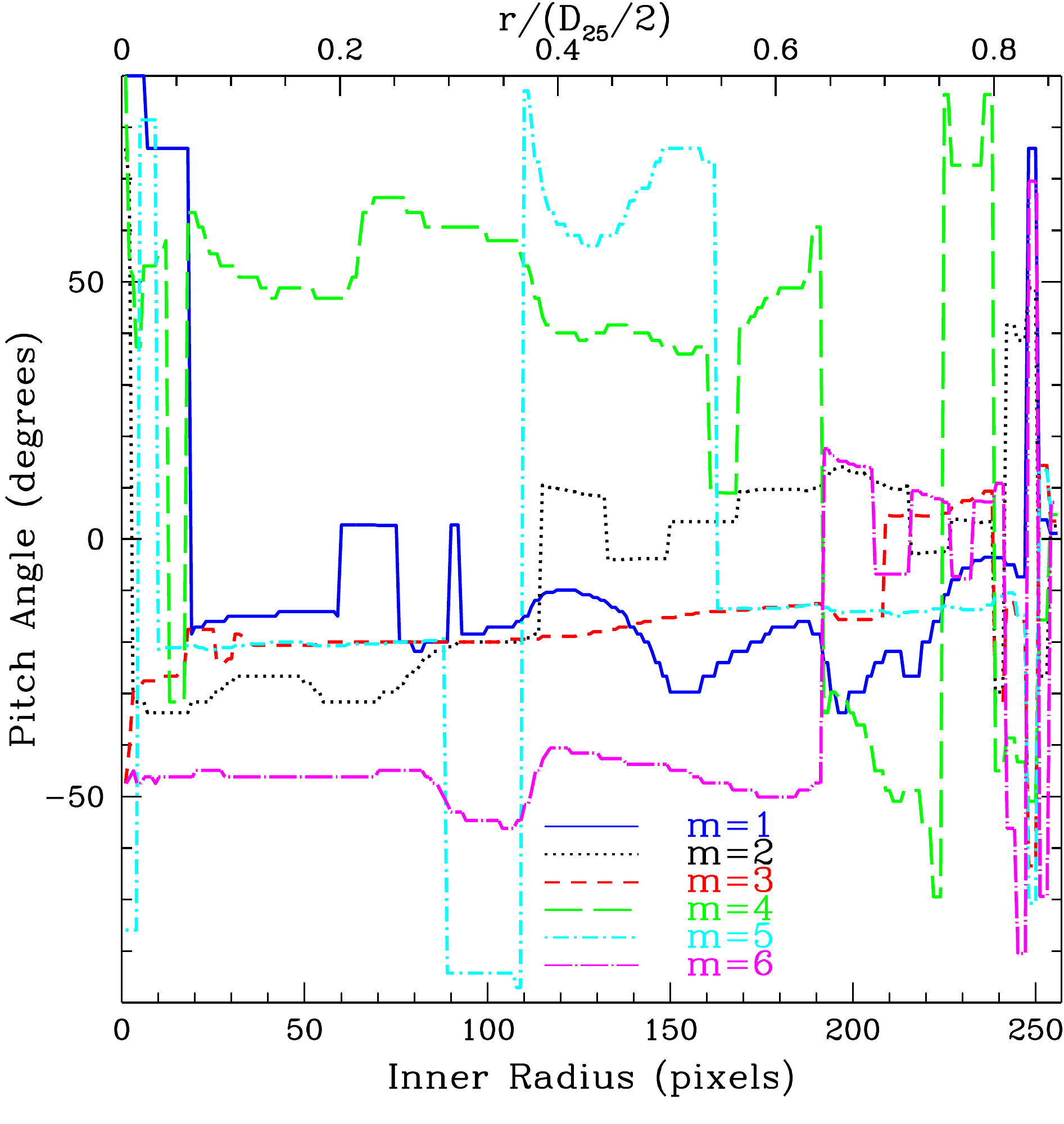}
\includegraphics[width=8.97cm]{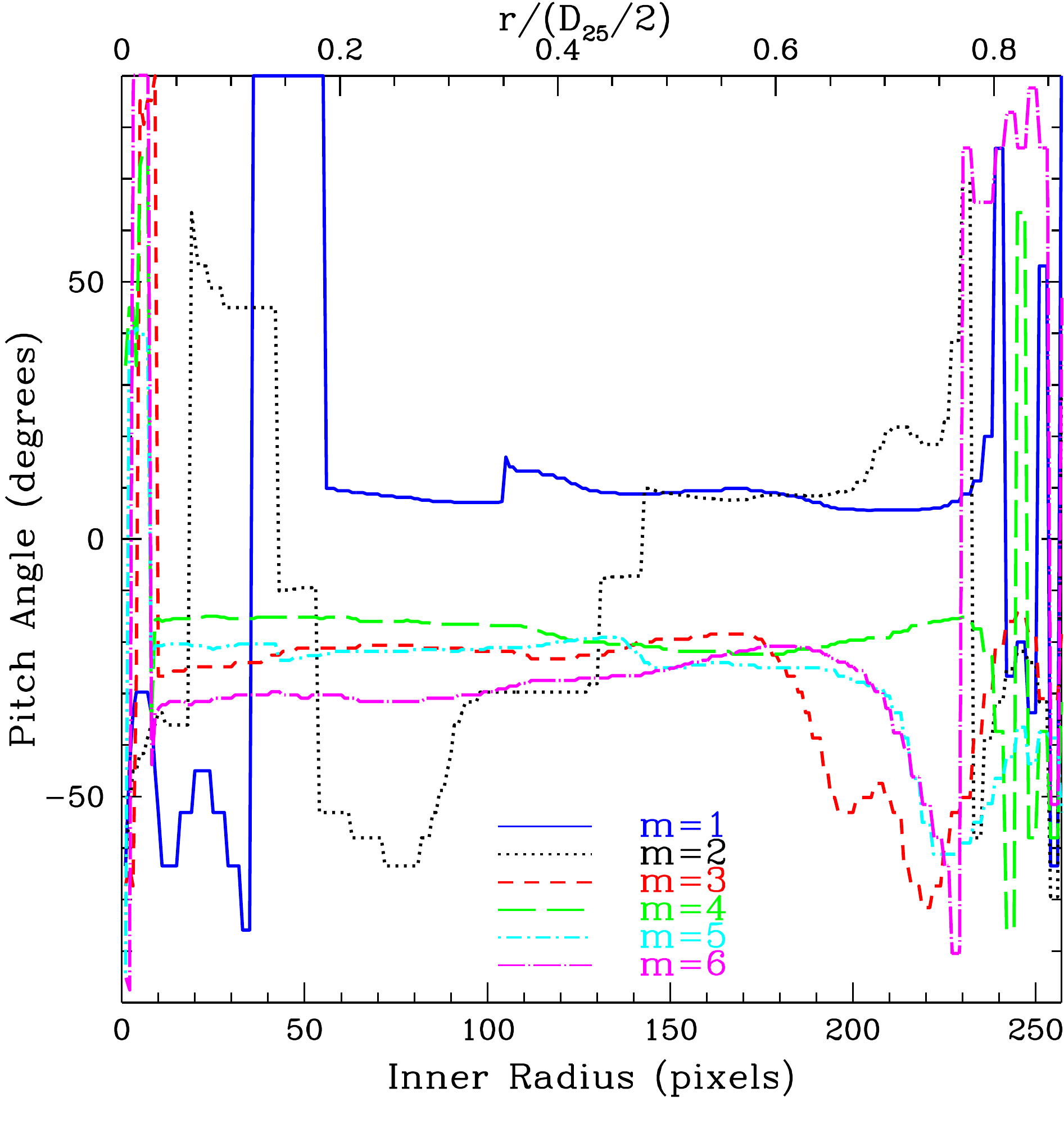}
\caption{IC  4538  B-band pitch angle  as  a  function  of inner  radius  for
  deprojected ($PA = 50^{\circ}$ \& $\alpha = 39.65^{\circ}$) images before  (left) and  after (right)  Gaussian star
  subtraction was performed. $r/(D_{25}/2)$ is plotted on the secondary $x$-axis with $D_{25}$ (major axis at the 25.0 B-mag/sq arcsec isophote) from the Third Reference Catalogue of Bright Galaxies \citep[RC3;][]{RC3}. {\it  Fig. \ref{fig9}a (left)} - a stable
  mean pitch angle  of $-17.52^{\circ}$ is determined for  the $m = 3$
  harmonic mode from  a minimum inner  radius of 36  pixels (9.32\arcsec) to a  maximum inner
  radius  of 208  pixels (53.9\arcsec), with  an outer  radius of  258  pixels (66.8\arcsec). This
  stable  region  of 172  pixels (44.5\arcsec) occupies 67\%  of the galactic disc. 
  Equation \ref{Error} yields $E_{\phi} = 3.17^{\circ}$ with $\lambda = 172$ pixels (44.5\arcsec), $\beta = 196$ pixels (50.8\arcsec), $\sigma = 2.75^{\circ}$, and $\epsilon_3 = 0.47^{\circ}$. The final determination of pitch  angle is therefore
  $-17.52^{\circ} \pm 3.17^{\circ}$.  {\it Fig. \ref{fig9}b (right)} -
  a stable mean pitch angle  of $-17.98^{\circ}$ is determined for the
  $m = 4$  harmonic mode from a minimum  inner radius of 9 pixels (2.33\arcsec) to a maximum
  inner radius of 235 pixels (60.9\arcsec), with an outer radius of 264 pixels (68.4\arcsec). This
  stable  region  of 226  pixels (58.5\arcsec) occupies 86\%  of the galactic disc. 
  Equation \ref{Error} yields $E_{\phi} = 2.61^{\circ}$ with $\lambda = 226$ pixels (58.5\arcsec), $\beta = 229$ pixels (59.3\arcsec), $\sigma = 2.56^{\circ}$, and $\epsilon_4 = 0.35^{\circ}$. The final determination of pitch  angle is therefore
  $-17.98^{\circ} \pm  2.61^{\circ}$. This result  is barely different
  from the result without star subtraction; the main difference is the
  redetermination  of the  dominant harmonic mode.   The percent  difference in
  mean pitch angle is $2.59\%$ with a $17.67\%$ reduction in error from
  the original.\label{fig9}}
\end{center}
\end{figure*}

In addition,  it is important to  consider the resolution  of the {\it
  2DFFT}   code   due   to   a   discrete  step   size   (see   Figure
\ref{fig6}). 
{\it 2DFFT}  captures $-50  \leq  p \leq  50$ values  in
discrete   steps    of   $0.25$   for   six    harmonic   modes   ($1 \leq m \leq 6$). Therefore,  only discrete values of pitch  angle are produced
by the subsequent conversion of  $p \rightarrow \phi$. The step size of the discrete Fourier transform is the analog of the frequency stepsize in One-Dimensional (1-D) discrete Fourier Transforms, the smallest measurable frequency. This leads to a
necessarily higher precision in the lower regime of pitch angle absolute
values and in the higher  order harmonic modes. The quantized error of
the mean  pitch angle due  to the resolution  of the code  (see Figure
\ref{fig7}) 
is  added  in  quadrature to  the  previously  determined
standard deviation of the mean pitch angle to give a total error. The final error is therefore
\begin{equation}\label{Error}
E_{\phi} =\sqrt{\left (\beta \sigma/\lambda \right)^{2}+\epsilon_m^2 }
\end{equation}
where $E_{\phi}$ is the total pitch angle error, $\epsilon_m$ is the quantized error for the dominant harmonic mode, $\sigma$ is the standard deviation about the mean pitch angle, $\beta$ is the distance (e.g., in pixels) from the innermost stable spiral structure (i.e., beyond the influence of a bulge or bar) to $90\%$ of the selected outer radius of the galaxy ($0.9r_{max}$), and $\lambda$ is the length (in the same units as used for $\beta$) of the stable range of radii over which the pitch angle is averaged. Figure \ref{fig9} 
serves as a good example of our error determination and its subsequent reduction by the use of star subtraction.


Equation \ref{Error} reflects the fact that in our method, a human researcher rather than a computer makes the final selection of pitch angle. That is, a balance of two main principles governs Equation \ref{Error}: the fluctuation across and the length of a chosen stable region of pitch angle as a function of inner radius. Our process ensures that the error about the mean pitch angle is appropriate, based on the choices made by the user. For example, an erratic ``stable" region or a short stable region will both be punished with appropriately high errors. Thus, a careful selection of stable region is required so as not to produce substantial errors.

\subsection{Inclination Angle and Galactic Center Position Errors}\label{subsect3.2}

The problem of making a poor choice of inner radius was addressed by
altering the code so that it calculates a pitch angle for all possible
inner radii. Other user-defined parameters have little impact on our results, and thus do not require such measures. The most important step is
deprojection, which requires the user to measure the galaxy's inclination
angle, presuming that the galaxy's disc is inherently circular. Tests with
a synthetic two-armed spiral with pitch angle of $-20^{\circ}$ (see Figure \ref{Syn_figs}b) demonstrate that measurement of pitch angle
is correct for any even number of arms and for inner radii up to $\approx 90\%$ of the outer radius (see
Figure \ref{Syn_figs}f). For a one-armed synthetic spiral, all harmonic modes are in agreement (see Figure \ref{Syn_figs}e).

\begin{figure*}
\begin{center}
\includegraphics[width=8.97cm]{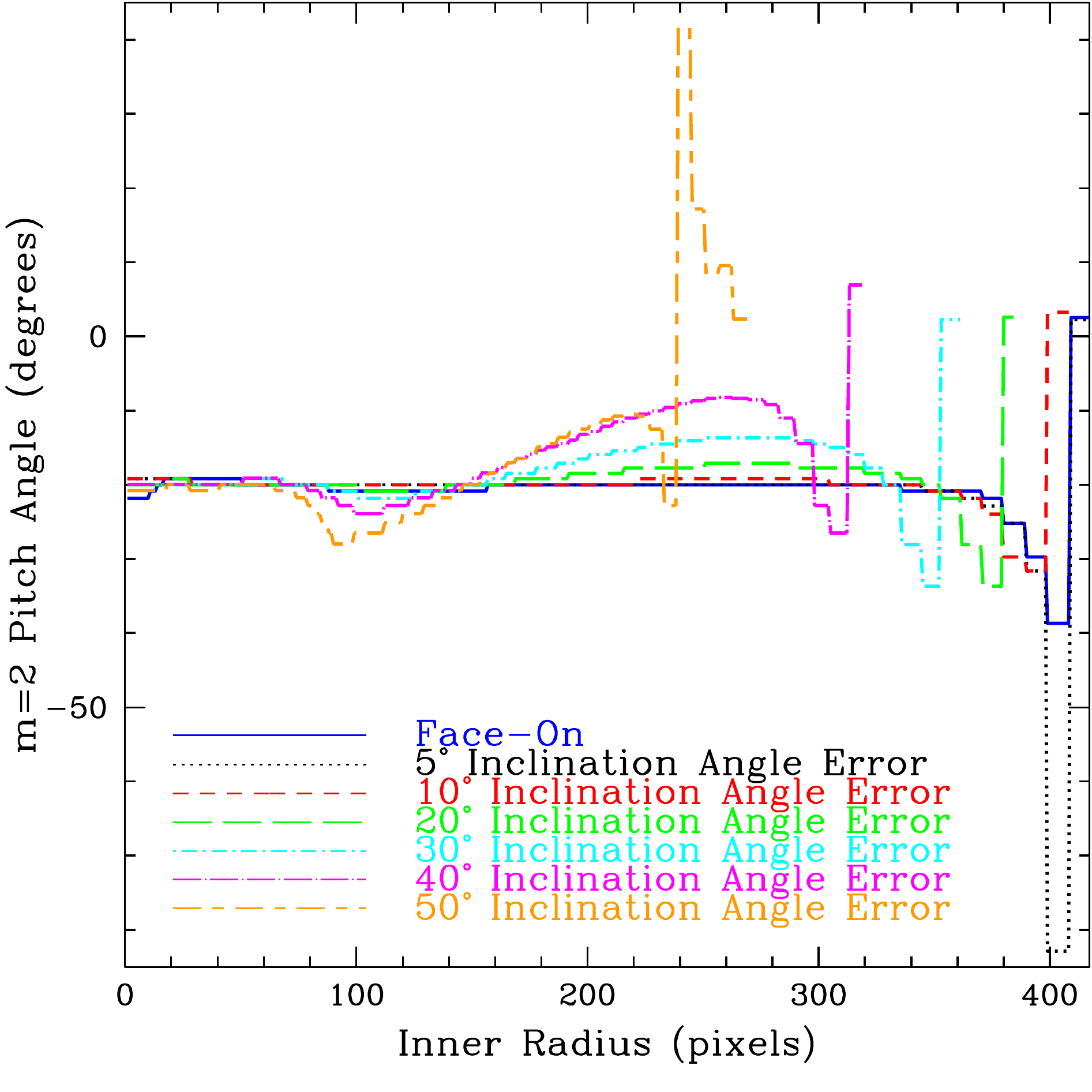}
\includegraphics[trim=0mm 8mm 0mm 0mm, clip, width=8.97cm]{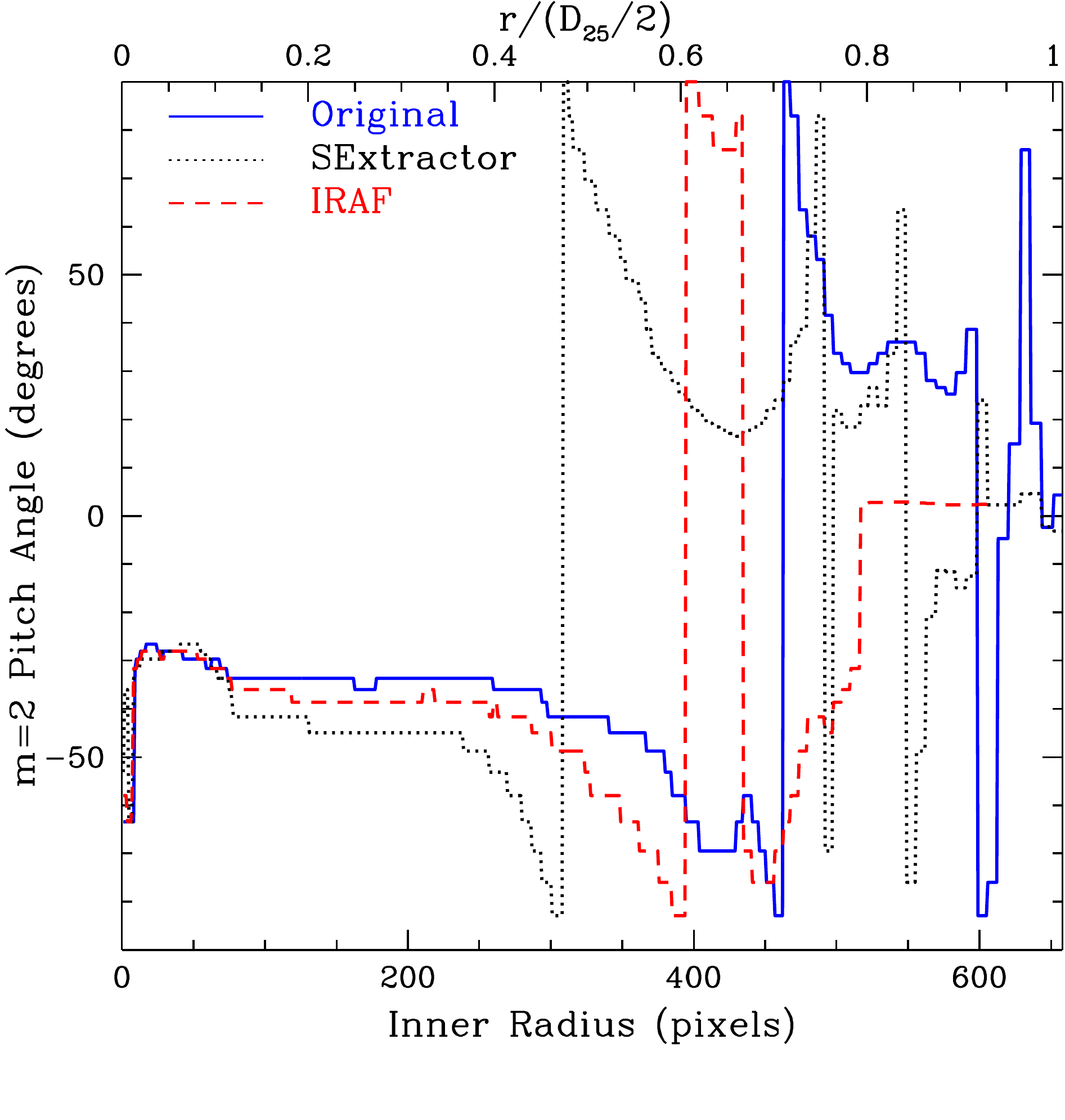}
\caption{Inclination angle tests on pitch angle output for a synthetic
  two-armed logarithmic  spiral   (left)  and   a  real two-armed galaxy (right).  {\it
    Fig.  \ref{fig11}a (left)}  -  Pitch angle  results for  different
  errors   in  inclination   angle   for  the   synthetic two-armed spiral   in
  Fig. \ref{Syn_figs}b.  Even at a  high degree  of simulated
  inclination angle error, the  mean pitch angles remain approximately
  the same  despite a gradually  shrinking stable region  across inner
  radii.  {\it Fig.  \ref{fig11}b (right)}  - Real two-armed galaxy inclination
  test  using  NGC  5247  (see  Fig. \ref{fig20}b).  Three  angles  of
  inclination  are  tested: Original  - the galaxy  before
  deprojection,  {\it SExtractor}  - incorporates  the  deprojection
  according  to   {\it SExtractor} ($PA=30.4^{\circ}$; $\alpha=36.97^{\circ}$),  and  IRAF  - incorporates  the
  deprojection  according  to IRAF ($PA=38.71^{\circ}$; $\alpha=25.18^{\circ}$).  Pitch  angle  results over  their
  respective stable regions are within each other's error bars. I-band
  pitch   angle   results   are   -   Original:   $-37.6^{\circ}   \pm
  5.69^{\circ}$,  {\it SExtractor}: $-35.62^{\circ}  \pm  10.96^{\circ}$, and
  IRAF: $-36.16^{\circ} \pm 9.25^{\circ}$.\label{fig11}}
\end{center}
\end{figure*}

\begin{figure*}
\begin{center}
\includegraphics[width=8.97cm]{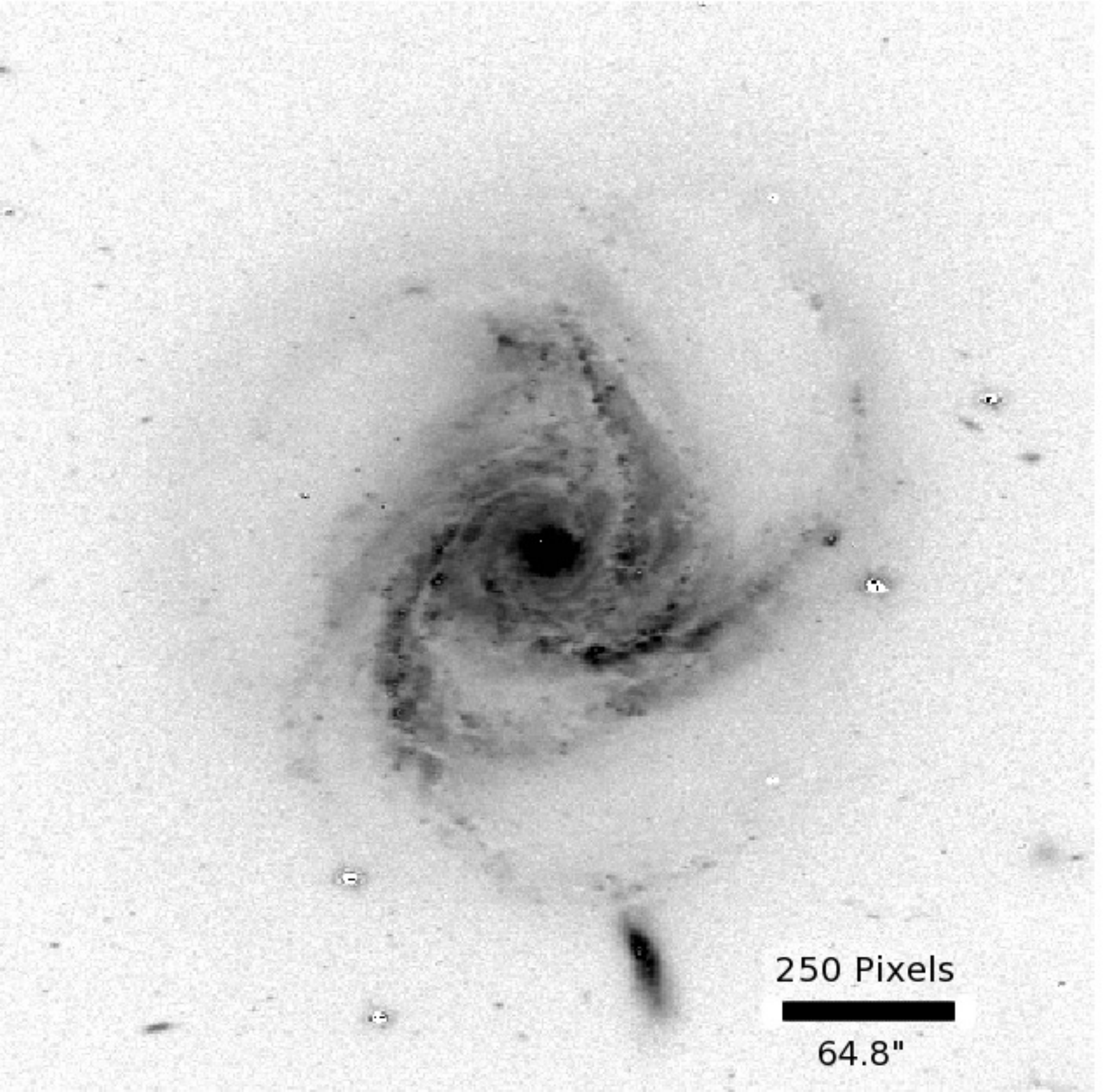}
\includegraphics[width=8.97cm]{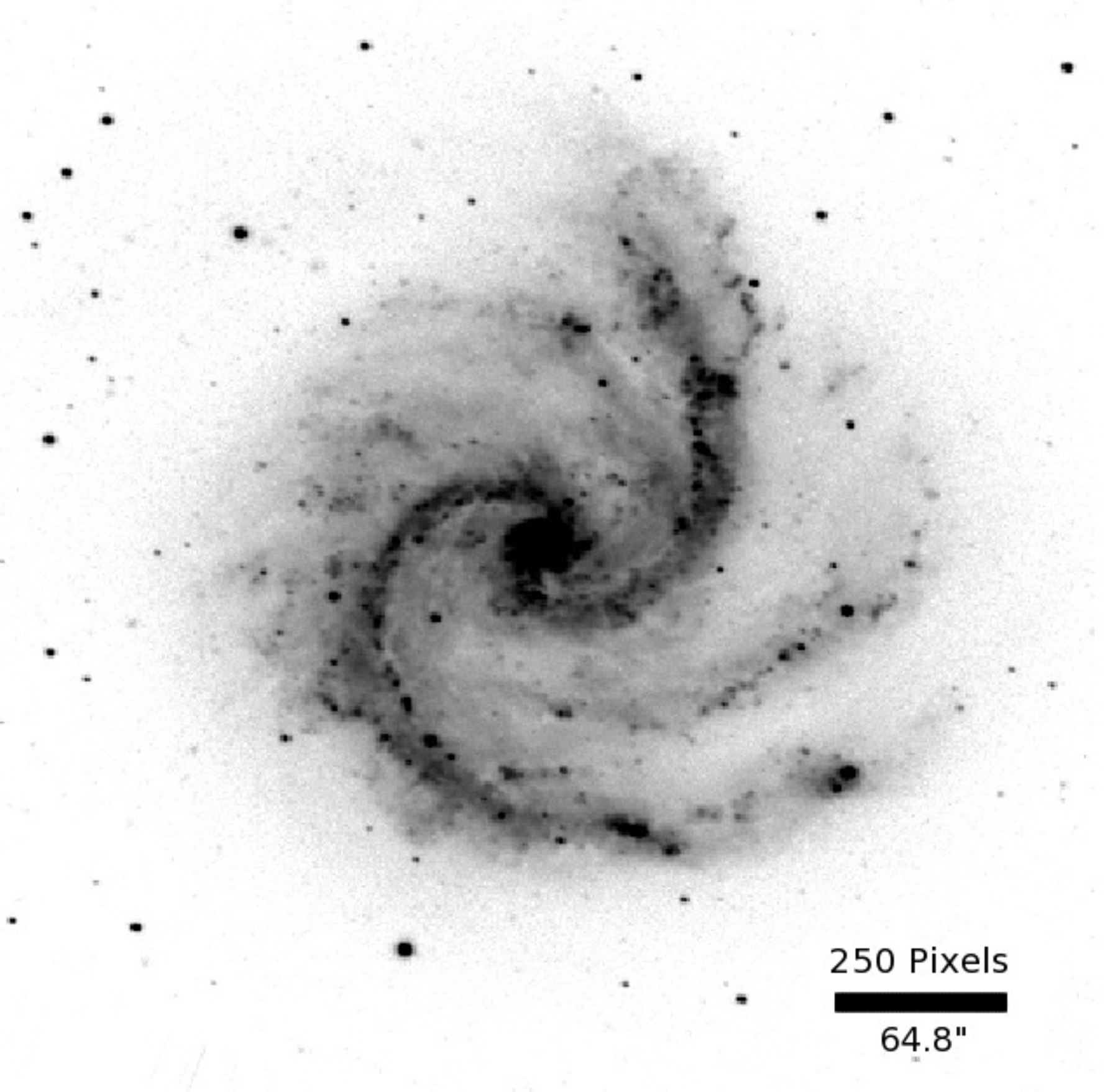}
\caption{{\it Fig. \ref{fig20}a (left)} - Deprojected ($PA = 160^{\circ}$ \& $\alpha = 53.84^{\circ}$) B-band   (inverted color) image of NGC 5054. {\it Fig. \ref{fig20}b (right)} - Deprojected ($PA = 20^{\circ}$ \& $\alpha = 28.36^{\circ}$) B-band (inverted color) image of NGC 5247.\label{fig20}}
\end{center}
\end{figure*}

When the synthetic two-armed spiral (see Figure \ref{Syn_figs}b) is shrunk along one axis incrementally to simulate
an increasingly inaccurate deprojection, the results show that there is
still a stable region of inner radii with the correct measure of
pitch angle (see Figure \ref{fig11}a). 
Similarly with a real two-armed galaxy, NGC 5247
(see Figure \ref{fig20}b), 
it is of interest that an incorrect choice of inclination
angle merely causes a gradual reduction in the length of the stable region over which
the selected inner radii yield the correct pitch angle (see Figure \ref{fig11}b). Thus, 
deprojection is still an important step, but is unlikely to be a significant
source of error when using the script, which calculates pitch angle for
a wide variety of possible inner radii.

\begin{figure*}
\begin{center}
\includegraphics[width=8.97cm]{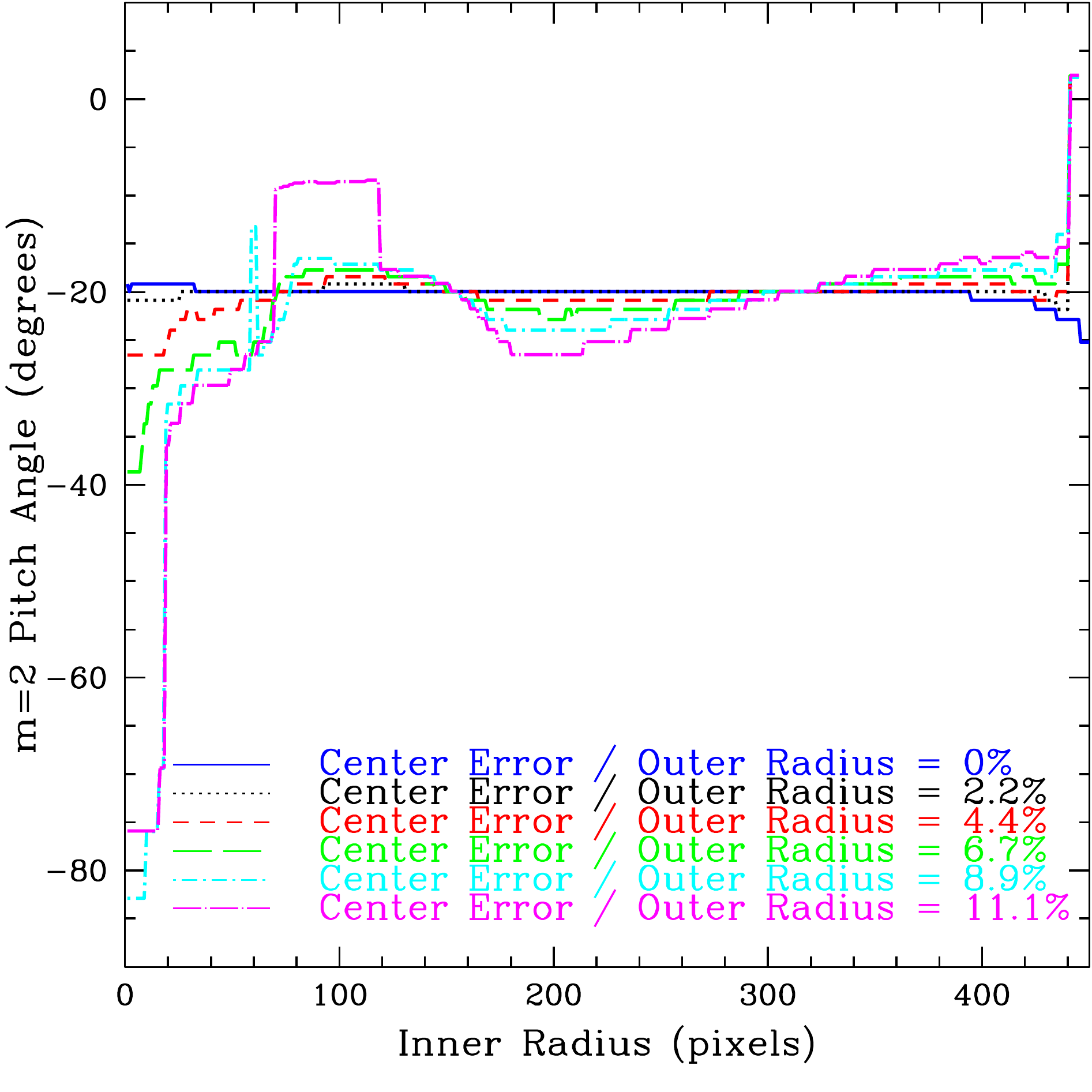}
\includegraphics[trim=0mm 7mm 0mm 0mm, clip, width=8.97cm]{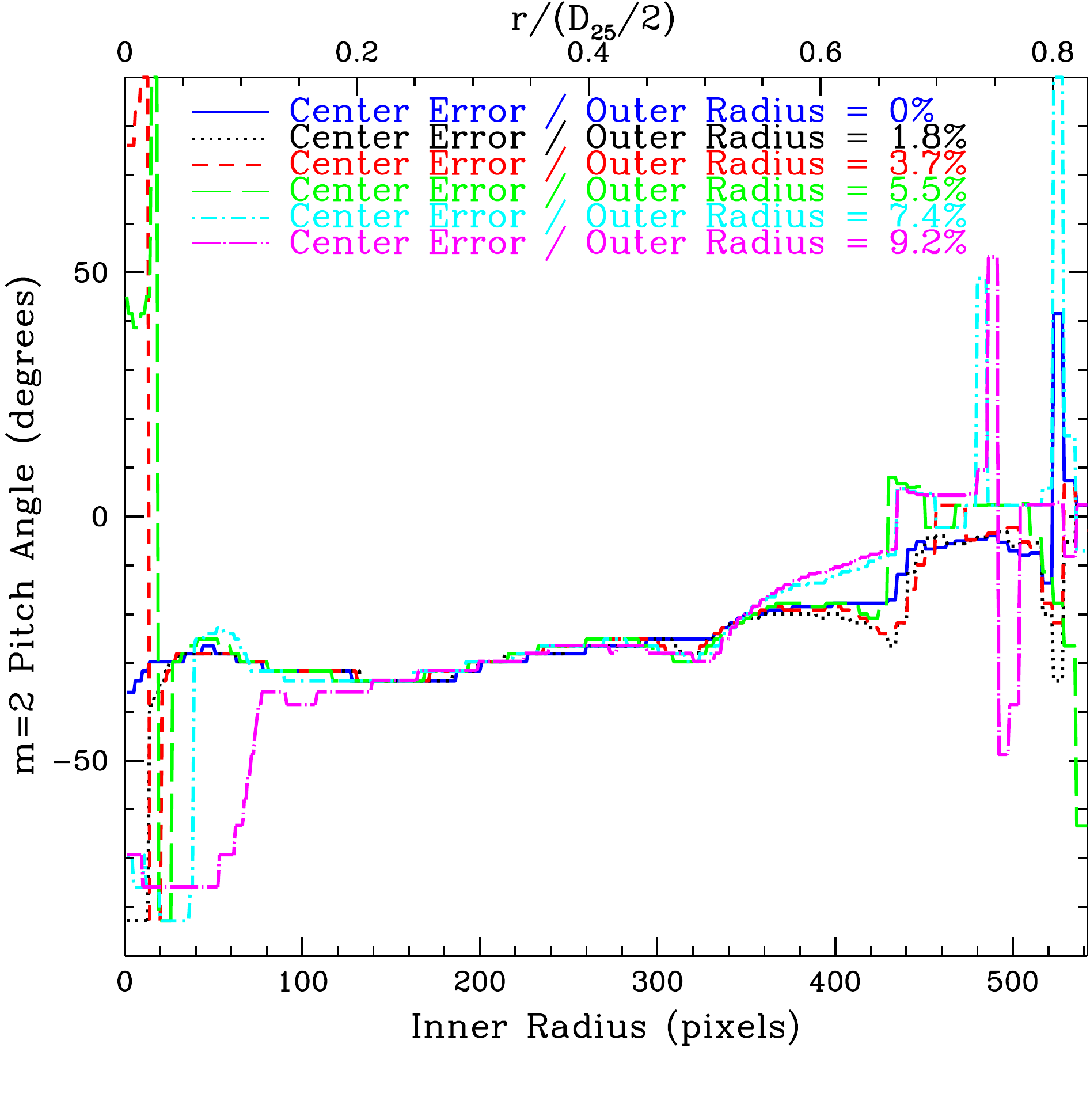}
\caption{Center error tests on pitch angle for a synthetic two-armed logarithmic
  spiral (left) and  a real two-armed galaxy (right). Errors of  10, 20, 30, 40,
  and 50  pixels from  the previously determined  center are  used for
  both.  {\it Fig.  \ref{fig12}a  (left)} -  Pitch  angle results  for
  different errors in center determination for the synthetic two-armed spiral in
  Fig. \ref{Syn_figs}b.  As  the error increases,  the stable
  region  gradually decreases,  yet the  approximate mean  pitch angle
  remains  about the  same.  {\it Fig.  \ref{fig12}b  (right)} -  Real
  two-armed galaxy  center test  using a B-band image of NGC  5247 (see  Fig.  \ref{fig20}b) after
  deprojection ($PA = 38.71^{\circ}$ \& $\alpha = 25.18^{\circ}$) was  performed. The same case is  true for
  the real galaxy image; the mean pitch angle remains constant despite
  a decreasing stable region with increasing error.\label{fig12}}
\end{center}
\end{figure*}

Similarly, when choosing the center of the galaxy image, tests with a
synthetic two-armed spiral (see Figure \ref{Syn_figs}b) and a real two-armed galaxy (NGC 5247, see Figure \ref{fig20}b) suggest that incremental errors in centering only gradually reduce the
stable region without affecting the actual measure of pitch angle 
(provided the stable region of roughly constant pitch angle remains
lengthy enough to be found, see Figure \ref{fig12}). 
Overall, these tests are
a testament to the robustness of the {\it 2DFFT} algorithm.


\subsection{Bulges and Bars}\label{subsect3.4}

\begin{figure*}
\begin{center}
\includegraphics[width=8.97cm]{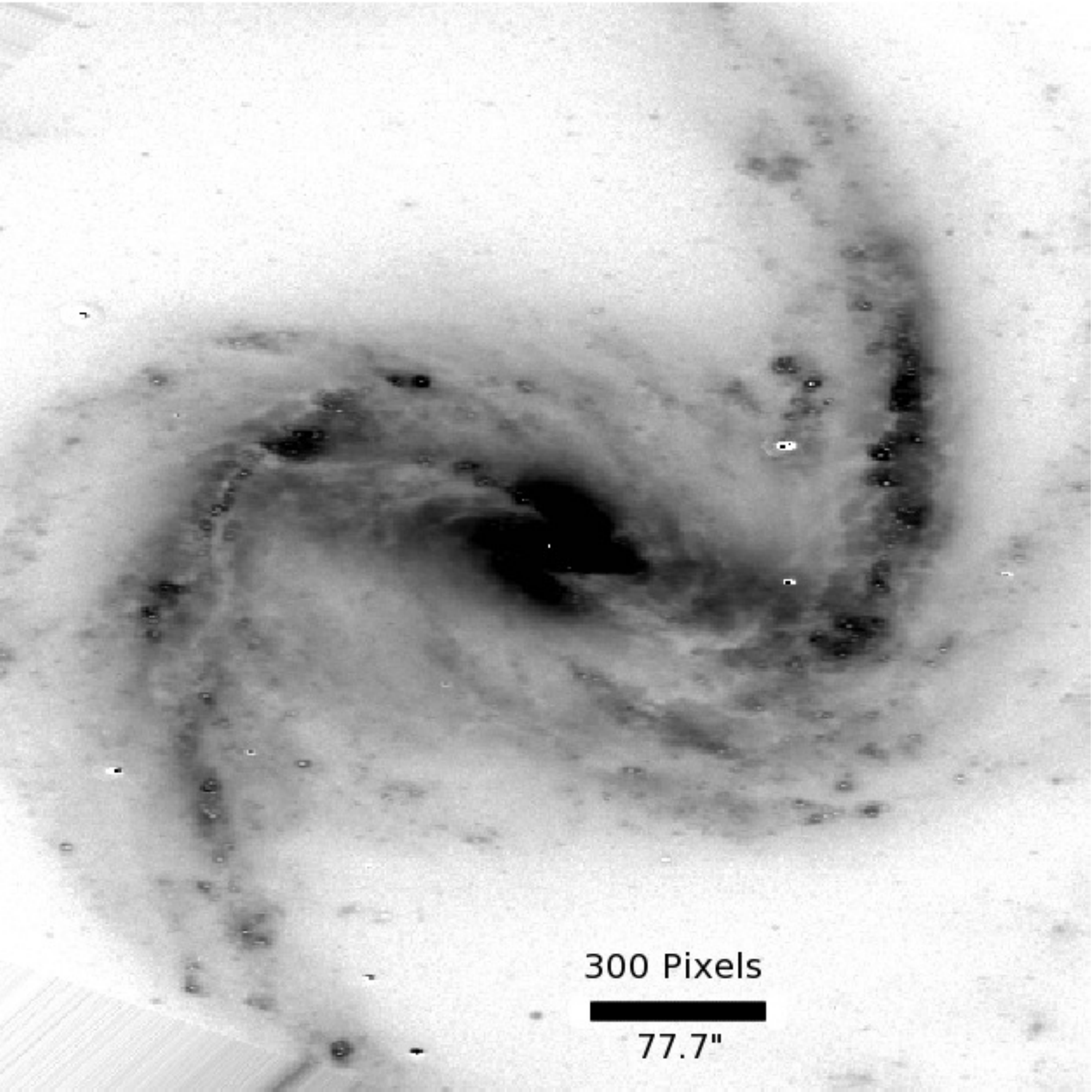}
\includegraphics[trim=0mm 12mm 0mm 0mm, clip, width=8.97cm]{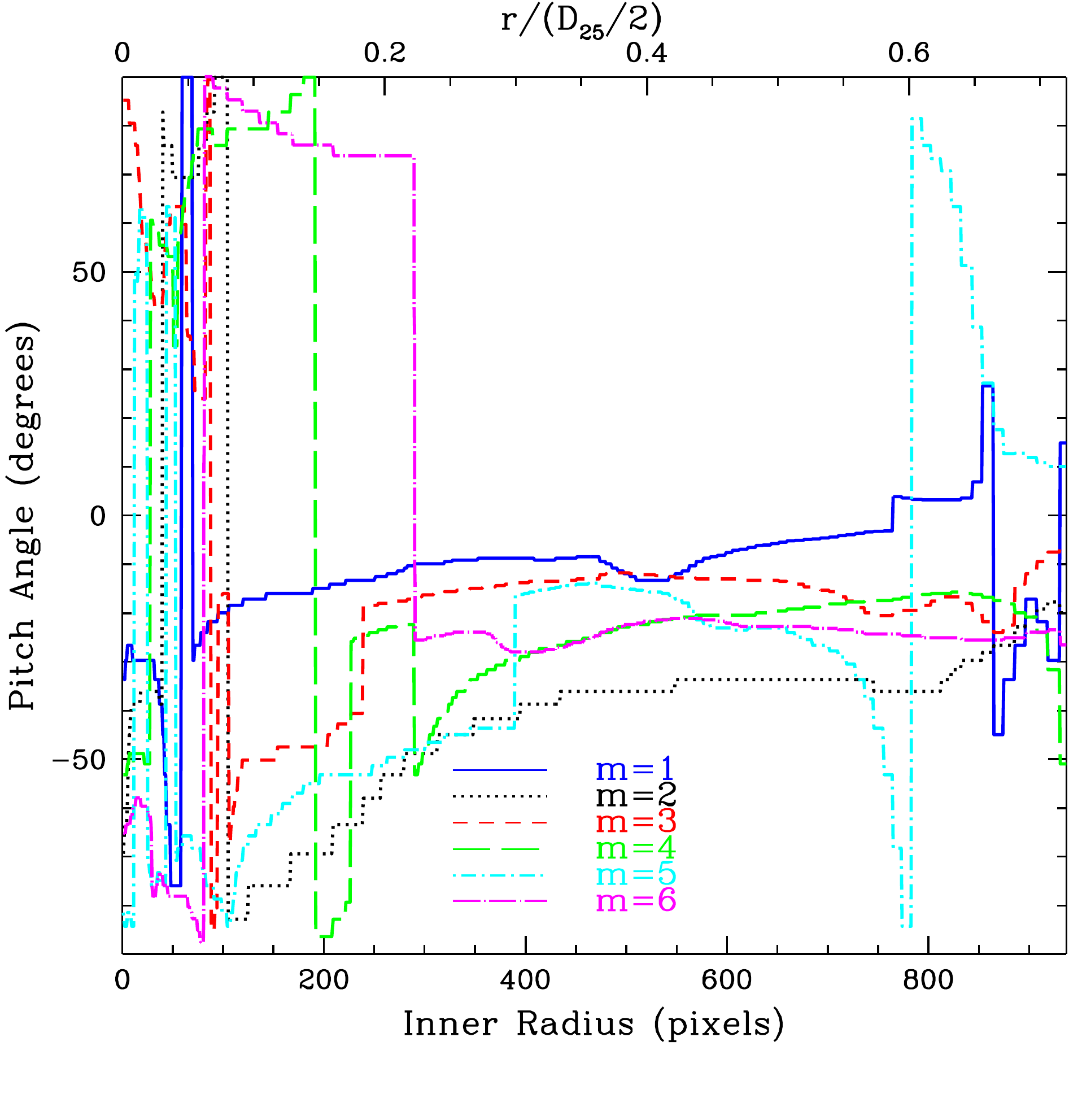}
\caption{{\it  Fig.   \ref{fig25}a  (left)}  -   Star-subtracted  and
  deprojected  B-band  (inverted color) image of  NGC  1365: $PA  = 32^{\circ}$  and
  $\alpha =  56.63^{\circ}$. {\it  Fig. \ref{fig25}b (right)}  - Pitch
  angle as  a function  of inner  radius for NGC  1365. A  stable mean
  pitch angle of  $-34.81^{\circ}$ is determined for the  $m = 2$ harmonic mode
  from a minimum  inner radius of 451 pixels (117\arcsec) to  a maximum inner radius
  of 812 pixels (210\arcsec),  with an outer radius of 938  pixels (243\arcsec). This stretch of
  361 pixels (93.5\arcsec) occupies  38\% of  the galactic disc. 
  Equation \ref{Error} yields $E_{\phi} = 2.80^{\circ}$ with $\lambda = 377$ pixels (97.6\arcsec), $\beta = 462$ pixels (120\arcsec), $\sigma = 1.17^{\circ}$, and $\epsilon_2 = 2.40^{\circ}$. The final  determination  of pitch  angle is  therefore
  $-34.81^{\circ} \pm 2.80^{\circ}$.  This galaxy demonstrates a large
  bar, approximately 34\%  of its outer radius. The  absolute value of
  the  pitch angle  can be  seen to  gradually decrease  from  $\phi =
  -82.87^{\circ}$  at an  inner radius  of  105 pixels (27.2\arcsec) until $\phi  =
  -36.03^{\circ}$ at an inner radius  of 435 pixels (113\arcsec), signaling the end
  of the bar.\label{fig25}}
\end{center}  
\end{figure*}

Our synthetic two-armed spiral was also used to study the effects of circular
bulges and bars in galactic nuclei on pitch angle measurements. When a circular bulge component is
added  to the  synthetic two-armed spiral  (see Figure  \ref{Syn_figs}c),  the even
numbered  harmonic modes are  unaffected, whereas  the odd  harmonic modes are
systematically different with the lower harmonic modes being the worst
(see  Figure \ref{Syn_figs}g). In contrast,  when a  bar component  is
added to the synthetic two-armed spiral (see Figure \ref{Syn_figs}d), the resulting
value of the measured pitch angle is significantly increased  at inner radii, with the correct
pitch angle value returning  after the inner radius is  beyond the extent of
the bar (see Figure \ref{Syn_figs}h). As an example, NGC 1365 (see Figure \ref{fig25}) 
displays a similar bar to the two-armed synthetic spiral with a bar added. These results confirm  that circular bulges  should not  affect pitch  angle, whereas
the barred geometry can significantly bias  pitch angle measurements towards higher values. 
We  are therefore confident  in the necessity of our efforts to systematically exclude barred nuclei
from the pitch angle measurement annulus.

\subsection{Problems with Underlying Presumptions}\label{subsect3.5}

So far we have presumed that the pitch angle of a logarithmic spiral is
a meaningful quantity to measure in images of disc galaxies. Certainly
there are very many disc galaxies for which logarithmic spiral patterns
are the most obvious feature of the disc, as the human eye perceives
it. Nevertheless, two important objections might be made concerning the measurement
of pitch angles as a useful characteristic of galaxies. One is that the
pitch angle may be different for the same galaxy when viewed at different
wavelengths. The other is that the pitch angle might vary with the radius of
the disc, in other words that the spiral is not truly logarithmic.

\subsubsection{The Effect of Wavelength on Pitch Angle}\label{Wavelength}

It is important to consider the
possibility of  different pitch angles arising  in different wavebands
of light and what physical processes that might imply. For instance, optical B-band images tend to trace
the bright massive star forming  regions of a galaxy and near-infrared
(NIR) images  tend to  trace the old  stellar populations  in galaxies
\citep{Seigar:James:1998,Eskridge:2002}.  The  old stellar  population
traces    the   spiral    density    wave   \citep{Seigar:James:1998}.
Furthermore,  a spiral  that appears  flocculent in  the B-band may
appear  to  have a  weak  grand  design  spiral in  the  near-infrared
\citep{Thornley:1996}.

\citet{Kendall:2011} used a 1-D FFT analysis on
optical and NIR images of grand design spiral galaxies to measure their $m =
2$ pitch  angles and concluded  that a good correlation  exists between
galaxies   being   grand  design   in   the   infrared   and  in   the
optical. \citet{Seigar:2006}  demonstrates that a  1:1 relation exists
between the  B and NIR band pitch  angles for a sample  of 66 galaxies
from  a combination  of the  CGS  \citep{Ho:2011} and  the Ohio  State
University      Bright     Spiral     Galaxy      Survey     \citep[OSUBSGS;][]{Eskridge:2002}. Alternatively, \citet{Grosbol:Patsis:1998} propose a contrary view. They find a systematic trend of arms being tighter in bluer colors by investigating five galaxies in B, V, I, and K$\arcmin$ filters. Admittedly, two of their five galaxies are tight spirals for which little or no change in pitch angle is observed, but it seems that more work with multiple filters is required.

\begin{figure}
\begin{center}
\includegraphics[width=8.6cm]{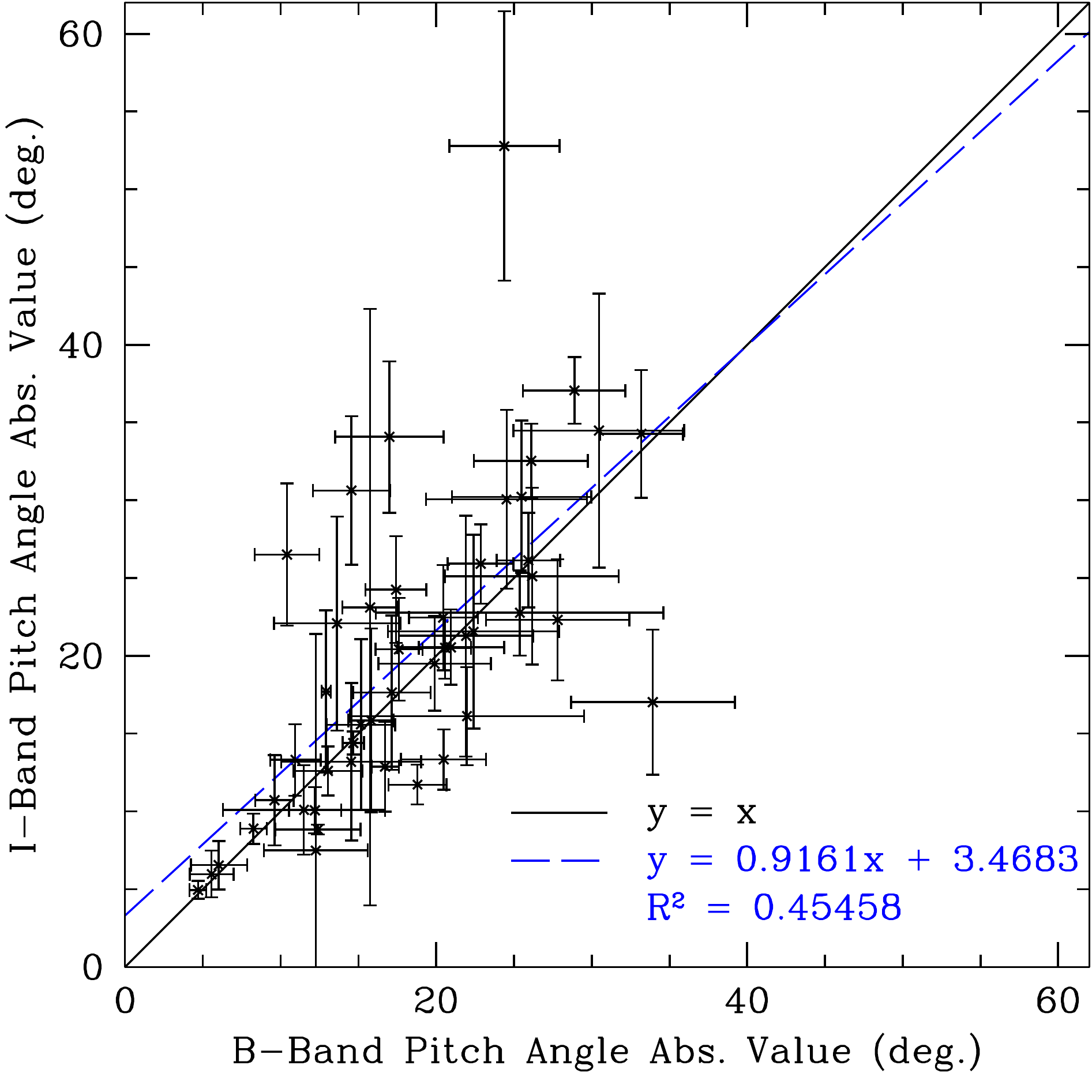}
\caption{Pitch angles for 47 spiral galaxies in both blue (B) and infrared (I) wavebands. The solid black line represents a 1:1 ratio. The dashed \textcolor{blue}{blue} line is a best-fit line, which is consistent with a 1:1 ratio within errors.\label{fig15}}
\end{center}
\end{figure}

Using  our method, we have remeasured a subset of 47 of the galaxies
appearing  in \citet[they used an earlier version of this method]{Seigar:2006} and have also identified a seemingly 1:1
relation  (see Figure \ref{fig15} and Table \ref{Galaxies}). 
Therefore,  despite  seemingly
small-scale differences  between spiral arms  in different wavelengths
of the optical-NIR spectrum,  the overall structure of the spiral arms, and thus the proposed density wave, is consistent across the optical-NIR
spectrum. This is in opposition to the prediction of the density wave theory that different pitch angles are expected for spirals when observed in different bands \citep{Hozumi:2003}. Although, the expected difference in pitch angle across wavelength is probably small enough that an extremely high precision fit would be necessary to falsify this prediction of density wave theory. 
\begin{deluxetable*}{llcrcrrcr}
\tablecolumns{9}
\tablecaption{Pitch Angle/Wavelength Comparison\label{Galaxies}}
\tablehead{
\colhead{} & \colhead{} & \colhead{} & \colhead{} & 
\multicolumn{2}{c}{B-Band} & \colhead{} & 
\multicolumn{2}{c}{I-Band}  \\
\cline{5-6} \cline{8-9}
\colhead{Galaxy Name} & \colhead{Morphology} & \colhead{$\alpha$ (deg.)} & 
\colhead{} & \colhead{m} & \colhead{$\phi$ (deg.)} & \colhead{} & \colhead{m} & 
\colhead{$\phi$ (deg.)}
}
\startdata
ESO 121-026 & SB(rs)bc & 50.95 & & 2 & $10.06 \pm 1.30$ & & 3 & $11.24 \pm 3.05$  \\
ESO 582-012 & SAB(rs)c & 50.21 & & 2 & $21.45 \pm 2.32$ & & 2 & $23.54 \pm 3.54$  \\
IC 2522 & SB(s)c pec & 44.77 & & 3 & $-26.73 \pm 4.69$ & & 5 & $-31.70 \pm 5.15$  \\
IC 2537 & SAB(rs)c & 49.46 & & 4 & $27.37 \pm 3.84$ & & 4 & $34.11 \pm 2.51$  \\
IC 3253 & SA(s)c: & 67.05 & & 4 & $-17.53 \pm 0.93$ & & 4 & $-13.50 \pm 3.03$  \\
IC 4538 & SAB(s)c: & 39.65 & & 4 & $-17.98 \pm 2.61$ & & 3 & $-18.49 \pm 5.20$  \\
IC 4831\tablenotemark{a} & (R')SA(s)ab & 77.47 & & 2 & $-22.93 \pm 2.41$ & & 2 & $-16.07 \pm 1.46$  \\ 
NGC 150 & SB(rs)b: & 60.00 & & 2 & $14.29 \pm 4.26$ & & 1 & $23.15 \pm 7.21$  \\ 
NGC 157 & SAB(rs)bc & 50.21 & & 3 & $8.66 \pm 0.89$ & & 3 & $9.32 \pm 1.02$  \\
NGC 289 & SB(rs)bc & 44.77 & & 5 & $19.71 \pm 1.95$ & & 3 & $12.29 \pm 1.36$  \\
NGC 578 & SAB(rs)c & 50.95 & & 3 & $16.51 \pm 1.88$ & & 2 & $24.24 \pm 20.11$  \\
NGC 613 & SB(rs)bc & 40.54 & & 3 & $21.57 \pm 1.77$ & & 3 & $21.50 \pm 2.06$  \\
NGC 908 & SA(s)c & 64.53 & & 3 & $15.26 \pm 2.61$ & & 2 & $32.12 \pm 5.01$  \\
NGC 1187 & SB(r)c & 41.41 & & 4 & $-21.96 \pm 3.61$ & & 4 & $-21.55 \pm 2.54$  \\
NGC 1232 & SAB(rs)c & 28.36 & & 3 & $-25.71 \pm 5.43$ & & 6 & $-31.51 \pm 6.03$  \\
NGC 1292 & SA(s)c & 64.53 & & 3 & $-15.89 \pm 2.30$ & & 3 & $-16.34 \pm 5.75$  \\
NGC 1337\tablenotemark{a} & SA(s)cd & 77.41 & & 2 & $-16.53 \pm 2.40$ & & 3 & $-19.58 \pm 2.45$  \\ 
NGC 1353 & SB(rs)b: & 65.80 & & 4 & $13.68 \pm 2.31$ & & 4 & $13.21 \pm 1.65$  \\
NGC 1365 & SB(s)b & 56.63 & & 2 & $-34.81 \pm 2.80$ & & 2 & $-35.94 \pm 4.31$  \\
NGC 1559 & SB(s)cd & 55.25 & & 2 & $-26.61 \pm 9.69$ & & 2 & $-23.87 \pm 2.87$  \\
NGC 1566 & SAB(s)bc & 36.87 & & 2 & $-17.81 \pm 3.67$ & & 2 & $-35.73 \pm 5.10$  \\
NGC 1792 & SA(rs)bc & 60.00 & & 3 & $-20.86 \pm 3.79$ & & 3 & $-20.45 \pm 3.20$  \\
NGC 1964 & SAB(s)b & 67.67 & & 2 & $-12.86 \pm 3.49$ & & 2 & $-7.85 \pm 14.61$  \\
NGC 2082 & SB(r)b & 19.95 & & 3 & $23.05 \pm 7.90$ & & 3 & $16.91 \pm 3.31$  \\
NGC 2090 & SA(rs)c & 60.00 & & 4 & $4.91 \pm 0.56$ & & 4 & $5.19 \pm 0.60$  \\
NGC 2280 & SA(s)cd & 60.66 & & 4 & $21.47 \pm 2.87$ & & 2 & $13.98 \pm 2.02$  \\
NGC 2835 & SB(rs)c & 47.93 & & 3 & $-23.97 \pm 2.22$ & & 3 & $-27.17 \pm 2.68$  \\
NGC 2935 & (R$'$)SAB(s)b & 38.74 & & 2 & $-15.24 \pm 4.72$ & & 2 & $-13.82 \pm 5.32$  \\
NGC 3052 & SAB(r)c: & 49.46 & & 3 & $-18.45 \pm 1.59$ & & 2 & $-21.41 \pm 3.46$  \\
NGC 3054 & SAB(r)b & 52.41 & & 3 & $12.80 \pm 1.77$ & & 3 & $10.56 \pm 1.55$  \\
NGC 3223 & SA(s)b & 52.41 & & 4 & $-10.92 \pm 2.17$ & & 6 & $-27.79 \pm 4.79$  \\
NGC 3261 & SB(rs)b & 40.54 & & 6 & $15.38 \pm 0.71$ & & 6 & $15.09 \pm 0.78$  \\
NGC 3318 & SAB(rs)b & 57.32 & & 3 & $35.58 \pm 5.53$ & & 3 & $17.85 \pm 4.89$  \\
NGC 3450 & SB(r)b & 28.36 & & 6 & $-13.55 \pm 0.31$ & & 2 & $-18.57 \pm 5.48$  \\
NGC 3513 & SB(rs)c  & 37.81 & & 1 & $5.84 \pm 1.46$ & & 1 & $6.26 \pm 1.57$  \\
NGC 3887 & SB(r)bc & 40.54 & & 4 & $-29.16 \pm 4.82$ & & 4 & $-23.40 \pm 4.08$  \\
NGC 4027 & SB(s)dm & 41.41 & & 1 &$-12.06 \pm 5.47$ & & 1 & $-10.58 \pm 3.01$  \\
NGC 4030 & SA(s)bc & 44.77 & & 3 & $23.48 \pm 5.76$ & & 3 & $22.60 \pm 6.52$  \\
NGC 4050 & SB(r)ab & 47.16 & & 1 & $-6.32 \pm 1.90$ & & 1 & $-6.85 \pm 1.63$  \\
NGC 4930 & SB(rs)b & 34.92 & & 3 & $30.29 \pm 3.45$ & & 6 & $38.86 \pm 2.24$  \\
NGC 4939 & SA(s)bc & 59.34 & & 6 & $11.48 \pm 1.71$ & & 6 & $13.95 \pm 2.41$  \\
NGC 4995 & SAB(rs)b & 50.21 & & 2 & $13.00 \pm 2.88$ & & 6 & $9.27 \pm 0.32$  \\
NGC 5054 & SA(s)bc & 53.84 & & 3 & $-25.57 \pm 3.72$ & & 3 & $-55.33 \pm 9.07$  \\
NGC 5247 & SA(s)bc & 25.18 & & 2 & $-31.94 \pm 5.75$ & & 2 & $-36.16 \pm 9.25$  \\
NGC 5483 & SA(s)c & 23.07 & & 2 & $-22.98 \pm 4.52$ & & 2 & $-22.31 \pm 8.12$  \\
NGC 5967 & SAB(rs)c: & 53.84 & & 3 & $18.26 \pm 2.05$ & & 3 & $25.42 \pm 3.60$  \\
NGC 6215 & SA(s)c & 30.68 & & 4 & $-27.43 \pm 5.85$ & & 4 & $-26.34 \pm 5.95$  \\
NGC 6221 & SB(s)c & 44.77 & & 6 & $-27.18 \pm 2.14$ & & 6 & $-27.41 \pm 3.19$  \\
NGC 6300 & SB(rs)b & 47.93 & & 4 & $-16.58 \pm 1.52$ & & 4 & $-16.61 \pm 6.19$  \\
\enddata
\tablecomments{Col. (1) galaxy name; col. (2) morphological type from the RC3 \citep{RC3}; col. (3) inclination angle; col. (4) B-band dominant harmonic mode; col. (5) B-band pitch angle; col. (6) I-band dominant harmonic mode; and col. (7) I-band pitch angle.}
\tablenotetext{a}{Not in \citet{Seigar:2006} or plotted in Figure \ref{fig15}.}
\end{deluxetable*}

From our experience, we have become accustomed to preferring B-band images in general due to their characteristic clarity of galactic stellar components. However, our comparison of pitch angles in different wavebands has convinced us that we can typically measure pitch angle across a wide range of electromagnetic wavelengths. In that regard, we have successfully measured pitch angles of galaxies in the extreme cases of far-ultraviolet and 21 cm radio wavelength images when no other imaging data was available.

\subsubsection{Variable Pitch Angle with Galactic Radius}\label{Variable}

Occasionally, spiral arms may appear to change pitch angle
in the outer region of the disc, sometimes discontinuously. These are
more the exception than the rule and we have generally preferred to measure
the inner part of the disc in such cases or use more elaborate processing methods (see \S \ref{subsect4.1}) to mitigate the severity of pitch angle variability. Considering the case where a dichotomy exists between the pitch angles measured in the inner and outer regions of a galactic disc, the code can be made to run iteratively for two separate regions of the galaxy and average the results to yield an average pitch angle for the disc. However, if the pitch angle results are subsequently used for building relationships to processes in the nucleus of a galaxy \citep[e.g.,][]{Seigar:2008}, pitch angles for the innermost portion of a galaxy perhaps make the most physical sense and are furthermore not as susceptible to extragalactic interaction. It is also likely that the entire extent of a galaxy might not display logarithmic spirals. If so, our stable regions are selected to only highlight clearly logarithmic sections of spiral arms.

\begin{figure*}
\begin{center}
\includegraphics[width=5.95cm]{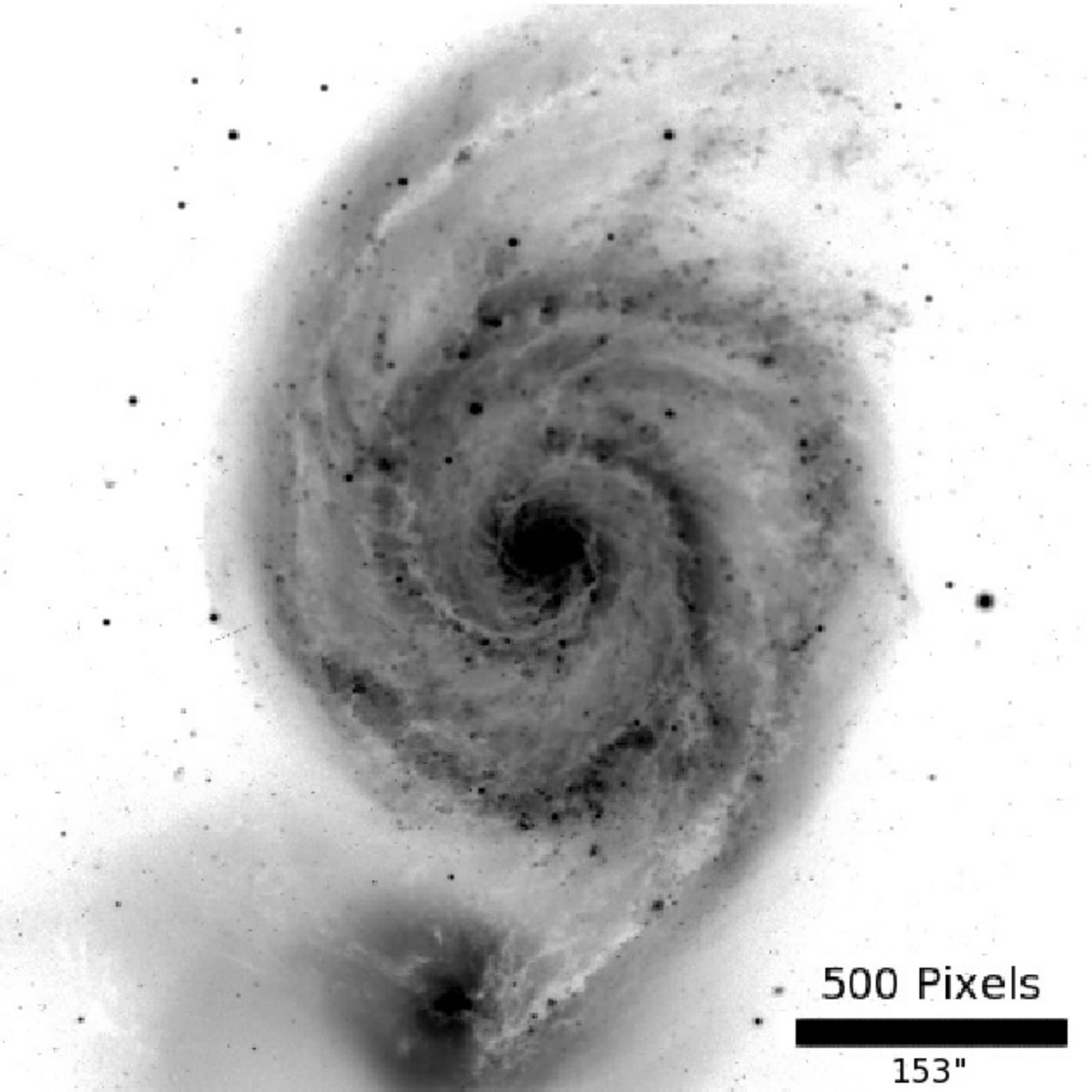} 
\includegraphics[trim = 0mm 10mm 0mm 0mm, clip, width=5.95cm]{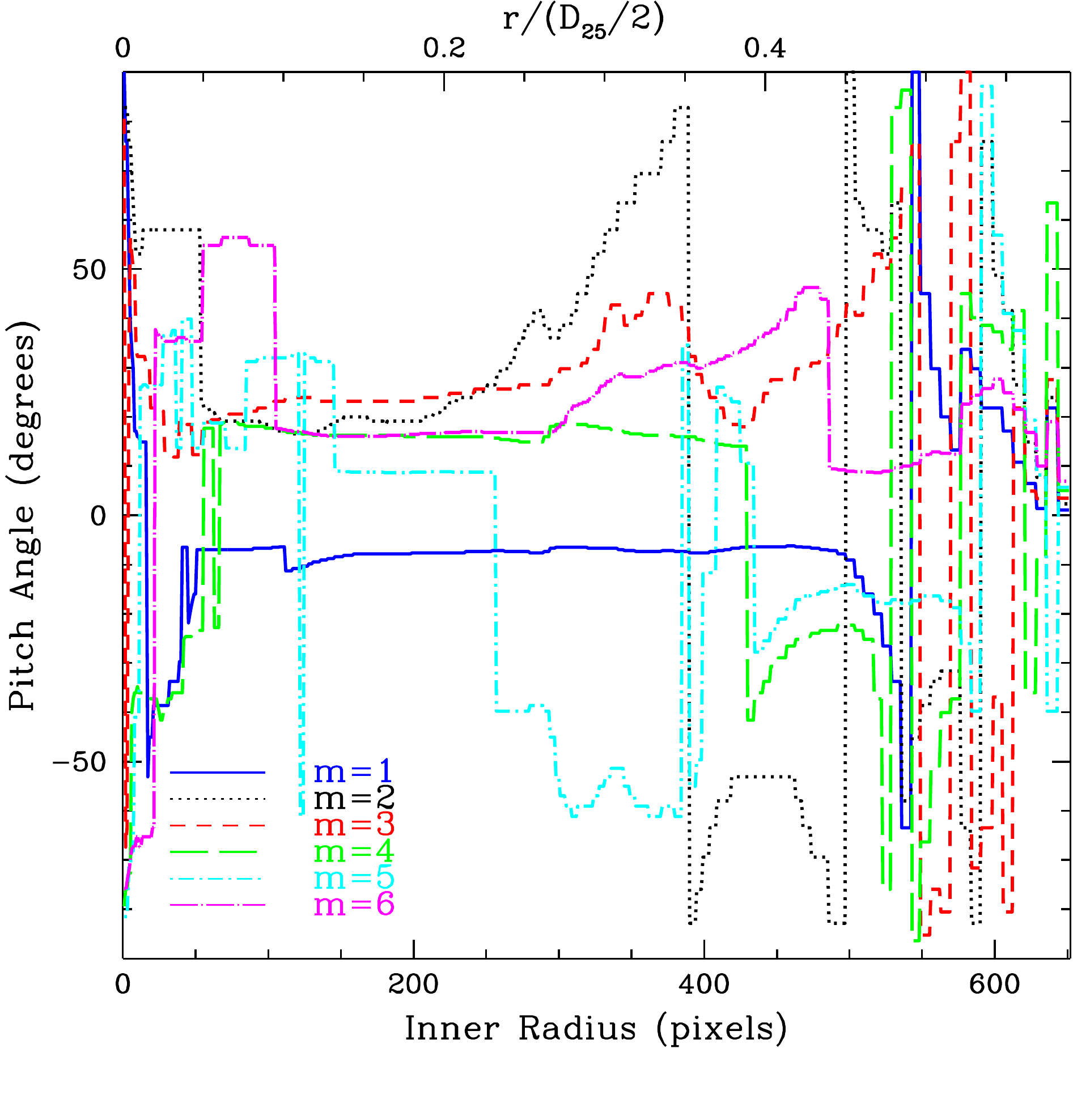}
\includegraphics[trim = 0mm 10mm 0mm 0mm, clip, width=5.95cm]{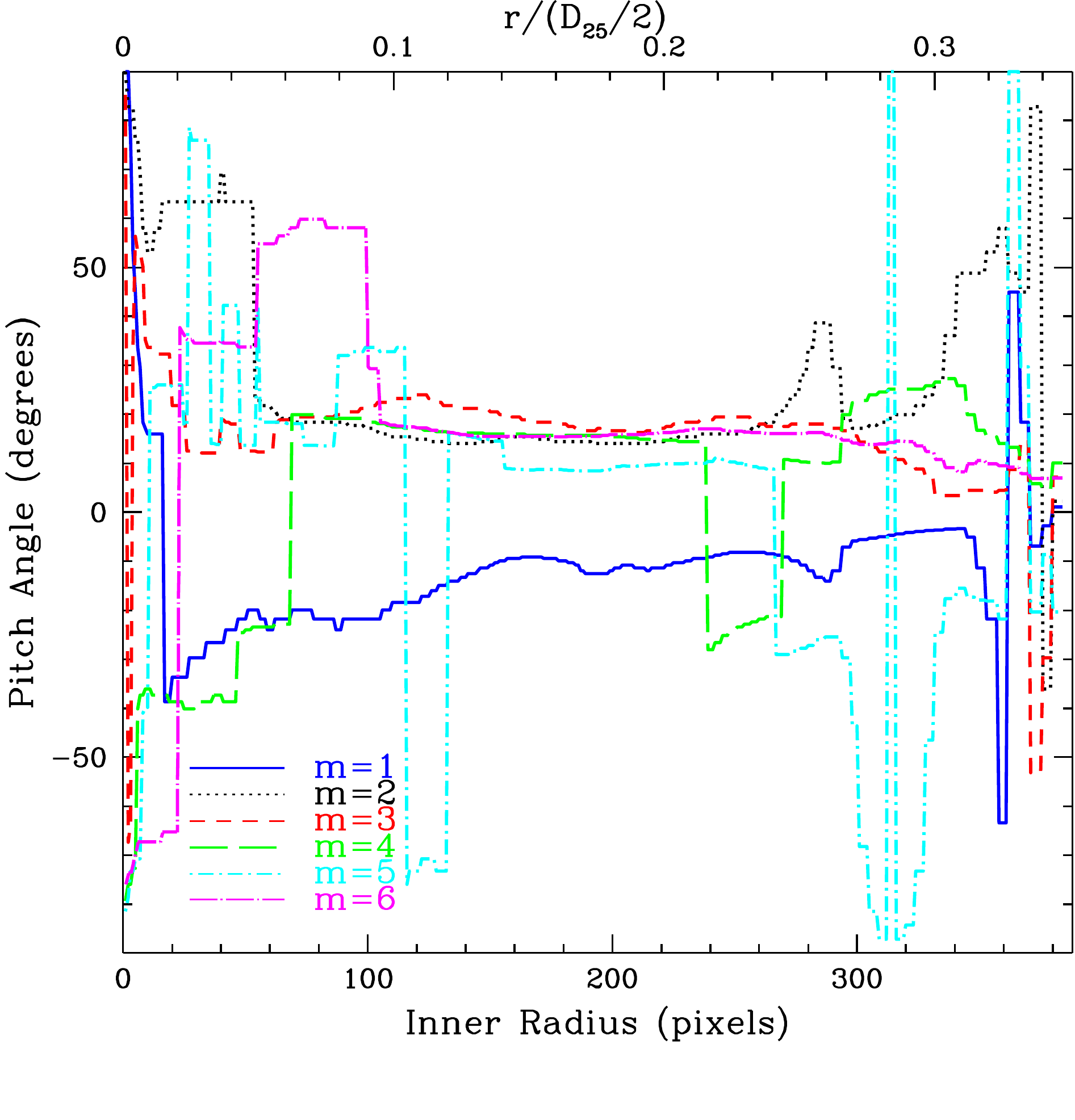}
\caption{{\it  Fig.  \ref{M51}a (left)} - Deprojected ($PA = -154.1^{\circ}$ \& $\alpha = 19.09^{\circ}$) B-band (inverted color) image of M51 (M51a and its dwarf companion galaxy M51b) acquired from NED (imaging from KPNO 2.1 m CFIM with a pixel scale of 0.305\arcsec$\,$pixel$^{-1}$). {\it Fig. \ref{M51}b (middle)} - A  stable mean
  pitch angle of  $19.13^{\circ}$ is determined for the  $m = 2$ harmonic mode
  from a minimum  inner radius of 54 pixels (16.5\arcsec) to  a maximum inner radius
  of 229 pixels (69.8\arcsec),  with an outer radius of 653  pixels (199\arcsec). This stretch of
  175 pixels (53.4\arcsec) occupies  27\% of  the galactic disc. Equation \ref{Error} yields $E_{\phi} = 4.76^{\circ}$ with $\lambda = 175$ pixels (53.4\arcsec), $\beta = 534$ pixels (163\arcsec), $\sigma = 1.54^{\circ}$, and $\epsilon_2 = 0.78^{\circ}$. The final  determination  of pitch  angle is  therefore $19.13^{\circ} \pm 4.76^{\circ}$. Due to the interaction with its companion galaxy, M51a shows a significant departure from a constant pitch angle in the outer regions of the galaxy. This is seen in Fig. \ref{M51}a and at the noticeable sign change in this plot at an inner radius of 389 pixels (119\arcsec). {\it Fig. \ref{M51}c (right)} - A  stable mean pitch angle of  $16.26^{\circ}$ is determined for the  $m = 2$ harmonic mode
  from a minimum  inner radius of 54 pixels (16.5\arcsec) to  a maximum inner radius
  of 276 pixels (84.2\arcsec),  with an outer radius of 389  pixels (119\arcsec). This stretch of
  222 pixels (67.7\arcsec) occupies  57\% of  the measurement annulus. Equation \ref{Error} yields $E_{\phi} = 3.20^{\circ}$ with $\lambda = 222$ pixels (67.7\arcsec), $\beta = 296$ pixels (90.3\arcsec), $\sigma = 2.36^{\circ}$, and $\epsilon_2 = 0.57^{\circ}$. The final  determination  of pitch  angle is  therefore $16.26^{\circ} \pm 3.20^{\circ}$. This alternate pitch angle measurement isolates the inner portion of the galaxy out to the clear break from constant pitch angle seen in Fig.\ref{M51}b. As a result, the unstable outer portion of the galaxy has been ignored and a more accurate pitch angle has been determined for the purer inner structure of this interacting galaxy.\label{M51}}
\end{center}  
\end{figure*}

To illustrate the case of measuring pitch angles of interacting galaxies, we have selected perhaps the most famous case of interacting galaxies, M51 (see Figure \ref{M51}a). 
M51 consists of M51a (NGC 5194) and its companion dwarf galaxy M51b (NGC 5195). Due to M51a's clear interaction with M51b, its well-defined spiral structure is seen to depart from regularity close to the companion. For this case, our typical method of measuring pitch angle across the entirety of the galactic disc knowingly samples the outer 40\% of the galaxy, which is clearly seen to be disrupted (see Figure \ref{M51}b). Just as in the case of iterative determination of pitch angle as a function of inner radius to omit interior regions, by alternatively selecting an outer radius interior to disrupted outer regions, we can confine our measurements to only the stable portions of M51a (see Figure \ref{M51}c) and other similar galaxies.


\subsubsection{Flocculence}\label{Flocculence}

Concerning the measurement of pitch angle of different types of spiral galaxies, flocculent spirals provide perhaps the biggest challenge. From our experience with flocculent galaxies, we find that our code most often finds them to have high-valued harmonic modes. Their characteristics can range from fragmented arms at best to chaos at worst. \citet{Elmegreen:1987} defined a system of arm classes and descriptions to categorize spirals into categories with varying degrees of flocculence. They defined 12 arm classes (classes 10 and 11 are no longer in use) with 12 having the most orderly spiral structure and 1 the least. Galaxies with arm classes 1-4 are considered flocculent, and those with arm classes 5-12 are grand design.

\begin{figure*}
\begin{center}
\includegraphics[width=8.97cm]{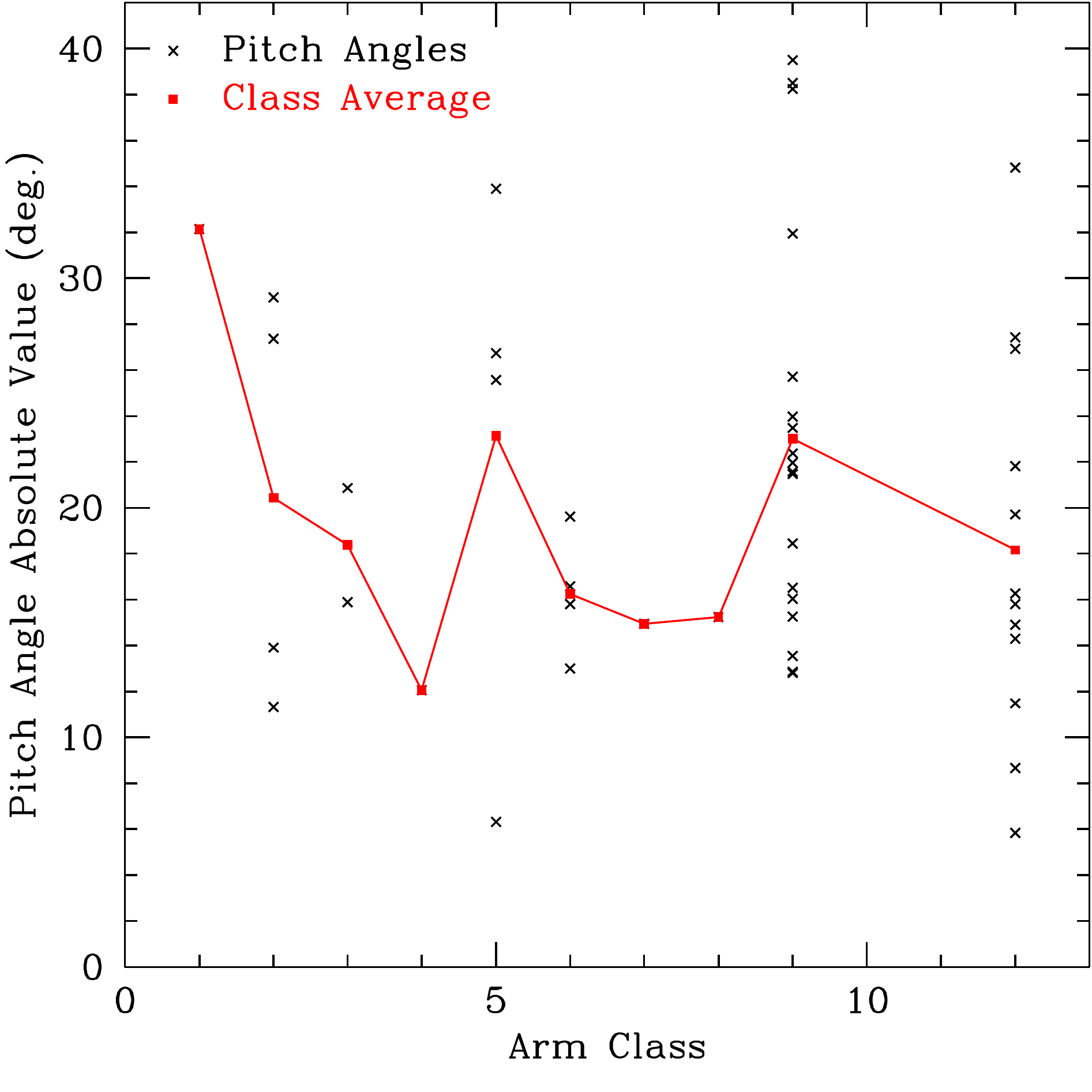}
\includegraphics[width=8.97cm]{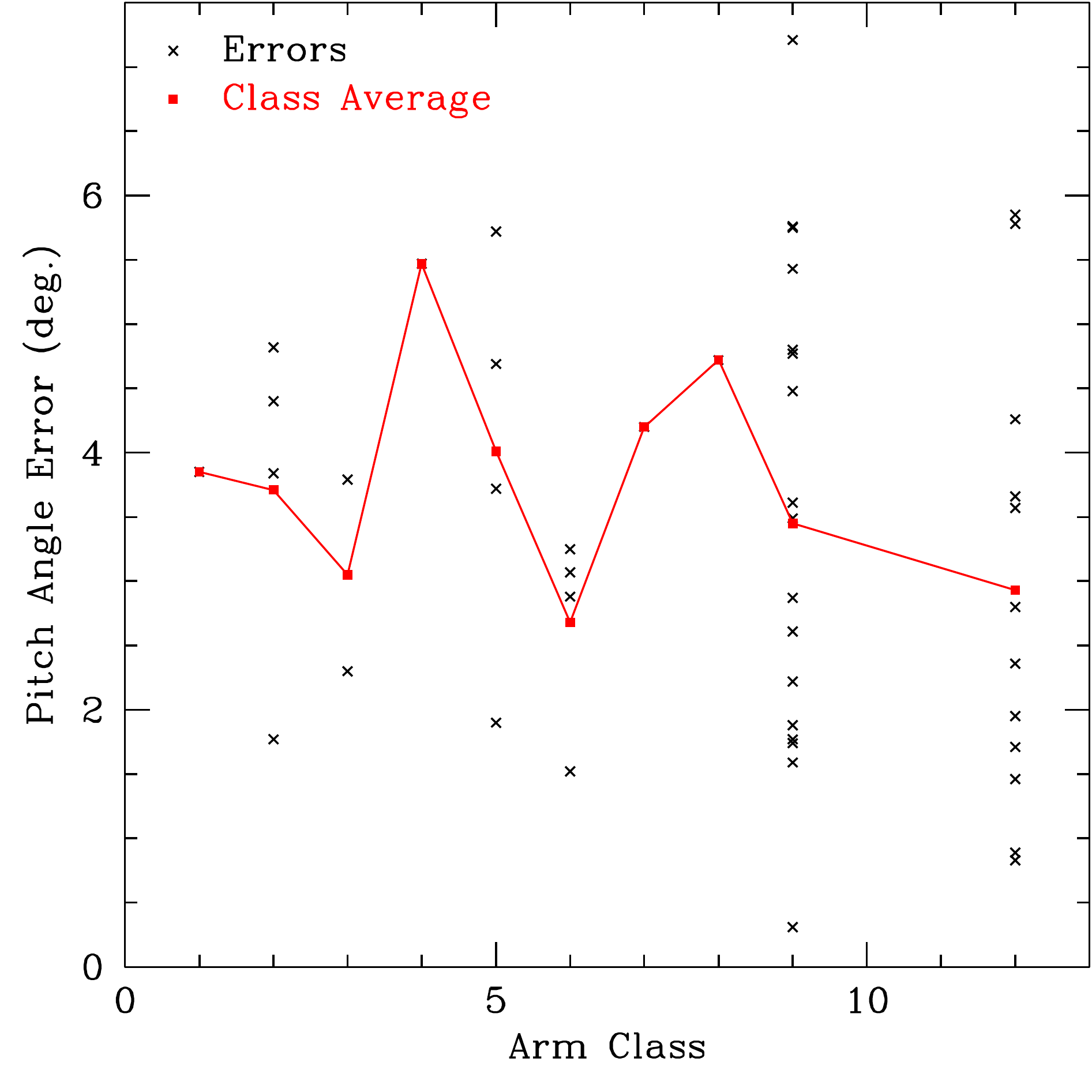}
\caption{Pitch angles (left) and their associated errors (right) sorted into their respective arm classes (classes 10 and 11 are no longer in use). {\it Fig. \ref{Class_Plot}a (left)} - Pitch angles (black crosses) arranged into their arm classes with binned averages (\textcolor{red}{red} squares connected by \textcolor{red}{red} line segments). No clear trend is recognizable between pitch angle and arm class. {\it Fig. \ref{Class_Plot}b (right)} - Pitch angle errors (black crosses) arranged into their arm classes with binned averages (\textcolor{red}{red} squares connected by \textcolor{red}{red} line segments). No clear trend is recognizable between pitch angle errors and arm class.\label{Class_Plot}}
\end{center}
\end{figure*}

We have subsequently ascertained the arm classes (based on blue images from the Palomar Observatory Sky Survey) given by \citet{Elmegreen:1987} for the galaxies we have measured for this paper and listed all available arm classes in Table \ref{Class_Table}. We have also created two plots (see Figure \ref{Class_Plot}) 
of pitch angle absolute value vs. arm class (see Figure \ref{Class_Plot}a) and pitch angle error vs. arm class (see Figure \ref{Class_Plot}a) in order to investigate possible dependencies on arm classes. No clear relationship can be found from either plot, thus measurement of flocculent spirals do not appear to be inherently less precise than grand design spirals. However, our method of measuring pitch angle is very much dependent on the visual inspection conducted by the user. When initially inspecting images of possible candidate spiral galaxies for subsequent pitch angle measurement, it is more natural to be drawn to grand design spirals. This selection bias can be seen in selection of our sample for this paper without prior knowledge of their arm classes. Of the 48 galaxies listed in Table \ref{Class_Table}, only 8 galaxies are classified as being flocculent, with the remaining galaxies all classified as being grand design spirals. It is our practice to only attempt pitch angle measurement on galaxies that display convincing evidence of definable spiral structure from image inspection. From our study of the CGS sample thus far, we have been able to convincingly measure pitch angles for 62\% of the spiral galaxies we have examined; 17\% of the galaxies were rejected due to their high angle of inclination and the remaining 21\% were omitted due to a lack of discernible spiral structure (of this 21\%, among those with arm classifications from \citet{Elmegreen:1987}, 60\% were classified as flocculent). 
\begin{deluxetable*}{llcrccr}
\tablecolumns{7}
\tablecaption{Pitch Angles and Arm Classes\label{Class_Table}}
\tablehead{
\colhead{Galaxy Name} & \colhead{Morphology} & \colhead{m} & \colhead{$\phi$ (deg.)} & \colhead{Band} & \colhead{Source} & \colhead{Arm Class}
}
\startdata
IC 2522 & SB(rs)bc & 3 & $-26.73 \pm 4.69$ & B & 1 & 5  \\
IC 2537 & SAB(rs)c & 4 & $27.37 \pm 3.84$ & B & 1& 2  \\
M51a & SA(s)bc pec & 2 & $16.26 \pm 2.36$ & B & 3 & 12  \\
NGC 45 & SA(s)dm & 3 & $-32.13 \pm 3.85$ & B & 1 & 1  \\
NGC 150 & SB(rs)b: & 2 & $14.29 \pm 4.26$ & B & 1 & 12  \\
NGC 157 & SAB(rs)bc & 3 & $8.66 \pm 0.89$ & B & 1 & 12  \\
NGC 210 & SAB(s)b & 2 & $-15.81 \pm 3.25$ & B & 1 & 6  \\
NGC 289 & SB(rs)bc & 5 & $19.71 \pm 1.95$ & B & 1 & 12  \\
NGC 578 & SAB(rs)c & 3 & $16.51 \pm 1.88$ & B & 1 & 9  \\
NGC 598 & SA(s)cd & 2 & $-33.90 \pm 5.72$ & 6450 \AA\tablenotemark{a} & 2 & 5  \\
NGC 613 & SB(rs)bc & 3 & $21.57 \pm 1.77$ & B & 1 & 9  \\
NGC 895 & SA(s)cd & 2 & $-38.50 \pm 4.77$ & I & 1 & 9  \\
NGC 908 & SA(s)c & 3 & $15.26 \pm 2.61$ & B & 1 & 9  \\
NGC 1042 & SAB(rs)cd & 4 & $39.50 \pm 4.48$ & R & 1 & 9  \\
NGC 1097\tablenotemark{b} & SB(s)b & 2 & $15.80 \pm 3.62$ & I & 1 & 12  \\
NGC 1187 & SB(r)c & 4 & $-21.96 \pm 3.61$ & B & 1 & 9  \\
NGC 1232 & SAB(rs)c & 3 & $-25.71 \pm 5.43$ & B & 1 & 9  \\
NGC 1292 & SA(s)c & 3 & $-15.89 \pm 2.30$ & B & 1 & 3  \\
NGC 1300 & SB(rs)bc & 2 & $-12.71 \pm 1.99$ & B & 1 & 12  \\
NGC 1365 & SB(s)b & 2 & $-34.81 \pm 2.80$ & B & 1 & 12  \\
NGC 1398 & (R$'$)SB(r)ab & 4 & $19.61 \pm 3.07$ & V & 1 & 6  \\
NGC 1566 & SAB(s)bc & 2 & $-17.81 \pm 3.67$ & B & 1 & 12  \\
NGC 1792 & SA(rs)bc & 3 & $-20.86 \pm 3.79$ & B & 1 & 3  \\
NGC 1964 & SAB(s)b & 2 & $-12.86 \pm 3.49$ & B & 1 & 9  \\
NGC 2280 & SA(s)cd & 4 & $21.47 \pm 2.87$ & B & 1 & 9  \\
NGC 2442 & SAB(s)bc pec & 2 & $14.95 \pm 4.20$ & V & 1 & 7  \\
NGC 2835 & SB(rs)c & 3 & $-23.97 \pm 2.22$ & B & 1 & 9  \\
NGC 2935 & (R$'$)SAB(s)b & 2 & $-15.24 \pm 4.72$ & B & 1 & 8  \\
NGC 3052 & SAB(r)c: & 3 & $-18.45 \pm 1.59$ & B & 1 & 9  \\
NGC 3054 & SAB(r)b & 3 & $12.80 \pm 1.77$ & B & 1 & 9  \\
NGC 3450 & SB(r)b & 6 & $-13.55 \pm 0.31$ & B & 1 & 9  \\
NGC 3513 & SB(rs)c & 1 & $5.84 \pm 1.46$ & B & 1 & 12  \\
NGC 3783 & (R$'$)SB(r)ab & 2 & $10.71 \pm 0.64$ & B & 1 & 9  \\
NGC 3887 & SB(r)bc & 4 & $-29.16 \pm 4.82$ & B & 1 & 2  \\
NGC 3938 & SA(s)c & 4 & $-22.37 \pm 7.21$ & B & 3 & 9  \\
NGC 4027 & SB(s)dm & 1 & $-12.06 \pm 5.47$ & B & 1 & 4  \\
NGC 4030 & SA(s)bc & 3 & $23.48 \pm 5.76$ & B & 1 & 9  \\
NGC 4050 & SB(r)ab & 1 & $-6.32 \pm 1.90$ & B & 1 & 5  \\
NGC 4321 & SAB(s)bc & 5 & $21.81 \pm 3.57$ & R & 3 & 12  \\
NGC 4939 & SA(s)bc & 6 & $11.48 \pm 1.71$ & B & 1 & 12  \\
NGC 4995 & SAB(rs)b & 2 & $13.00 \pm 2.88$ & B & 1 & 6  \\
NGC 5054 & SA(s)bc & 3 & $-25.57 \pm 3.72$ & B & 1 & 5  \\
NGC 5085 & SA(s)c & 2 & $-11.32 \pm 1.77$ & 4680 \AA\tablenotemark{c} & 4 & 2  \\
NGC 5236 & SAB(s)c & 6 & $-16.04 \pm 1.74$ & B & 1 & 9  \\
NGC 5247 & SA(s)bc & 2 & $-31.94 \pm 5.75$ & B & 1 & 9  \\
NGC 5861 & SAB(rs)c & 2 & $-14.91 \pm 0.83$ & V & 1 & 12  \\
NGC 6215 & SA(s)c & 4 & $-27.43 \pm 5.85$ & B & 1 & 12  \\
NGC 6300 & SB(rs)b & 4 & $-16.58 \pm 1.52$ & B & 1 & 6  \\
NGC 7793 & SA(s)d & 2 & $13.91 \pm 4.40$ & B & 1 & 2  \\
\enddata
\tablecomments{Col. (1) galaxy name; col. (2) morphological type from the RC3 \citep{RC3}; col. (3) dominant harmonic mode; col. (4) pitch angle; col. (5) waveband/wavelength; col. (6) telescope/survey imaging source; and col. (7) arm class from \citet{Elmegreen:1987}. Source (1) CGS; source (2) Palomar 48 inch Schmidt; source (3) KPNO 2.1 m CFIM; and source (4) UK 48 inch Schmidt.}
\tablenotetext{a}{103aE emulsion.}
\tablenotetext{b}{In addition to spiral arms in the disc of the galaxy, NGC 1097 displays rare $m = 2$ nuclear spiral arms in the bulge. These arms display an opposite chirality to the disc arms with $\phi = -30.60^{\circ} \pm 2.68^{\circ}$.}
\tablenotetext{c}{IIIaJ emulsion.}
\end{deluxetable*}

\section{Image Analysis}\label{sect4}


The sign  of the  pitch angle  and the  number of  harmonic modes are very
important for correct image analysis. As the pitch angle is calculated
over  all possible  values of  inner  radii for  a galaxy,  it is  not
uncommon   for  pitch   angle   to  vary   drastically  in   different
harmonic modes. Different  harmonic modes will have  different signs of pitch  angle and
even  across one  harmonic mode,  sign  changes may  occur.  The most  apparent
feature  to  the  human  eye,  for discernable  spiral  arms,  is  the
chirality of the  spiral arms.  As a result, harmonic modes that favor opposing
chirality can immediately be ruled out after a quick visual inspection
of the image. 

\begin{figure*}
\begin{center}
\includegraphics[width=8.97cm]{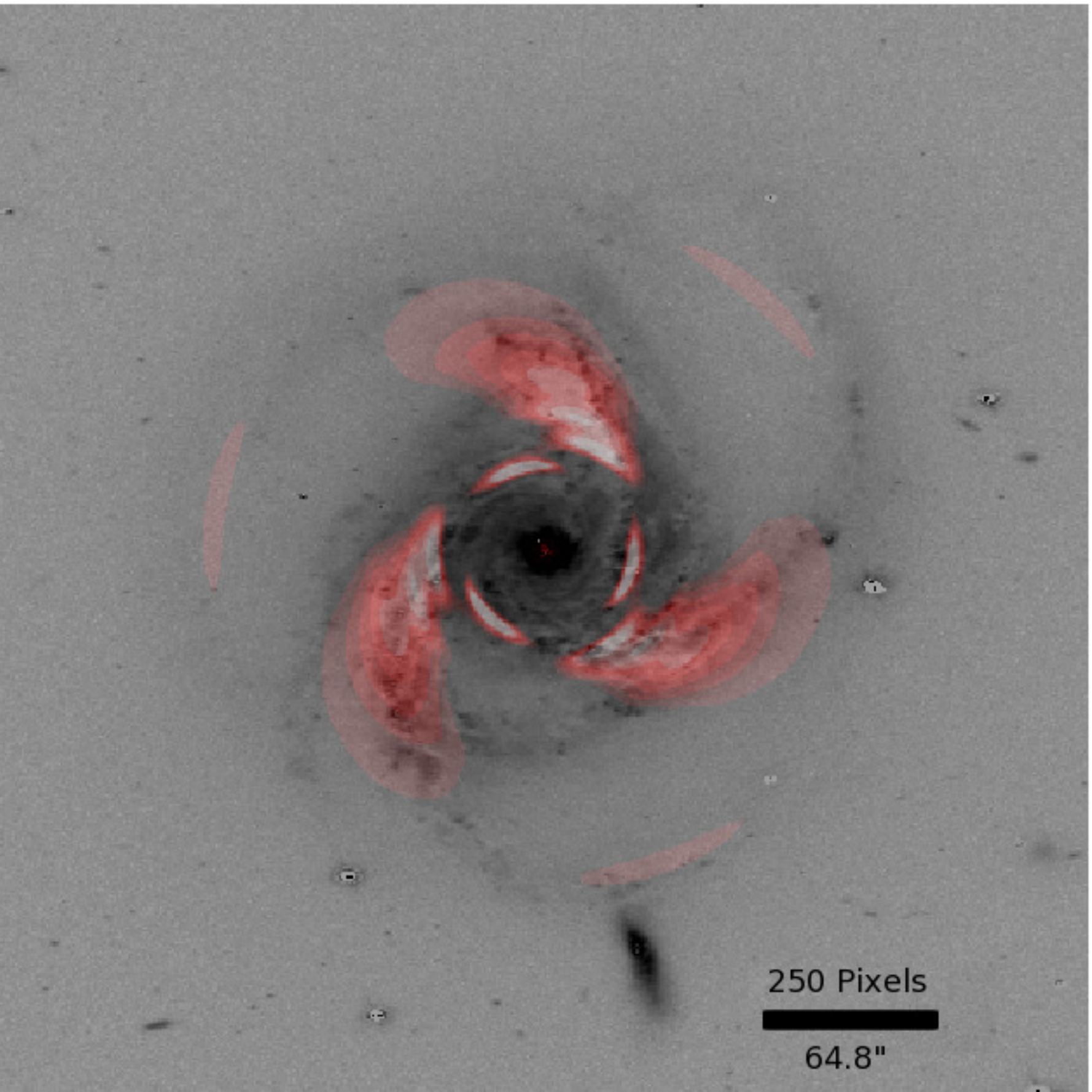}
\includegraphics[trim = 0mm 10mm 0mm 0mm, clip, width=8.97cm]{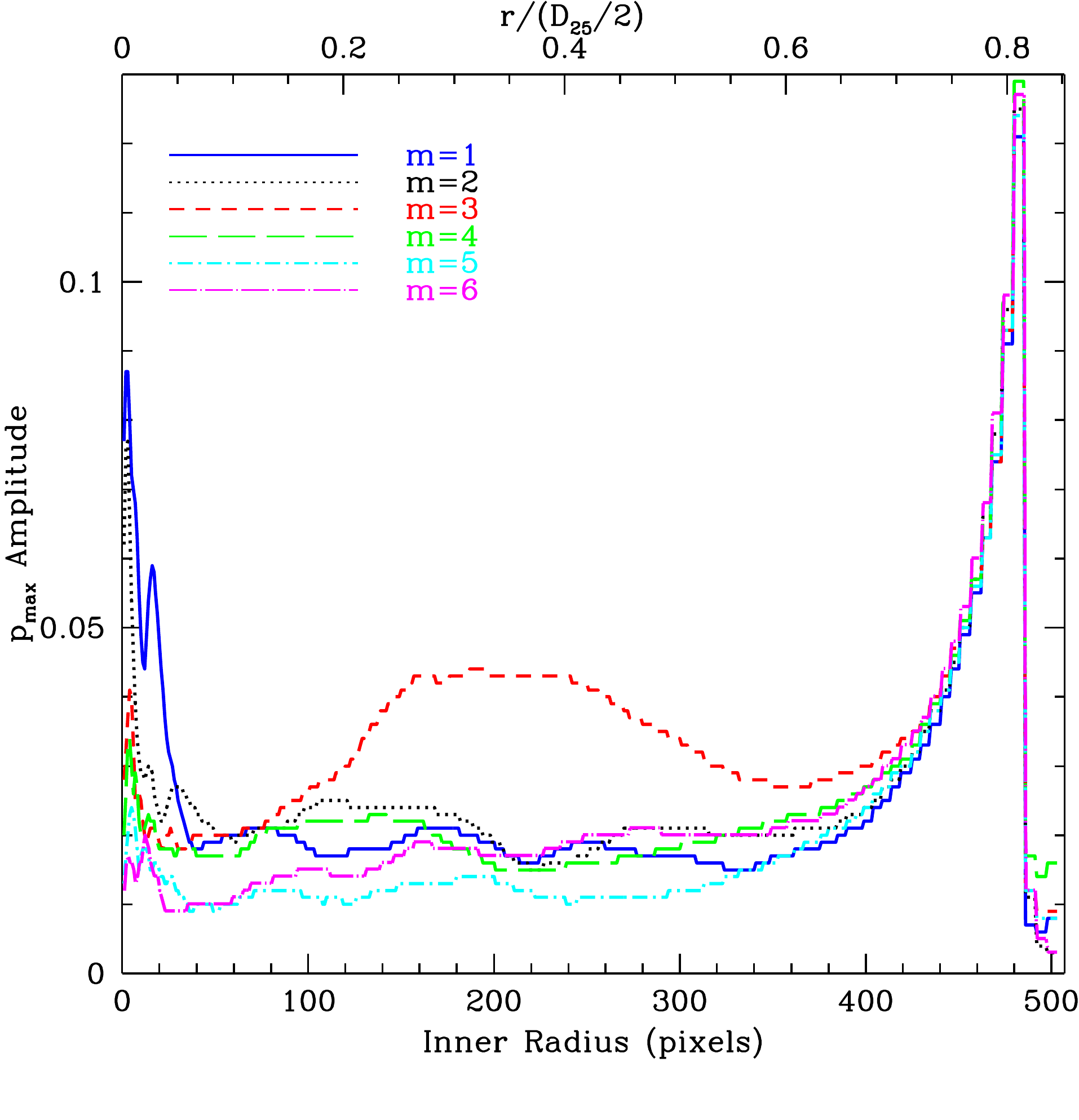}
\caption{{\it  Fig.   \ref{fig16}a  (left)}  -   Star-subtracted  and
  deprojected ($PA = 160^{\circ}$ \& $\alpha = 53.84^{\circ}$) B-band (inverted color) image   of   NGC   5054   (see
  Fig. \ref{fig20}a),  overlaid with
  the contours of  the Inverse 2-D FFT for  the $m = 3$
  harmonic mode (in \textcolor{red}{red}) conducted  with an inner radius of 160 pixels (41.4\arcsec)
  and an outer radius of 508 pixels (132\arcsec). The contours are the real part of
  the  complex  spatial function  of  Equation  \ref{eqn6}  with $m  =
  3$. The  contours illustrate the  different levels of  amplitude for
  the  $m =  3$  harmonic  mode. The  Inverse  2-D FFT
  displays a single value pitch angle of $-40.60^{\circ}$ (as shown in
  Fig. \ref{fig1}b). However,  the pitch angle can be  seen to tighten
  (decrease)  as the  inner radius  increases. {\it  Fig. \ref{fig16}b
    (right)} - Plot of the amplitude of $p_{max}$ as a function of inner radius for NGC
  5054, indicating the $m = 3$ component as the dominant harmonic mode
  for the galaxy.  The $m = 3$ harmonic mode  is dominant from an inner radius of 77 to 456 pixels (19.9\arcsec$\,$to 118\arcsec), constituting about $75\%$ of the galaxy's
  radius.\label{fig16}}
\end{center}  
\end{figure*}

For  galaxies with  visually  distinctive spiral  arms,  it is  simple
enough to count the number of spiral arms by eye and adopt that number
of arms as the correct  harmonic mode. However, in flocculent galaxies
or galaxies  with galactic arm spurs,  it maybe necessary  to adopt other
methods  in  selecting  the  correct harmonic  mode.   Typically,  the
harmonic mode  with the  largest region of  stable pitch  angle across
inner radii is  the most valuable for our purposes. Nonetheless, other
aspects  of the  code  can lend  a  hand in  identifying the  dominant
harmonic  mode. The  easiest method  is by  plotting the  amplitude of
$p_{max}$  as a  function of inner  radius. This  will help  identify the
harmonic  mode with the  strongest amplitude  over the  largest radial
range of the galaxy. For NGC 5054 (see Figure \ref{fig16}b), 
the $m = 3$
harmonic mode is dominant \citep{Block:1999} over the outer $84.8\%$ of the galaxy's
radius. In many cases, several harmonic modes agree reasonably well as to the pitch angle. In addition, producing an image of the Inverse FFT of a harmonic mode can help visually identify the ``correct" harmonic mode (see \S \ref{subsect4.2}).

\subsection{Symmetrical Component Significance}\label{subsect4.1}

\begin{figure*}
\begin{center}
\includegraphics[trim = 0mm 8mm 0mm 0mm, clip, width=8.97cm]{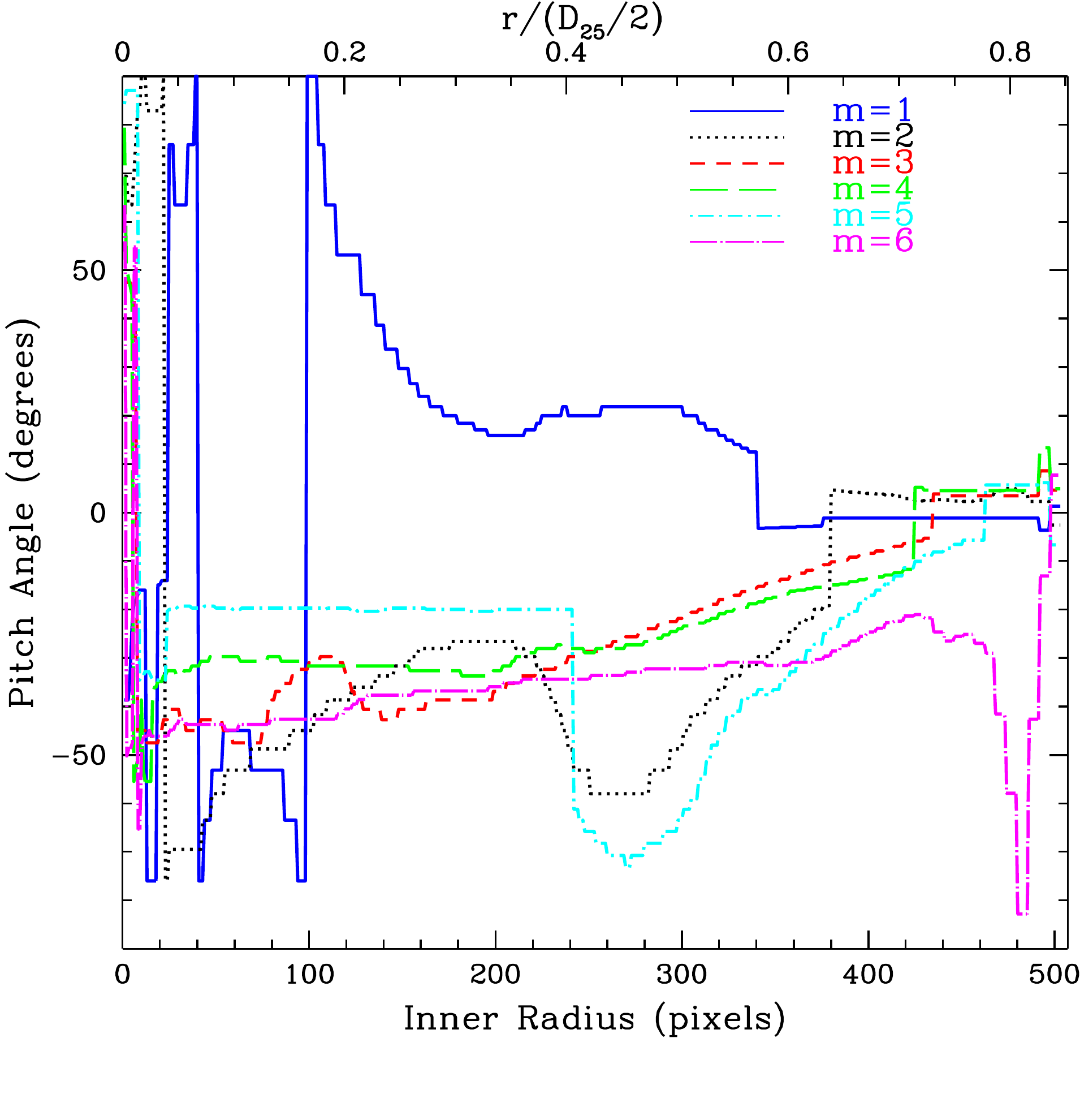}
\includegraphics[trim = 0mm 10mm 0mm 0mm, clip, width=8.97cm]{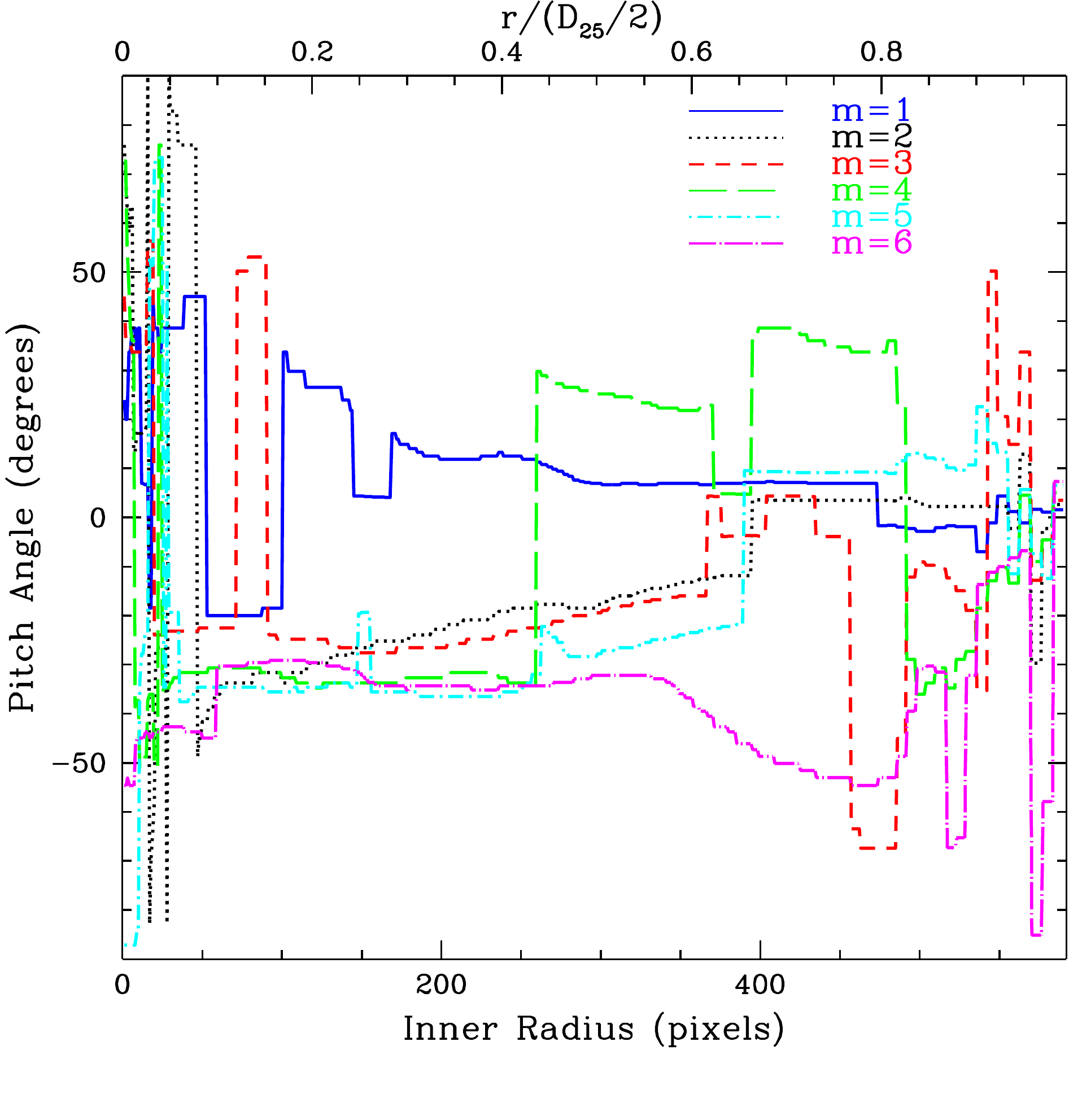}
\caption{{\it Fig. \ref{fig21}a (left)} - B-band pitch angle as a function of
  inner radius  for NGC 5054,  after star subtraction  and deprojection ($PA = 160^{\circ}$ \& $\alpha = 53.84^{\circ}$)
  were performed (see Fig. \ref{fig20}a). A stable  mean pitch angle
  is difficult to define since  the pitch angle is seen to continually
  decrease from an  inner radius of about 200  to one of about 425. A rough mean pitch  angle of $-24.52^{\circ}$ (a significant departure from the single value, non-iterative measurement of $-40.60^{\circ}$, see Fig. \ref{fig1}b) is determined for
  the $m  = 3$ harmonic mode  from a  minimum inner radius  of 123 pixels (31.9\arcsec) to a
  maximum  inner radius of  434 pixels (112\arcsec),  with an  outer radius  of 508
  pixels (132\arcsec).  This  stretch of  311  pixels (80.5\arcsec) occupies 61\%  of the galactic disc. 
  This measurement has a significant error  due to  the unstable
  pitch  angle. Equation \ref{Error} yields $E_{\phi} = 12.84^{\circ}$ with $\lambda = 311$ pixels (80.5\arcsec), $\beta = 334$ pixels (86.5\arcsec), $\sigma = 11.92^{\circ}$, and $\epsilon_3 = 0.92^{\circ}$. The final  determination of  pitch angle  is therefore
  $-24.52^{\circ} \pm 12.84^{\circ}$.  {\it Fig. \ref{fig21}b (right)}
  -   For  the   pure   symmetrical  component   of   NGC  5054   (see
  Fig. \ref{fig23}a), the B-band pitch angle (same deprojection parameters as Fig. \ref{fig21}a) as a function of inner radius is
  far more  stable. A stable  mean pitch angle of  $-25.57^{\circ}$ is
  determined for  the $m = 3$ harmonic mode  from a minimum inner  radius of 91
  pixels (23.6\arcsec) to a maximum inner radius of 253 pixels (65.5\arcsec), with an outer radius
  of 593 pixels (154\arcsec). This stretch of 162 pixels (42.0\arcsec) occupies 27\%  of the galactic disc. 
  Equation \ref{Error} yields $E_{\phi} = 3.72^{\circ}$ with $\lambda = 162$ pixels (42.0\arcsec), $\beta = 443$ pixels (115\arcsec), $\sigma = 1.31^{\circ}$, and $\epsilon_2 = 1.00^{\circ}$. The final determination of pitch  angle is therefore
  $-25.57^{\circ} \pm 3.72^{\circ}$,  a percent difference of $4.19\%$
  in mean  pitch angle  with a $72.03\%$  reduction in error  from the
  original.\label{fig21}}
\end{center}
\end{figure*}

\begin{figure*}
\begin{center}
\includegraphics[width=8.97cm]{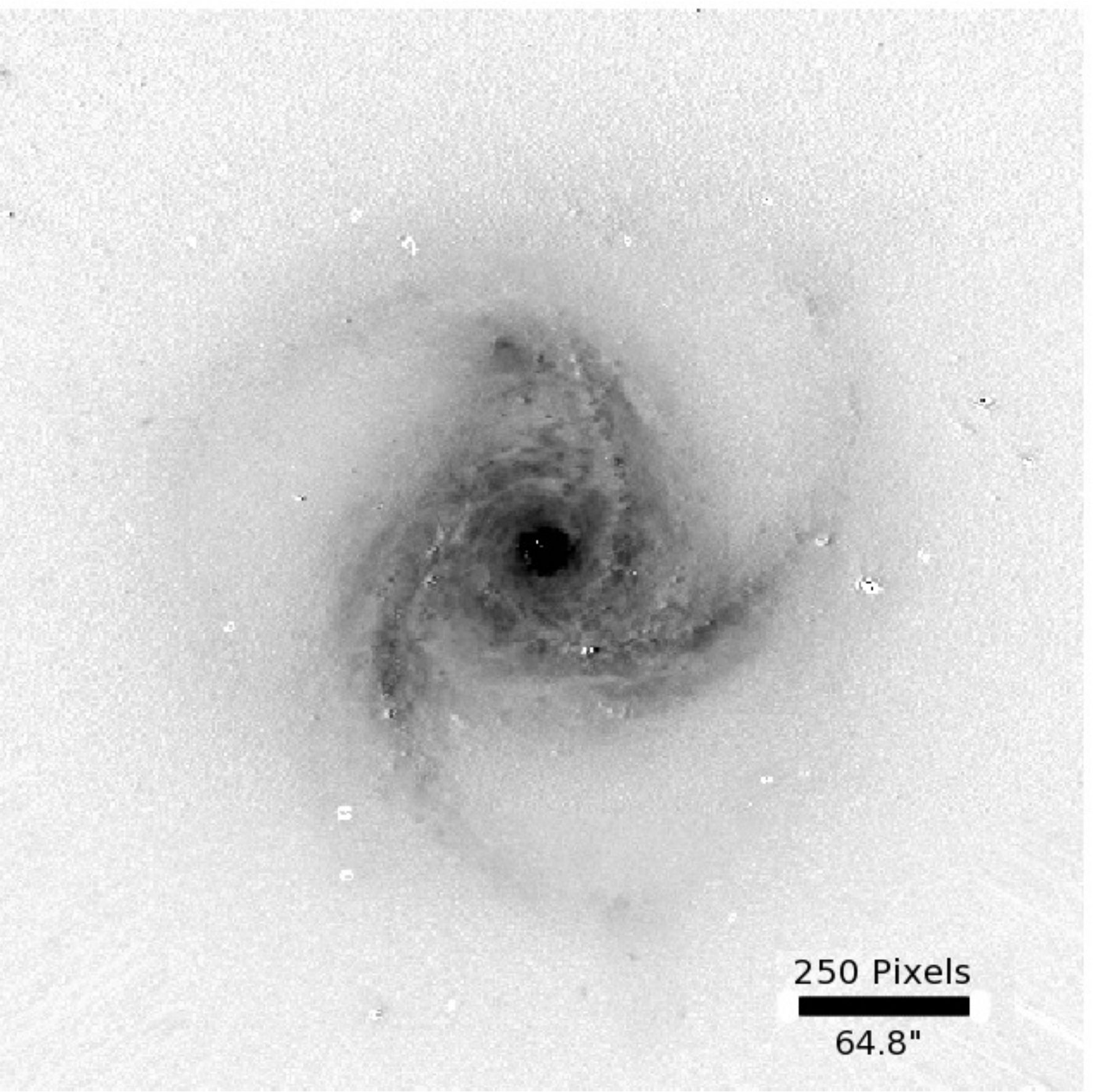}
\includegraphics[width=8.97cm]{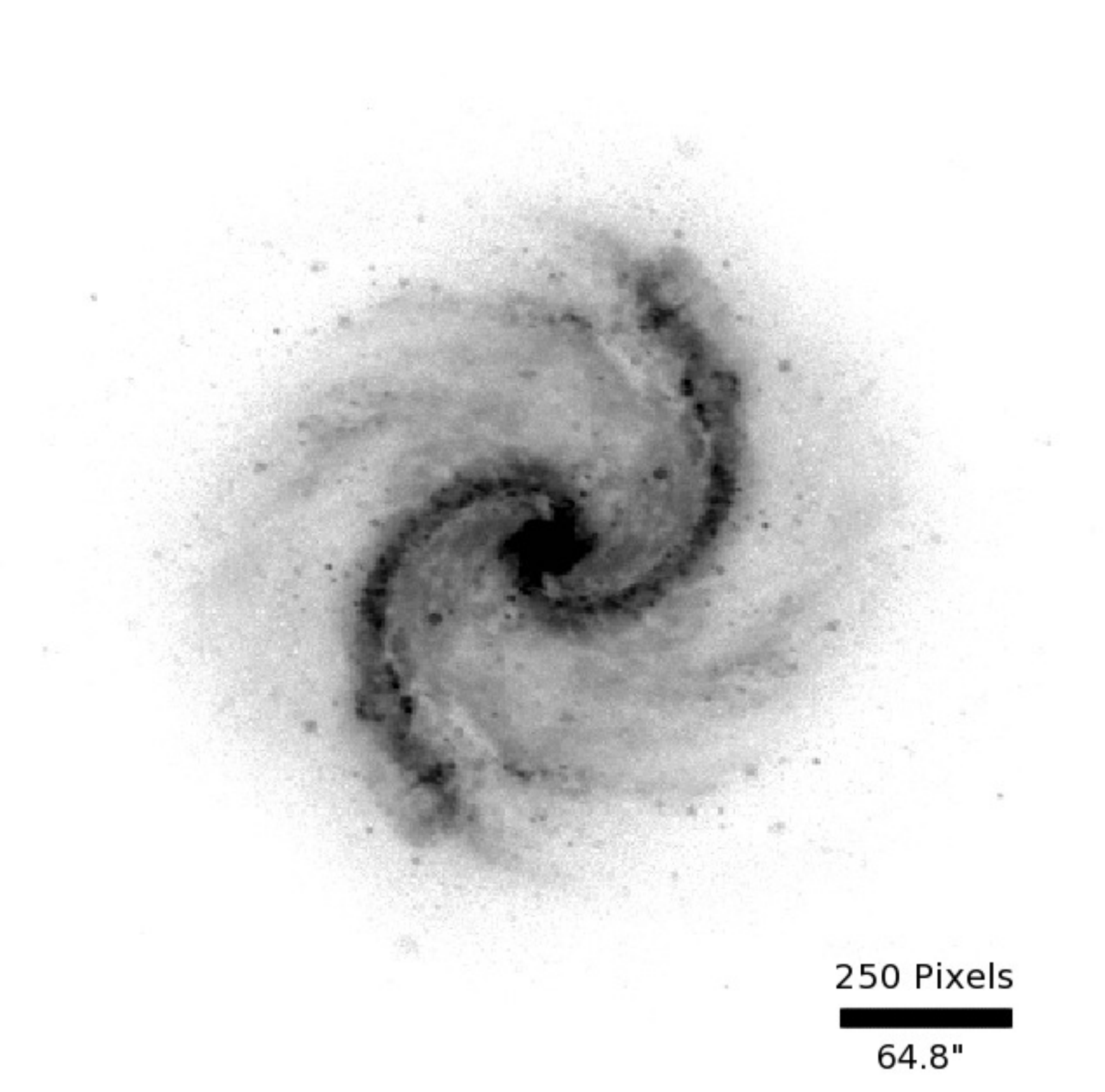}
\caption{{\it Fig. \ref{fig23}a (left)} - $m = 3$ symmetrical component (inverted color and with the same deprojection parameters as Fig. \ref{fig20}a) of NGC
  5054  (see Fig.  \ref{fig20}a).  {\it Fig.  \ref{fig23}b (right)}  -
  $m = 2$ symmetrical       component       (inverted color and with the same deprojection parameters as Fig. \ref{fig20}b) of       NGC       5247       (see
  Fig. \ref{fig20}b).\label{fig23}}
\end{center}
\end{figure*}

\begin{figure*}
\begin{center}
\includegraphics[trim = 0mm 0mm 0mm 0mm, clip, width=8.97cm]{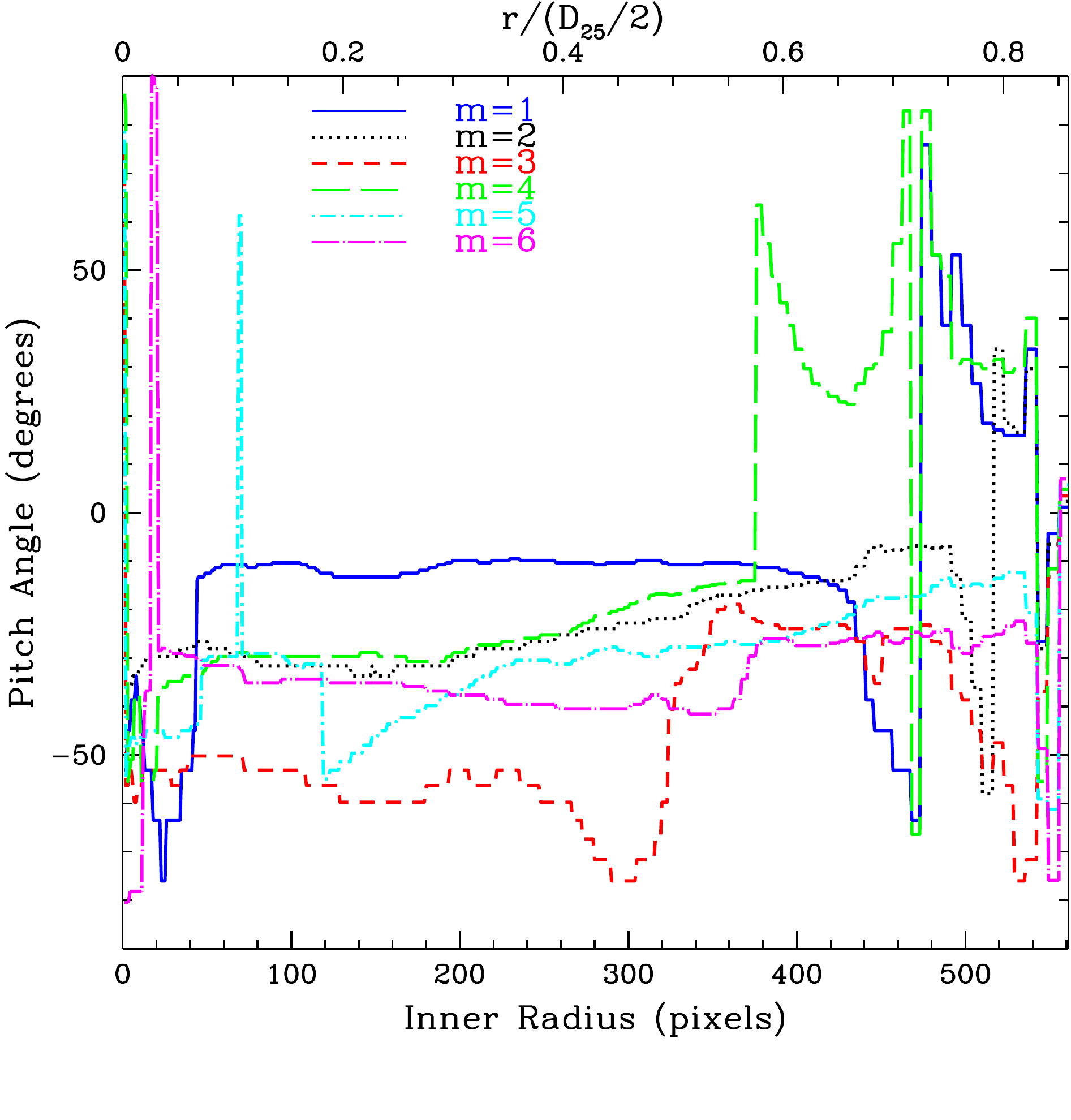}
\includegraphics[trim = 0mm 0mm 0mm 0mm, clip, width=8.97cm]{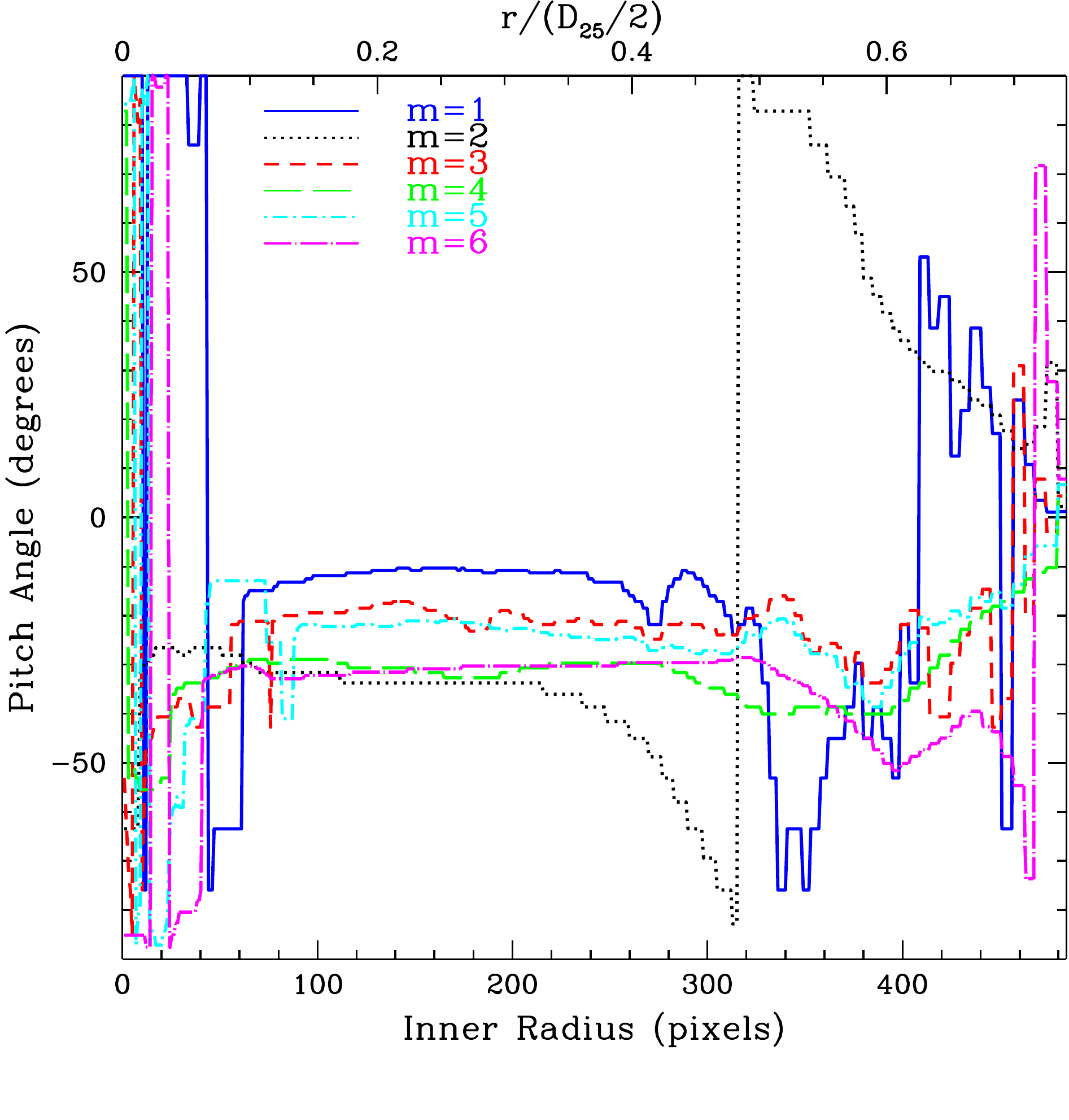}
\caption{{\it Fig. \ref{fig22}a (left)} - B-band pitch angle as a function of
  inner  radius for NGC  5247 after  deprojection ($PA = 20^{\circ}$ \& $\alpha = 28.36^{\circ}$) was  performed (see
  Fig. \ref{fig20}b). A stable mean pitch angle of $-28.76^{\circ}$ is
  determined for  the $m =  2$ harmonic mode from  a minimum inner radius  of 1
  pixel (0.259\arcsec) to a maximum inner radius  of 331 pixels (85.7\arcsec), with an outer radius
  of 565 pixels (146\arcsec). This stretch of 330 pixels (85.5\arcsec) occupies 58\%  of the galactic disc. 
  Equation \ref{Error} yields $E_{\phi} = 5.73^{\circ}$ with $\lambda = 330$ pixels (85.5\arcsec), $\beta = 508$ pixels (132\arcsec), $\sigma = 3.56^{\circ}$, and $\epsilon_2 = 1.70^{\circ}$. The final determination of pitch  angle is therefore
  $-28.76^{\circ} \pm 5.73^{\circ}$. {\it Fig. \ref{fig22}b (right)} -
  For   the   pure   symmetrical    component   of   NGC   5247   (see
  Fig. \ref{fig23}b), a stable mean B-band pitch angle (same deprojection parameters as Fig. \ref{fig22}a) of $-31.94^{\circ}$ is
  determined for  the $m =  2$ harmonic mode from  a minimum inner radius  of 9
  pixels (2.33\arcsec) to a maximum inner radius of 235 pixels (60.9\arcsec), with an outer radius
  of 486 pixels (126\arcsec). This stretch of 226 pixels (58.5\arcsec) occupies 47\%  of the galactic disc. 
  Equation \ref{Error} yields $E_{\phi} = 5.75^{\circ}$ with $\lambda = 226$ pixels (58.5\arcsec), $\beta = 428$ pixels (111\arcsec), $\sigma = 2.83^{\circ}$, and $\epsilon_2 = 2.06^{\circ}$. The final determination of pitch  angle is therefore
  $-31.94^{\circ} \pm 5.75^{\circ}$, a percent difference of $10.48\%$
  in mean  pitch angle  with essentially the same error as the
  original. As a characteristic example, NGC 5247 displays a similar pattern of agreement among even harmonic modes as the two-armed synthetic spiral with an added symmetrical bulge component (see Figure \ref{Syn_figs}g).\label{fig22}}
\end{center}
\end{figure*}

It is  a likely possibility that all  of the arms of  a spiral galaxy,
especially  galaxies  with  multiple   arms,  might  not  be  perfectly
symmetric. This could be the result of tidal disruption, galaxy
harassment, etc.  Whatever the reason, slight  imperfections should be
handled by the  robustness of the FFT.  A common trend
among galaxies we  have analyzed is that some  galaxies exhibit spiral
arms which gradually  tighten, or decrease in absolute  value of pitch
angle,  toward the  outer regions  of the  galaxy (see Figures \ref{fig20}a and \ref{fig21}a). 
For  galaxies with
drastically asymmetric spiral arms  or arms which demonstrate variable
pitch angle,  we use the  method of \citet{Elmegreen:1992}  to isolate
the  symmetrical component of  a galaxy  (see Figure \ref{fig23}) 
and  then we perform  a pitch angle determination  on the
symmetrical     component     (see     Figures    \ref{fig21}b     and
\ref{fig22}b). 
Symmetric  parts of galaxies are  illustrated by making
images from  successive rotations  and subtractions. The  procedure of
\citet{Elmegreen:1992} is
\begin{equation}
S_m(r, \theta) = (m - 1)F(r, \theta) - \sum_{j=1}^{m-1}[F(r, \theta) - F(r, \theta - \frac{2j\pi}{m})]_T\label{eqn11}
\end{equation}
where  for $m  \geq 2$,  $S_m$ is  the image  displaying  the $m$-fold
symmetric part of  a galaxy made from the original  image $F$, and the
subscript $T$ stands for truncation, meaning that pixels with negative
intensities  are set  to  zero. For  the  case of  a two-armed  spiral
galaxy,  the  $S_2$ image  consists  of  all  bright features  in  the
original image that have  equally bright features diametrically across
the  galaxy. This  procedure  highlights symmetric  emission, such  as
spiral  arm spurs,  star formation  regions, etc.,  but  it introduces
spurious absorption features. For example,  if there is a dust lane in
only one  arm, then only  the bright part  of that arm will  appear in
both arms  of the  $S_2$ image; this  gives the false  impression that
there is  a dust lane  in the other  arm also. This method  appears to
decrease  our  error  estimates  when  performed.  For  galaxies  with
apparent  initial  symmetry,  the  mean  pitch angle  is  not  changed
significantly; e.g.,  NGC 5247 (see Figure  \ref{fig22}),  the percent
difference in mean pitch angle  is $10.48\%$ with essentially the same error. This tool  seems  most useful  for galaxies  that
display variable pitch angle (see Figure \ref{fig21}). Error estimates
can be reduced drastically for these cases; e.g., NGC 5054, for which the percent
difference in mean  pitch angle is $4.19\%$ with  a $72.03\%$ decrease
in error.  Moreover, this process  can act as an effective substitute
for star subtraction.

\subsection{Two-Dimensional Inverse Fast Fourier Transform}\label{subsect4.2}

One of the most powerful tools provided  by {\it 2DFFT} is the ability to
run an Inverse FFT. After having deprojected the images
and  identified the  dominant  harmonic modes,  we  can calculate  the
inverse of the transforms according to \citet{Seigar:Block:2005}. The
inverse transform can be written as
\begin{equation}
S(u, \theta) = \sum_{m}S_m(u)e^{im\theta}
\label{eqn6}
\end{equation}
where
\begin{equation}
S_m(u) = \frac{D}{e^{2u}4\pi^2}\int_{p_-}^{p_+}G_m(p)A(p, m)e^{ipu}dp.
\label{eqn7}
\end{equation}
$G_m(p)$      is     a      high-frequency     filter      used     by
\citet{Puerari:Dottori:1992}.  For  the  logarithmic spiral governed  by  Equation
\ref{eqn2}, it has the form
\begin{equation}
G_m(p) = e^{-\frac{1}{2}(\frac{p - p_{max}}{25})^2}.
\label{eqn9}
\end{equation}
This  filter is  also used  to smooth  the $A(p,  m)$ spectra  at the
interval  ends  ($p_-  =  -50$  and  $p_+ =  50$  with  $dp  =  0.25$)
\citep{Puerari:Dottori:1992}. Equation \ref{eqn6} is designed as such,
to  allow  the user  to  create an  inverse transform for  a
selected  number  of  harmonic  components. For  example,  the inverse transform can be calculated  for one component, e.g.,  $m=2$, or
any number of components can  be combined to yield a composite result, e.g., $m = 2, 3, \& \: 4$.

\begin{figure*}
\begin{center}
\includegraphics[width=8.97cm]{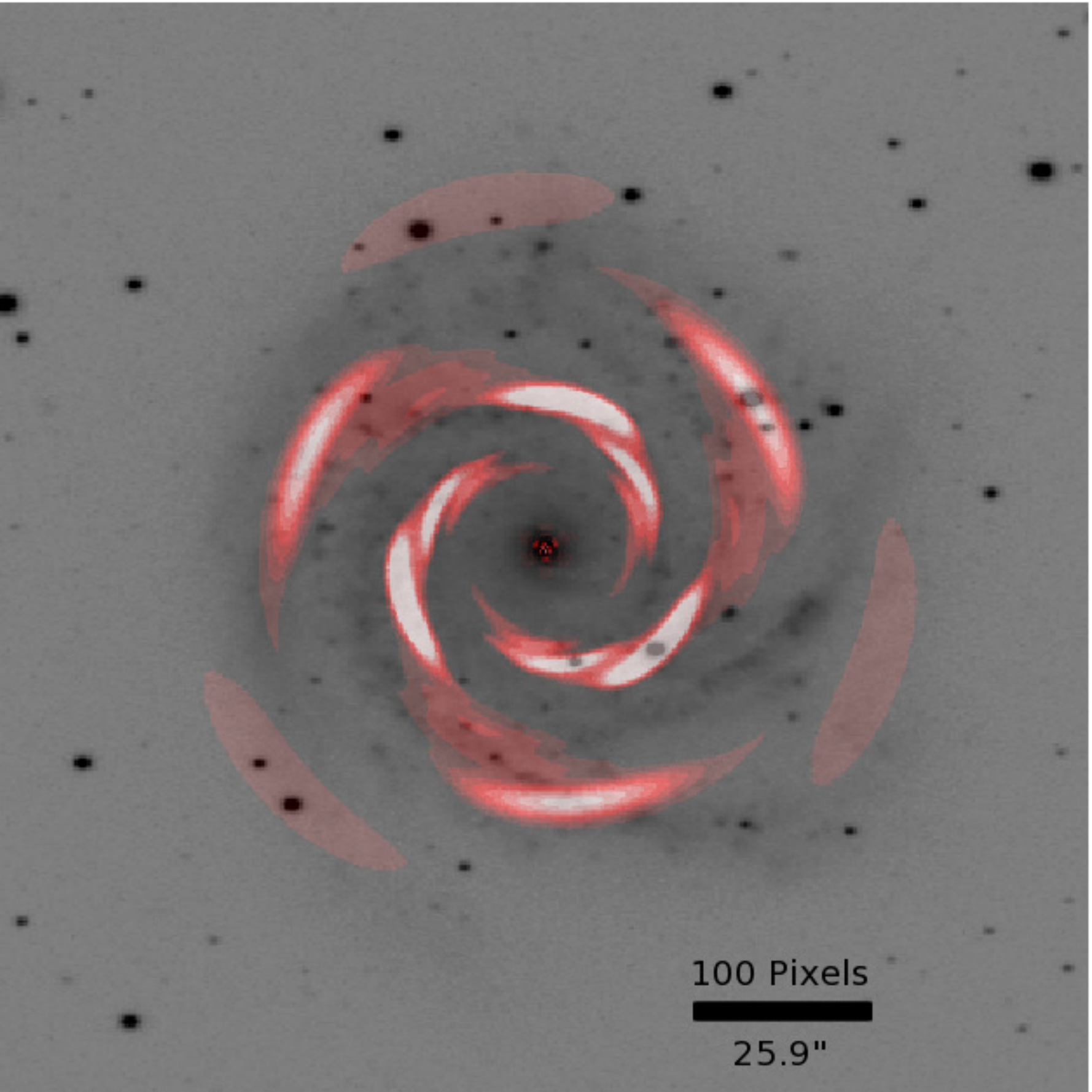}
\includegraphics[width=8.97cm]{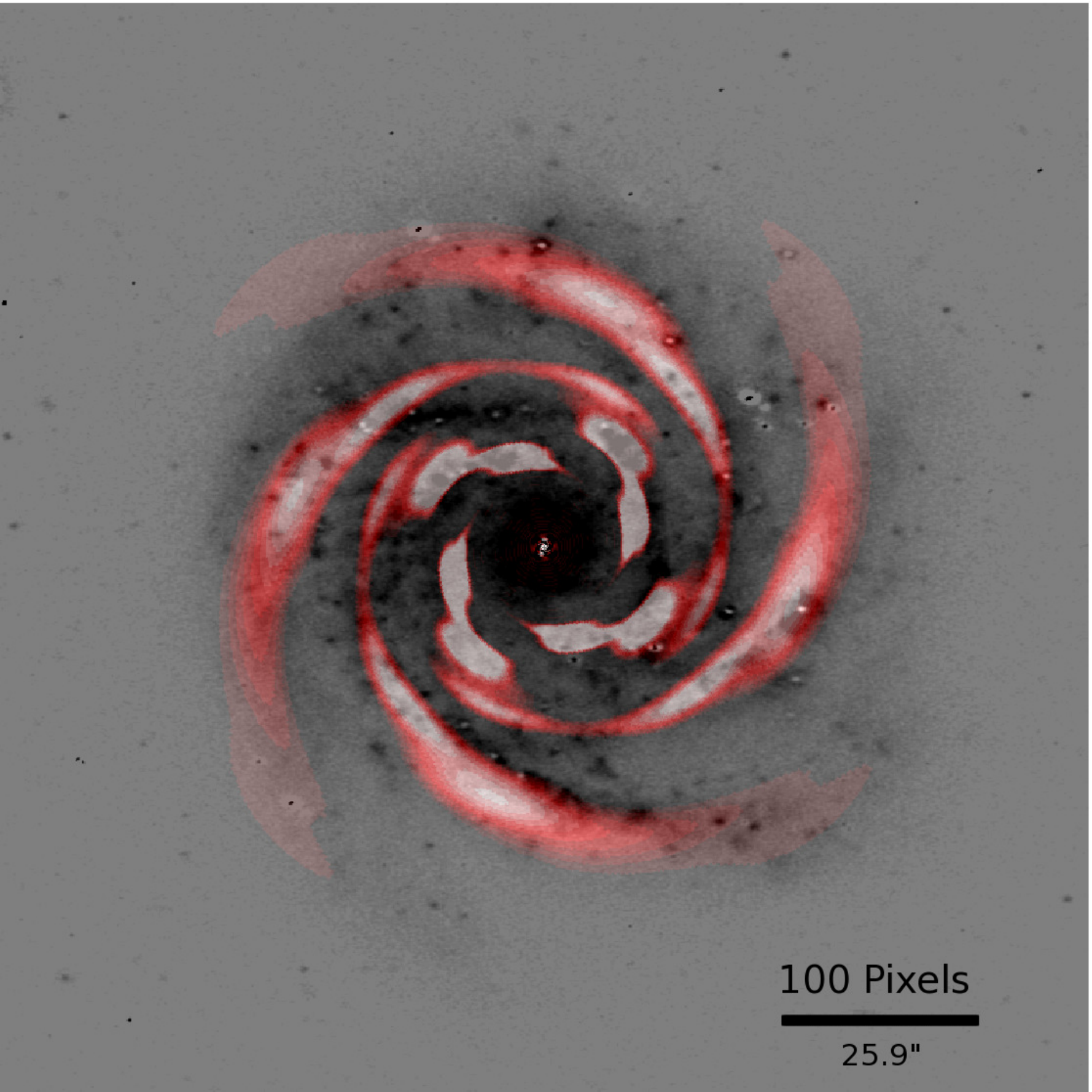}
\caption{{\it  Fig.  \ref{fig19}a  (left)}  -  Deprojected ($PA = 50^{\circ}$ \& $\alpha = 39.65^{\circ}$) B-band
  inverse image of  IC 4538, before star subtraction,  overlaid with the
  contours  of the  Inverse FFT  for the  $m  = 3$
  harmonic mode (in \textcolor{red}{red}), conducted with an inner  radius of 54 pixels (14.0\arcsec)
  and an  outer radius of 258  pixels (66.8\arcsec), demonstrating a  pitch angle of
  $-19.98^{\circ}$.  The contours  are the  real part  of  the complex
  spatial function of  Equation \ref{eqn6} with $m =  3$. The contours
  illustrate  the  different levels  of  amplitude  for  the $m  =  3$
  harmonic  mode.  The  overlaid  Inverse  FFT  tends   to  track  the  bright   foreground  stars.  {\it
    Fig. \ref{fig19}b (right)} -  Star-subtracted and deprojected (with the same parameters as Fig. \ref{fig19}a) B-band  inverse image of  IC 4538,  overlaid with  the contours  of the
  Inverse FFT  for the $m  = 4$ harmonic  mode (in \textcolor{red}{red}), conducted with an inner radius of 54 pixels (14.0\arcsec) and an outer radius
  of 264 pixels (68.4\arcsec), demonstrating  a pitch angle of $-15.42^{\circ}$. The
  contours  are the  real  part  of the  complex  spatial function  of
  Equation  \ref{eqn6}  with $m  =  4$.  The  contours illustrate  the
  different levels of amplitude for  the $m = 4$ harmonic mode. In  the absence  of bright foreground  stars, all  four visible
  arms of the galaxy are tracked by the Inverse FFT
  overlay.\label{fig19}}
\end{center}
\end{figure*}

Once  an Inverse  FFT  is  created, it  can be  directly
compared  to the  deprojected  image  of the  galaxy,  allowing us  to
effectively observe  what the code  is seeing. Figure  \ref{fig16}a and
Figure  \ref{fig19} 
show images  of  spiral  galaxies overlaid  with
contours  representing  the   results  of  Inverse  FFTs of the same galaxy.  The contours  are  the  real part  of  the complex  spatial
function  of Equation  \ref{eqn6}.  The use  of these images  to analyze a galaxy can  lead  to more  confident  determination of  pitch
angle.

\section{discussion and Future Work}\label{sect5}

Our modified version of {\it  2DFFT} is a powerful tool for accurately
measuring   galactic spiral  arm  pitch   angle. Our software, combined with careful image and data inspection, comparative pitch angle selection, and a self-regulating error determination allows for reliable pitch angle measurements.  We hope that quantitative determinations of spiral arm pitch angle will aid in galaxy classification, in the indirect study of central black hole masses and more generally in our understanding of galactic morphology and its evolution. One important advantage of this means of describing galaxies is its relative ease of acquisition, since only  imaging data is required to measure it. Also, it provides us with great opportunity to test competing theories behind galactic spiral arm genesis \citep{Martinez-Garcia:2011}.

\subsection{Comparison to Other Methods}\label{Others}

Our method adds one dimension to recently published FFT methods \citep[e.g.,][]{Kendall:2011}. In doing so, we are able to effectively use more of the inherent information in the images. Alternatively, 1-D methods identify radial peaks in intensity azimuthally about the galactic center by use of radial search segments that ultimately do not utilize the full resolution of the image. Admittedly, our method comes at a higher price in terms of computing power to analyze the full resolution of an image, but with modern computational power, this computational expense is trivial and is easily handled by modern processors. Ultimately, users of both 1-D and 2-D FFT methods are still obliged to visually inspect images. As for flocculent galaxies, 1-D FFT methods are admittedly only equipped to handle grand design spirals. Our 2-D FFT, though it may occasionally encounter trouble with high degrees of flocculence, will perform adequately with flocculent galaxies and with the support of additional image analysis methods (see \S\ref{sect4} and its subsections) and multi-wavelength imaging, it can confidently approach any galaxy with hints of spirality. Additionally, our pitch angle measurements are unique in that fact that we measure and quote pitch angles for multi-armed galaxies (dominant harmonic modes $m > 2$). Other researchers may have analyzed and discussed the influences from higher order harmonic modes, but in the end always publish pitch angles resulting from harmonic modes $m \leq 2$.


In order to compare the results of our method to other published methods, we have chosen a well-studied sample of galaxies whose pitch angles have been determined independently, using different techniques in the literature. For this sample, we have selected the results of \citet{Martinez-Garcia:2011}, \citet{Kendall:2011}, \citet{Ma:2001}, \citet{Grosbol:Patsis:1998}, and \citet{Kennicutt:1981} as references with which to compare our measurements (see Table \ref{Compare}). These five references provide a nice spread in measurement methods. \citet{Kennicutt:1981} used $H\alpha$ photographic plates and subsequent by-hand geometric measurements of nearby Sa-Sc galaxies to calculate average pitch angles determined from the two main arms in each galaxy\footnote{\citet{Savchenko:2011} remeasured pitch angles for 46 of the spiral galaxies measured by \citet{Kennicutt:1981} with two distinct methods: by-hand and 1-D FFT analysis. Results of both methods show good mutual agreement with the mean differences between measurements less than a few degrees in both cases.}; \citet{Grosbol:Patsis:1998} utilized accurate surface photometry and derivation of axisymmetric components to calculate $m = 2$ pitch angles for five galaxies from the residuals of intensive image processing and Fourier components of the azimuthal intensity variations; \citet{Ma:2001} visually selected points along spiral arms in CCD images of galaxies and fit logarithmic spirals to the points; \cite{Kendall:2011} employed 1-D FFT decomposition to calculate $m = 2$ pitch angles for a sample of grand design spiral galaxies; and \citet{Martinez-Garcia:2011} used both a ``slope method" and adopted a 2-D FFT algorithm similar to our own, but without our analysis of pitch angle as a function of inner radius.

Table \ref{Compare} compares 38 of our pitch angle measurements against available measurements from five independent sources. The mean difference between the measurements for the same galaxies are as follows: $\Delta\phi_{1} = -5.13^{\circ} \pm 19.41^{\circ}$ \citep[the difference between our measured pitch angles and those of][]{Martinez-Garcia:2011} or $\Delta\phi_{1} = -0.11^{\circ} \pm 7.38^{\circ}$ if the outlying measurement of NGC 4995 is disregarded, $\Delta\phi_{2} = 1.73^{\circ} \pm 3.58^{\circ}$ \citep[the same for][]{Kendall:2011}, $\Delta\phi_{3} = 0.66^{\circ} \pm 9.97^{\circ}$ \citep[the same with the average arm measurement from][]{Ma:2001}, $\Delta\phi_{4} = 2.93^{\circ} \pm 1.71^{\circ}$ \citep[the same for][]{Grosbol:Patsis:1998}, and $\Delta\phi_{5} = 5.15^{\circ} \pm 8.66^{\circ}$ \citep[the same for][]{Kennicutt:1981}. These differences are comparable to the mean error in our measurements for this sample: $\phi_{E} = 3.34^{\circ} \pm 1.94^{\circ}$.

\subsubsection{NGC 7083}

\begin{figure*}
\begin{center}
\includegraphics[width=5.95cm]{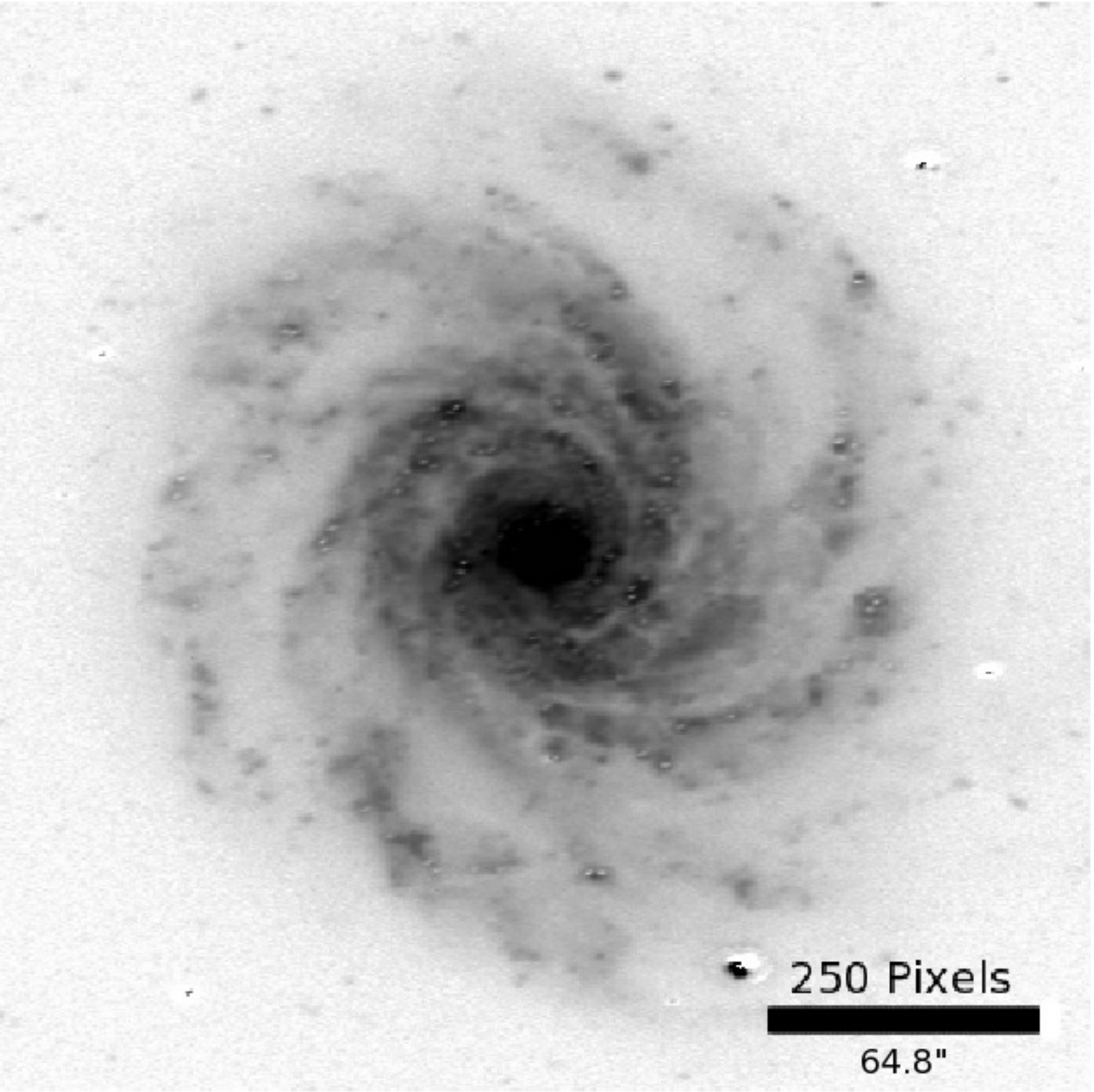} 
\includegraphics[trim = 0mm 10mm 0mm 0mm, clip, width=5.95cm]{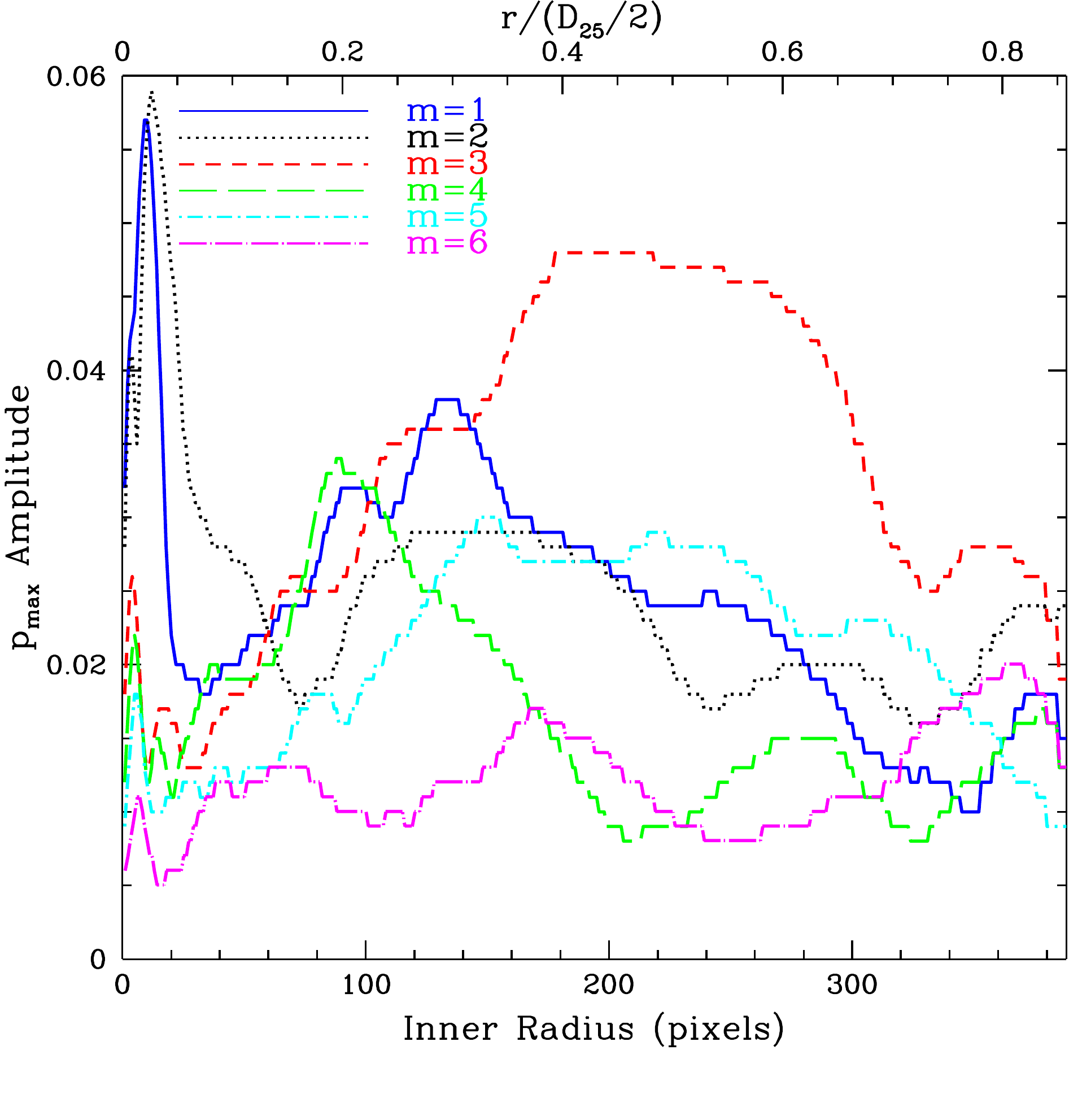}
\includegraphics[trim = 0mm 10mm 0mm 0mm, clip, width=5.95cm]{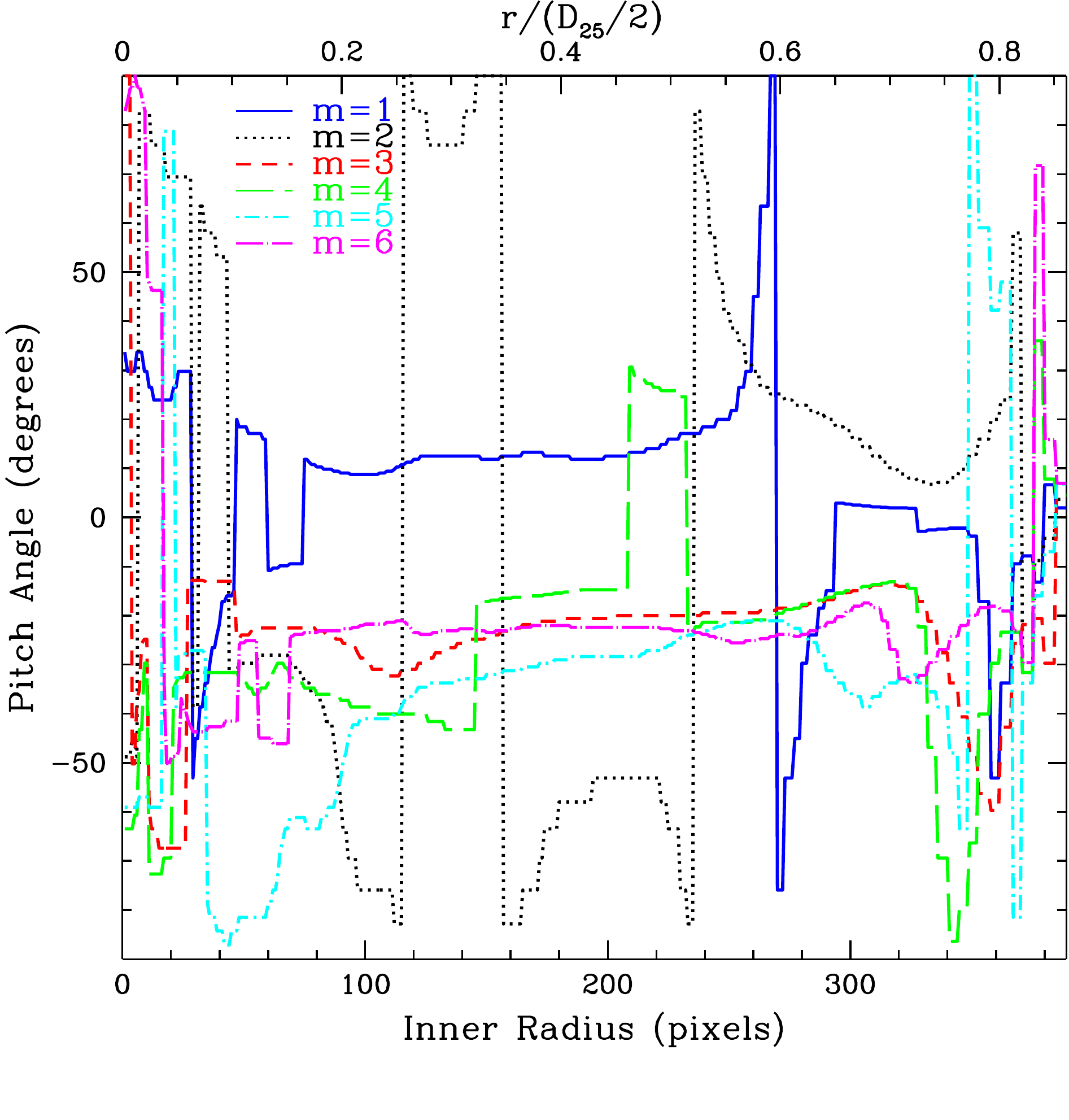}
\caption{{\it Fig. \ref{ngc7083}a (left)} - Star-subtracted and deprojected ($PA = 5^{\circ}$ \& $\alpha = 53.84^{\circ}$) B-band inverse image NGC 7083. {\it Fig. \ref{ngc7083}b (middle)} - Plot of the amplitude of $p_{max}$ as a function of inner radius for NGC 7083, indicating the $m = 3$ component as the dominant harmonic mode for the galaxy. The $m = 3$ harmonic mode is dominant from an inner radius of 145 to 384 pixels (37.6\arcsec$\,$to 99.5\arcsec), constituting about $61\%$ of the galaxy's radius. {\it Fig. \ref{ngc7083}c (right)} - A  stable mean
  pitch angle of  $-19.44^{\circ}$ is determined for the  $m = 3$ harmonic mode
  from a minimum  inner radius of 143 pixels (37.0\arcsec) to  a maximum inner radius
  of 319 pixels (82.6\arcsec),  with an outer radius of 390  pixels (101\arcsec). This stretch of
  176 pixels (45.6\arcsec) occupies  45\% of  the galactic disc. Equation \ref{Error} yields $E_{\phi} = 3.21^{\circ}$ with $\lambda = 176$ pixels (45.6\arcsec), $\beta = 208$ pixels (53.9\arcsec), $\sigma = 2.67^{\circ}$, and $\epsilon_3 = 0.58^{\circ}$. The final  determination  of pitch  angle is  therefore $-19.44^{\circ} \pm 3.21^{\circ}$.
\label{ngc7083}}
\end{center}
\end{figure*}

For individual measurements, several differences can be explained by our selection of a different harmonic mode from that chosen by the other group. For example, our measurement of NGC 7083 differs from the measurement by \citet{Grosbol:Patsis:1998}; $-19.44^{\circ} \pm 3.21^{\circ}$ and $-15.0^{\circ} \pm 1.0^{\circ}$, respectively. This can be explained by our selection of the $m = 3$ harmonic mode and their selection of the $m = 2$ harmonic mode (see Figure \ref{ngc7083}). 
In short, we believe this is a three-armed galaxy, not two-armed, a finding supported by the strength of our code's $m = 3$ harmonic mode and by visual inspection. It is of note that we find our measurement of the $m = 4$ harmonic mode's pitch angle to be $-15.38^{\circ} \pm 2.97^{\circ}$ (see Figure \ref{ngc7083}c), which coincides with the measurement of \citet{Grosbol:Patsis:1998}. However, our measurement of the $m = 2$ harmonic mode is not possible due to its chiral instability. Despite the nice agreement of pitch angle between our $m = 4$ pitch angle and their $m = 2$ pitch angle, we find strong indications that the dominant harmonic mode is $m = 3$ (see Figure \ref{ngc7083}b).

\subsubsection{NGC 4995}

NGC 4995 depicts a significant outlier from the pitch angle measurement of \citet{Martinez-Garcia:2011}; $78.3_{-8.3}^{+5.4}$ degrees compared to our measurement of $13.00^{\circ} \pm 2.88^{\circ}$. Their enormously high measured pitch angle is most sensibly unphysical, along with any measurement of $\phi \apgt 60^{\circ}$.

\subsubsection{NGC 1365}

\begin{figure}
\begin{center}
\includegraphics[width=8.6cm]{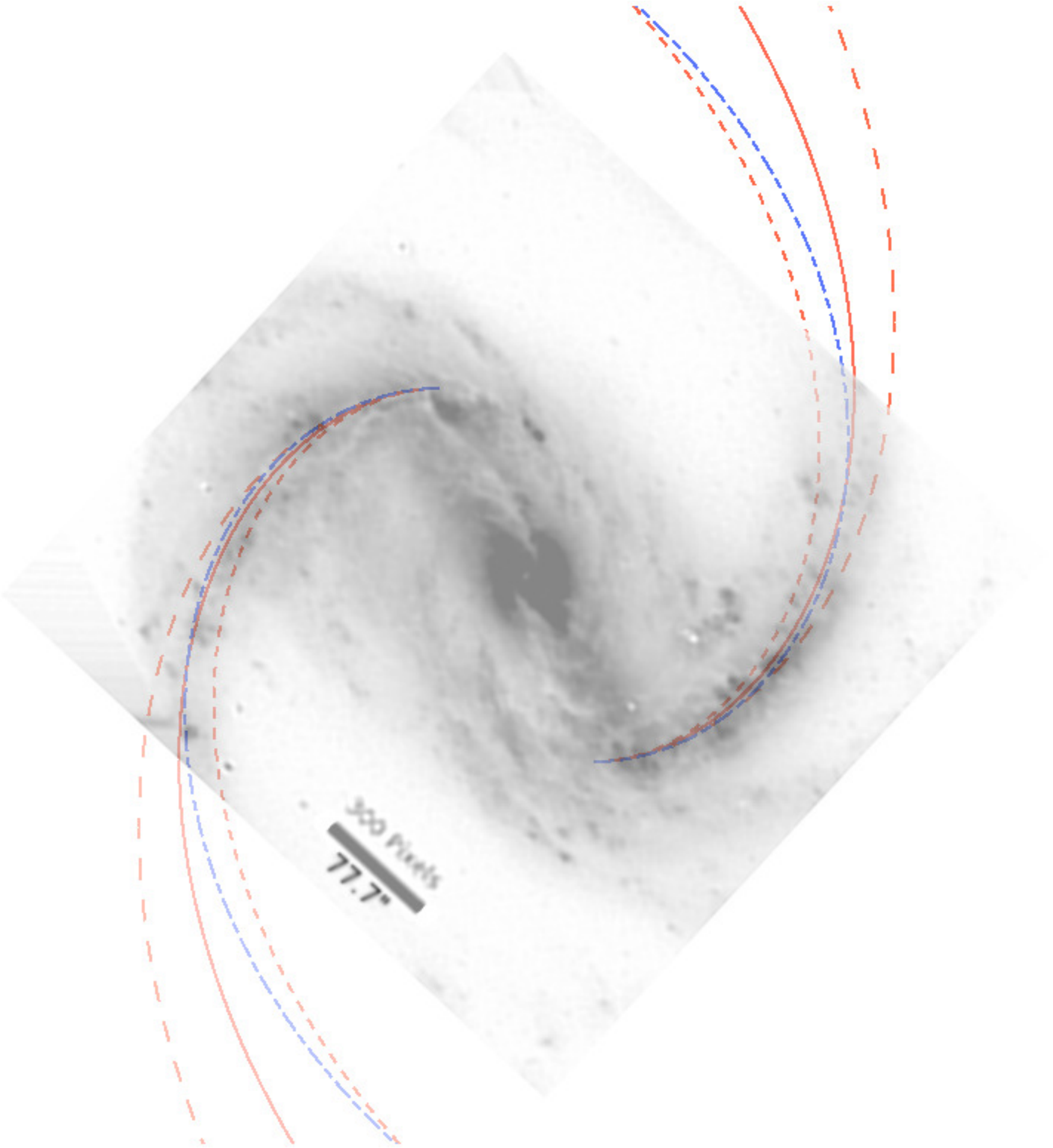}
\caption{NGC 1365 (see Figure \ref{fig25}a) overlaid with a $m = 2$ spiral with $\phi = -34.81^{\circ}$ (solid \textcolor{red}{red} lines) representing our best-fit pitch angle measurement (see Figure \ref{fig25}b), $\phi = -32.01^{\circ}$ (short dashed \textcolor{red}{red} lines) representing our lower limit fit, $\phi = -37.61^{\circ}$ (long dashed \textcolor{red}{red} lines) representing our upper limit fit, and $\phi = -16.5^{\circ}$ (alternating short-long dashed \textcolor{blue}{blue} lines) representing the average fit of \citet{Ma:2001} and \citet{Kennicutt:1981}.\label{ngc1365}}
\end{center}
\end{figure}

A big discrepancy can be seen in comparison of our measurement of NGC 1365, $-34.81^{\circ} \pm 2.80^{\circ}$, to the pitch angle absolute value measurements of \citet{Ma:2001}, $13.8^{\circ}$ \& $17.8^{\circ}$, and \citet{Kennicutt:1981}, $18^{\circ} \pm 3^{\circ}$. We find strong indications both from our code (see Figure \ref{fig25}) and visually (see Figure \ref{ngc1365}) 
that the pitch angle beyond the large bar is on the high side. In order to visually compare the fit of two logarithmic spirals with different pitch angles, the scaling must be adjusted. According to Equation \ref{eqn5}, the radius of a logarithmic spiral with a higher pitch angle will grow much more rapidly than a logarithmic spiral with a lower pitch angle. In order to allow the radius of the $\phi = 16.5^{\circ}$ logarithmic spiral to grow at the same rate as the three higher pitch angle logarithmic spirals in Figure \ref{ngc1365}, we multiplied its resulting radius by a factor of 2.8. When optimally scaled, our high pitch angle measurement and their low pitch angle measurements can be brought into rough agreement. The difficulty in accurately measuring pitch angle increases as the amount of angular wrapping around a galaxy decreases, i.e., $\theta_{max} \simeq \frac{\pi}{2}$ for NGC 1365 whereas the spirals do not become significantly different until $\theta \apgt \frac{3\pi}{4}$ at the edge of Figure \ref{ngc1365}. Furthermore, NGC 1365 is a well-documented case of variable pitch angle \citep{Ringermacher:2009} with high pitch angle near the arm-bar junction and low pitch angle in the outermost regions of the galaxy. Our code correctly identifies this high pitch angle near the arm-bar junction and provides us with our desired innermost stable pitch angle of the galaxy (see \S\ref{Variable}).

\subsubsection{NGC 3513}

NGC 3513 demonstrates a case where \citet{Ma:2001} measures dramatically different pitch angles between two arms ($18.2^{\circ}$ \& $8.1^{\circ}$). Our measurement of $5.84^{\circ} \pm 1.46^{\circ}$ for the $m = 1$ component is in near agreement with the smaller of Ma's two measurements. From inspection of the image of NGC 3513, it appears to have one arm with near-constant pitch angle and another, more ambiguous arm, with a clear kink in it. The forced inclusion of the non-logarithmic arm might also be responsible for making Mart\'inez-Garc\'ia's 2-D FFT measurement of $24.2_{-0.7}^{+1.7}$ degrees even higher than Ma's individual measurement of the arms.

\subsection{Pitch Angle - SMBH Relation}\label{Pitch-SMBH}

Strong evidence suggests that SMBHs reside
in         the         nuclei         of         most         galaxies
\citep{Kormendy:Richstone:1995,Kormendy:Gebhardt:2001}.  Additionally,
it has recently been shown that a correlation exists between the pitch
angle   of   spiral   arms    and   SMBH   mass   in   disc   galaxies
\citep{Seigar:2008}.   The relation  is such  that more  massive SMBHs
reside in galaxies with low  pitch angle spiral arms (i.e., those that
are tightly wound)  and the least massive SMBHs  are found in galaxies
with high  pitch angle spiral arms.   Given that a significant fraction of galaxies in the
Universe have spiral  or barred spiral morphologies \citep{Buta:1989},
we  wish to  improve upon  existing methods  for measuring  spiral arm
pitch angle in order to  quantify their structure. This measure can in
turn be used to estimate the central black hole mass.

One of the current widely used relationships to SMBH mass is stellar velocity  
dispersion              of              the             bulge/spheroid
\citep{Gebhardt:2000a,Ferrarese:Merritt:2000}. This technique requires
spectroscopy of  the galactic nucleus. Pitch  angle determination only
requires optical imaging. Compared to simple optical imaging, which is
widely and  readily available,  spectroscopy is time  intensive. Other
methods such  as Reverberation Mapping  \citep{Gebhardt:2000b} require
long-term  campaigns   to  obtain  multi-epoch   spectra  and  require
significant telescope  time and allocation. Even  black hole estimates
from    single-epoch    spectra    \citep{Vestergaard:2002}    require
spectroscopy.  Other  techniques such  as  bulge luminosity  estimates
\citep{Kormendy:1993,Kormendy:Richstone:1995,Magorrian:1998,Marconi:Hunt:2003,Haring:Rix:2004}
require  bulge decomposition. One  specific bulge  luminosity estimate
incorporates   S\'ersic  Index  \citep{Sersic:1963}   measurements  of
elliptical galaxies and  the bulges of disc galaxies  and relates them
to SMBH mass \citep{Graham:Driver:2007}.


\subsection{Evolution of Pitch Angle with Redshift}\label{Redshift}

No  matter  how  a  logarithmic  spiral  is  scaled,  pitch  angle  is
unaffected. This allows pitch  angle measurements for distant galaxies
to  be considered  equally  valid  as those  for  local galaxies.  For
distant  galaxies,  as long  as  spiral  arms  are detectable,  it  is
possible to measure a pitch  angle. Unlike other methods, details such
as distance,  extinction, etc.  do not  need to be  known in  order to
measure  pitch angle.  Additionally, the measurement of pitch angle is independent of cosmological assumptions. Upon examining  the GOODS  \citep[Great Observatories
Origins  Deep Survey;][]{GOODS} fields,  we have  identified 224
spiral galaxies with spectroscopic \citep{Barger:2008} and photometric
\citep{Wolf:2004}  redshift  $(z)$  data  for GOODS  North  and  South,
respectively. Of  these 224 spiral  galaxies, 179 galaxies lie  in the
range $z  \leq 1$, 43 galaxies  in the range $1  < z \leq  2$, and two
galaxies with redshift greater than $z  = 2$. So far, we have measured
their pitch angles  using the previous version of  {\it 2DFFT} and are
planning on remeasuring the sample  using the new iterative version of
the code. This work demonstrates that it is not uncommon to be able to
measure pitch  angle for  galaxies beyond a  redshift of  one. Current
work also includes artificially redshifting \citep{FERENGI} this GOODS  
sample of  spiral galaxies  in order to  test the completeness  of the  
GOODS  fields \citep{Shields:2012}. Artificial  redshifting allows  us to  predict  at what
distance spiral  arms are  no longer visible  and thus pitch  angle is
immeasurable. Results thus far show no indications of a relationship between pitch angle and redshift \citep{Shields:2010}, but this matter will be further explored.

\subsection{Continuing Efforts}\label{Continuing}

It is encouraging to notice increasing interest in the measurement of galactic spiral arm pitch angle in the astronomical community, as evidenced even in this past year \citep{Kendall:2011,Martinez-Garcia:2011}. Recent involvement from the computer science community has also been initiated with the use of computer vision techniques to measure galactic spiral arm pitch angle \citep{Davis:Hayes:2012}. We feel that our modification to the previously established Fourier methods for measuring galactic spiral arm pitch angle is a marked improvement and helps to satisfy a growing demand for its rapid and accurate measurement. Furthermore, we are glad to make the code publicly available to the benefit of anyone interested in furthering a much-needed exploration of spiral galaxy structure.



\acknowledgments

The authors gratefully acknowledge support for this work from NASA Grant NNX08AW03A. Data  presented  in this  paper  were  collected  as part  of  the Carnegie-Irvine             Galaxy             Survey            (CGS; \url{http://cgs.obs.carnegiescience.edu}), using  facilities at Las Campanas Observatory, Carnegie Institution for  Science.  The optical data were reduced independently  from those presented  in \citet{Ho:2011}.  We thank Luis Ho and Aaron Barth  for supplying the CGS dataset. We also thank an anonymous referee for comments that helped improve this paper. This  research has  made  use of  the NASA/IPAC Extragalactic  Database (NED) which  is operated by  the Jet Propulsion  Laboratory,  California  Institute  of  Technology,  under contract with the National Aeronautics and Space Administration. This research has made use of NASA's Astrophysics Data System. I.P. thanks the Mexican Foundation Conacyt for grants.

\vspace{15mm}

\bibliography{bibliography}

\clearpage

\begin{turnpage}
\begin{flushleft}
\begin{deluxetable*}{lccrcccclcccccrccr}
\tablecolumns{18}
\tablecaption{Pitch Angle Literature Comparison\label{Compare}}
\tabletypesize{\scriptsize}
\tablehead{
\colhead{} & \colhead{} & 
\multicolumn{4}{c}{This Work} & \colhead{} &
\multicolumn{2}{c}{Ref. 1} & \colhead{} &
\multicolumn{1}{c}{Ref. 2} & \colhead{} &
\multicolumn{1}{c}{Ref. 3} & \colhead{} &
\multicolumn{2}{c}{Ref. 4} & \colhead{} &
\multicolumn{1}{c}{Ref. 5}  \\
\cline{3-6} \cline{8-9} \cline{11-11} \cline{13-13} \cline{15-16} \cline{18-18}
\colhead{Galaxy} & \colhead{} & \colhead{m} & \colhead{$\phi$} & \colhead{Band} & \colhead{Source} & \colhead{} & \colhead{m} & \colhead{$\phi$ (deg.)} & \colhead{} & \colhead{$\phi$ (deg.)} & \colhead{} & \colhead{$\phi$ (deg.)} & \colhead{} & \colhead{$\phi$ (deg.)} & \colhead{Band} & \colhead{} & \colhead{$\phi$ (deg.)}
}
\startdata
M51a & & 2 & $16.26 \pm 3.20$ & B & 3 & & \nodata & \nodata & & 14 & & $16.7, 15.8$ & & \nodata & \nodata & & $15 \pm 2$  \\
NGC 150 & & 2 & $14.29 \pm 4.26$ & B & 1 & & 2 & $17.6_{-2.0}^{+1.3}$ & & \nodata & & $35.0$ & & \nodata & \nodata & & \nodata  \\
NGC 157 & & 3 & $8.66 \pm 0.89$ & B & 1 & & \nodata & \nodata & & \nodata & & \nodata & & \nodata & \nodata & & $19 \pm 4$  \\
NGC 210 & & 2 & $-15.81 \pm 3.25$ & B & 1 & & 2 & $15.7_{-0.5}^{+0.5}$ & & \nodata & & \nodata & & \nodata & \nodata & & $11 \pm 2$  \\
NGC 289 & & 5 & $19.71 \pm 1.95$ & B & 1 & & 2 & $17.2_{-1.8}^{+1.3}$ & & \nodata & & \nodata & & \nodata & \nodata & & $11 \pm 2$  \\
NGC 578 & & 3 & $16.51 \pm 1.88$ & B & 1 & & 2 & $23.0_{-1.3}^{+0.9}$ & & \nodata & & \nodata & & \nodata & \nodata & & \nodata  \\
NGC 598 & & 2 & $-33.90 \pm 5.72$ & 6450 \AA\tablenotemark{a} & 2 & & \nodata & \nodata & & \nodata & & \nodata & & \nodata & \nodata & & $31 \pm 5$  \\
NGC 895 & & 2 & $-38.50 \pm 4.77$ & I & 1 & & \nodata & \nodata & & \nodata & & \nodata & & \nodata & \nodata & & $22 \pm 3$  \\
NGC 1042 & & 4 & $39.50 \pm 4.48$ & R & 1 & & \nodata & \nodata & & \nodata & & \nodata & & \nodata & \nodata & & $13 \pm 3$  \\
NGC 1097\tablenotemark{b} & & 2 & $15.80 \pm 3.62$ & I & 1 & & \nodata & \nodata & & \nodata & & $10.6$ & & \nodata & \nodata & & $17 \pm 4$  \\
NGC 1187 & & 4 & $-21.96 \pm 3.61$ & B & 1 & & 2 & $19.4_{-1.0}^{+1.1}$ & & \nodata & & \nodata & & \nodata & \nodata & & $14 \pm 4$  \\
NGC 1232 & & 3 & $-25.71 \pm 5.43$ & B & 1 & & \nodata & \nodata & & \nodata & & \nodata & & \nodata & \nodata & & $15 \pm 4$  \\
NGC 1300 & & 2 & $-12.71 \pm 1.99$ & B & 1 & & 2 & $13.1_{-0.3}^{+7.7}$ & & \nodata  & & $12.1, 11.0$ & & \nodata & \nodata & & \nodata  \\
NGC 1365 & & 2 & $-34.81 \pm 2.80$ & B & 1 & & \nodata & \nodata & & \nodata & & $13.8, 17.8$ & & \nodata & \nodata & & $18 \pm 3$  \\
NGC 1398 & & 4 & $19.61 \pm 3.07$ & V & 1 & & \nodata & \nodata & & \nodata & & \nodata & & \nodata & \nodata & & $6 \pm 2$  \\
NGC 1566 & & 2 & $-17.81 \pm 3.67$ & B & 1 & & \nodata & \nodata & & 20 & & $19.1, 14.0$ & & \nodata & \nodata & & $22 \pm 2$  \\
NGC 2442 & & 2 & $14.95 \pm 4.20$ & V & 1 & & \nodata & \nodata & & \nodata & & \nodata & & \nodata & \nodata & & $28 \pm 3$  \\
NGC 2835 & & 3 & $-23.97 \pm 2.22$ & B & 1 & & \nodata & \nodata & & \nodata & & \nodata & & \nodata & \nodata & & $20 \pm 2$  \\
NGC 3198 & & 2 & $18.20 \pm 10.01$ & R & 3 & & \nodata & \nodata & & 15 & & \nodata & & \nodata & \nodata & & \nodata  \\
NGC 3223 & & 4 & $-10.92 \pm 2.17$ & B & 1 & & \nodata & \nodata & &  \nodata & & \nodata & & $-8.7 \pm 0.5$ & B & & \nodata  \\
NGC 3261 & & 6 & $15.38 \pm 0.71$ & B & 1 & & 1 & $10.2_{-0.4}^{+20.1}$ & & \nodata & & \nodata & & \nodata & \nodata & & $13 \pm 2$  \\
NGC 3513 & & 1 & $5.84 \pm 1.46$ & B & 1 & & 2 & $24.2_{-0.7}^{+1.7}$ & & \nodata & & $18.2, 8.1$ & & \nodata & \nodata & & \nodata  \\
NGC 3783 & & 2 & $10.71 \pm 0.64$ & B & 1 & & \nodata & \nodata & & \nodata & & \nodata & & \nodata & \nodata & & $7 \pm 1$  \\
NGC 3938 & & 3 & $-22.37 \pm 7.21$ & B & 3 & & \nodata & \nodata & & 15 & & \nodata & & \nodata & \nodata & & $12 \pm 3$  \\
NGC 4030 & & 3 & $23.48 \pm 5.76$ & B & 1 & & \nodata & \nodata & & \nodata & & \nodata & & \nodata & \nodata & & $12 \pm 2$  \\
NGC 4321 & & 5 & $21.81 \pm 3.57$ & R & 3 & & \nodata & \nodata & & 20 & & $21.0, 14.3$ & & \nodata & \nodata & & $15 \pm 3$  \\
NGC 4930 & & 3 & $30.29 \pm 3.45$ & B & 1 & & 1 & $13.9_{-2.0}^{+0.9}$ & & \nodata & & \nodata & & \nodata & \nodata & & \nodata  \\
NGC 4939 & & 6 & $11.48 \pm 1.71$ & B & 1 & & \nodata & \nodata & & \nodata & & $8.1, 10.8$ & & \nodata & \nodata & & $11 \pm 2$  \\
NGC 4995 & & 2 & $13.00 \pm 2.88$ & B & 1 & & 2 & $78.3_{-8.3}^{+5.4}$ & & \nodata & & \nodata & & \nodata & \nodata & & \nodata  \\
NGC 5085 & & 2 & $-11.32 \pm 1.77$ & 4680 \AA\tablenotemark{c} & 4 & & \nodata & \nodata & & \nodata & & \nodata & & $-10.9 \pm 0.5$ & B & & $12 \pm 2$  \\
NGC 5236 & & 6 & $-16.04 \pm 1.74$ & B & 1 & & \nodata & \nodata & & \nodata & & \nodata & & \nodata & \nodata & & $16 \pm 2$  \\
NGC 5247 & & 2 & $-31.94 \pm 5.75$ & B & 1 & & \nodata & \nodata & & \nodata & & \nodata & & $-27.4 \pm 0.7$ & B & & $28 \pm 4$  \\
NGC 5483 & & 2 & $-22.98 \pm 4.52$ & B & 1 & & 2 & $21.6_{-0.9}^{+1.4}$ & & \nodata & & \nodata & & \nodata & \nodata & & \nodata  \\
NGC 5861 & & 2 & $-14.91 \pm 0.83$ & V & 1 & & \nodata & \nodata & & \nodata & & $13.6, 12.7$ & & $-11.9 \pm 0.5$ & V & & \nodata  \\
NGC 6221 & & 6 & $-27.18 \pm 2.14$ & B & 1 & & 2 & $19.9_{-0.7}^{+1.6}$ & & \nodata & & \nodata & & \nodata & \nodata & & \nodata  \\
NGC 6300 & & 4 & $-16.58 \pm 1.52$ & B & 1 & & 2 & $20.3_{-1.1}^{+0.6}$ & & \nodata & & \nodata & & \nodata & \nodata & & \nodata  \\
NGC 7083 & & 3 & $-19.44 \pm 3.21$ & B & 1 & & \nodata & \nodata & & \nodata & & \nodata & & $-15.0 \pm 1.0$ & B & & \nodata  \\
NGC 7793 & & 2 & $13.91 \pm 4.40$ & B & 1 & & \nodata & \nodata & & 16 & & \nodata & & \nodata & \nodata & & \nodata  \\
\enddata
\tablecomments{Col. (1) galaxy name; col. (2) dominant harmonic mode found by this work; col. (3) pitch angle from this work; col. (4) waveband/wavelength used by this work; col. (5) telescope/survey imaging source used by this work; col. (6) ``Fourier Method" harmonic mode used by \citet{Martinez-Garcia:2011}; col. (7) B-band ``Fourier Method" pitch angle absolute value from \citet{Martinez-Garcia:2011}; col. (8) NIR and optical average $m = 2$ pitch angle absolute value from \citet{Kendall:2011}; col. (9) B-band pitch angle angle absolute value for individual arms from \citet{Ma:2001}; col. (10) $m = 2$ pitch angle from \citet{Grosbol:Patsis:1998}; col. (11) waveband used by \citet{Grosbol:Patsis:1998}; and col. (12) $H\alpha$ $m = 2$ pitch angle absolute value from \citet{Kennicutt:1981}. Source (1) CGS; source (2) Palomar 48 inch Schmidt; source (3) KPNO 2.1 m CFIM; and source (4) UK 48 inch Schmidt. Ref. (1) \citet{Martinez-Garcia:2011}; ref. (2) \citet{Kendall:2011}; ref. (3) \citet{Ma:2001}; ref. (4) \citet{Grosbol:Patsis:1998}; and ref. (5) \citet{Kennicutt:1981}.}
\tablenotetext{a}{103aE emulsion.}
\tablenotetext{b}{In addition to spiral arms in the disc of the galaxy, NGC 1097 displays rare $m = 2$ nuclear spiral arms in the bulge. These arms display an opposite chirality to the disc arms with $\phi = -30.60^{\circ} \pm 2.68^{\circ}$.}
\tablenotetext{c}{IIIaJ emulsion.}
\end{deluxetable*}
\end{flushleft}
\end{turnpage}

\end{document}